%% file: main.tex
\documentclass[12pt, letterpaper]{article}
\usepackage{amsfonts}

\usepackage[utf8]{inputenc}
\usepackage{amsmath}
\usepackage{bbm}
\usepackage{graphicx}
\usepackage{setspace}
\usepackage{natbib}
\usepackage{verbatim}
\usepackage{clipboard}

\usepackage[pdftex, 
				pdfauthor={Dominic Smith, Raphael Schoenle, and Sergio Ocampo},
                pdftitle={Robust Measures of Inflation},
                pdfsubject= {Inflation},
                pdfkeywords = {Inflation} {Core Inflation}{Median Inflation}{Trimmed Mean Inflation},
                pagebackref=true]{hyperref}
\renewcommand*\backref[1]{\ifx#1\relax \else (Cited on #1) \fi}

\usepackage{xcolor}
\usepackage{listings}
\usepackage[]{threeparttable}
\usepackage{multirow}
\usepackage{tikz}
\usepackage{siunitx}
\usepackage{csvsimple}
\usepackage{breqn}
\usepackage{enumitem}
\usepackage{caption}
\usepackage{subcaption}
\usepackage[section]{placeins}
\usepackage{nicefrac}
\usepackage{lipsum}
\usepackage{mathtools}
\usepackage{chngcntr}
\usepackage{url}
\usepackage{minitoc}
\usepackage[top=1.05in, bottom=1.0in, left=1.0in, right=1.0in, heightrounded]{geometry}
\usepackage[toc,page,header]{appendix}
\usepackage[normalem]{ulem}

\setcitestyle{authoryear, open={((},close={)}}
\emergencystretch=\maxdimen
\definecolor{red}{RGB}{252,141,89}
\definecolor{blue}{RGB}{101,134,153}%
\definecolor{cite_blue}{RGB}{51,102,153}

\hypersetup{
    colorlinks=true,
    linkcolor={cite_blue}, %
    citecolor={cite_blue}, %
    urlcolor={cite_blue}   %
}

\title{
    \vspace{-1cm}
Extending the Range of Robust PCE Inflation Measures\thanks{
    We appreciate feedback from Christian Bustamante, Stephen Cechetti, Gustavo Joaquim, Rory McGee, Luba Petersen, Robert Rich, Adam Shapiro, Randal Verbrugge, Steve Williamson, and Saeed Zaman.  
    We are also thankful to participants in various conferences and seminars, particularly those at the SGE Session at the 2024 ASSA meetings and the CEBRA Inflation Drivers and Dyanamics series.
    A previous version of this paper circulated as ``How Robust Are Robust Measures of PCE Inflation?''  First draft: May 2022.}
    }
\author{ \onehalfspacing
    Sergio Ocampo\thanks{
        Email: \url{socampod@uwo.ca}; website: \url{https://sites.google.com/site/sergiocampod/}.} 
    \\  {\small University of Western  Ontario }
    \and 
    Raphael Schoenle\thanks{
        Email: \url{schoenle@brandeis.edu}; website: \url{https://people.brandeis.edu/\~schoenle/}.
        } 
    \\ {\small Brandeis  University}
    \and 
    Dominic A. Smith\thanks{
        Email: \url{smith.dominic@bls.gov}; website: \url{https://www.bls.gov/pir/authors/smith.htm}.
        }
    \\ {\small Bureau of Labor  Statistics}
    }
\date{February 2025 \vspace{-0.5cm}} 
\begin{document}

\maketitle
\thispagestyle{empty}

\begin{abstract}
\onehalfspacing 
    \noindent
    Robust inflation measures gauge inflation behavior by excluding volatile expenditure categories from headline inflation.
    We evaluate the forecasting performance of a wide set of such measures between 1970 and 2024, including core, median, and trimmed mean personal-consumption-expenditure (PCE) inflation. 
    Core inflation performs significantly worse than official median and trimmed mean inflation.
    Among a set of alternative trimmed mean measures, there is no single best trim based on forecasting performance: 
    A wide set of trims generates statistically indistinguishable average errors. 
    Nonetheless, different trims imply different predictions for trend inflation in any given month, within a range of 0.5 to 1 percentage points. 
    In tracking trend inflation, this range and its midpoint outperform all trimmed mean inflation measures, suggesting the use of the \emph{range} of inflation implied by the set of \emph{near-optimal trims} as a valuable complement to any single inflation measure.

    

\end{abstract}

\setcounter{page}{0}
\clearpage

\input{numbers.tex}

\input{Sections_2.0/01n_introduction}

\input{Sections_2.0/02m_trimmed_mean}

\input{Sections_2.0/03m_Evaluation}

\input{Sections_2.0/04m_Additional_Trimmed_Mean}

\input{Sections_2.0/05m_Conclusions}

\clearpage
\singlespacing
\bibliographystyle{jpe} %
\bibliography{bibliography.bib}

\clearpage
\appendix

\counterwithin{figure}{section}
\counterwithin{table}{section}
\counterwithin{equation}{section}

\FloatBarrier

\input{Appendix/a01m_Replication}

\input{Appendix/a02m_Add_Figures}

\end{document}

%% file: numbers.tex
\newclipboard{results}

\input{Figures/01m-numbers}
\input{Figures/02m-numbers}
\input{Figures/26m-numbers}
\input{Figures/31m-numbers}
\input{Figures/36m-numbers}

\Copy{min_range_eqv_RMSE}{0.55}
\Copy{max_range_eqv_RMSE}{1.20}

%% file: Figures/01m-numbers.tex
\Copy{trim_cats}{177}
\Copy{median_cats}{200}

%% file: Figures/02m-numbers.tex
\Copy{ss_cats}{182} %
\Copy{ss_cats_nohousing}{178} %

%% file: Figures/26m-numbers.tex
\Copy{sign_match_core_pce}{74} %
\Copy{sign_match_trimmed_mean}{57} %
\Copy{sign_match_median_pce}{67} %

%% file: Figures/31m-numbers.tex
\Copy{range_infl_measures}{0.49}
\Copy{range_infl_low}{0.47}
\Copy{range_infl_med}{0.49}
\Copy{range_infl_high}{0.54}
\Copy{range_pre_2020}{0.48}
\Copy{range_post_2020}{0.67}
\Copy{range_infl_long}{0.49}
\Copy{max_mad_measures}{2.20}
\Copy{mad_2021_and_2022}{0.70}
\Copy{inflation_gt_5pp}{178}
\Copy{range_pre_2020_core_yoy}{0.41}
\Copy{range_post_2020_core_yoy}{0.56}
\Copy{range_infl_long_core_yoy}{0.43}

%% file: Figures/36m-numbers.tex
\Copy{range_90_10}{1.32}
\Copy{range_75_25}{0.45}

%% file: Sections_2.0/01n_introduction.tex
\section{Introduction}

The return of high and volatile inflation after the 2020 pandemic has renewed attention toward inflation measures that serve as indicators of current and future inflation trends.
The three most common inflation measures used for this purpose are ``core'' personal consumption expenditure (PCE) inflation, trimmed mean PCE inflation \citep{DallasTrimmed}, and median PCE inflation \citep{CleMedian}.
Core inflation excludes a fixed set of historically volatile categories, such as food and energy, from headline inflation, while the other two measures exclude expenditure categories with the highest and lowest inflation rates before computing mean inflation, varying the set of excluded categories every month. 
These measures are referred to as \textit{robust inflation measures} and are commonly used by central bankers to communicate the behavior of inflation to the public.\footnote{
    For instance, the president of the Federal Reserve Bank of Richmond has stated that ``\textit{I want to see inflation, and median and trimmed mean, compellingly headed back to our target}'' \citep{Bloomberg_1,Bloomberg_2}.
    The \citet{BoC} and the \citet{Riksbank} list trimmed mean and median inflation as part of their preferred measures of ``\textit{underlying}'' inflation, as do other central banks.
    }

In this paper, we contribute to understanding the behavior of robust inflation measures during high-inflation regimes when gauging trend inflation is most important for central bankers.
We do so by re-evaluating the forecasting performance of the previously mentioned robust inflation measures between 1970 and 2024 against standard benchmarks \citep[see,][]{bryan1994measuring,Clark2001,Dolmas2005}, and by further extending our analysis to alternative trimmed mean measures.
Our sample includes several episodes of high-inflation, including those in the 1970s and the recent post-pandemic inflation.

Several insights emerge when we evaluate the forecasting performance of a wide set of such robust measures of inflation.\footnote{
    There are alternative robust measures of inflation obtained from dynamic factor models \citep*[see, among others,][]{ForniHallinLippiReichlin,Euro_Core_Factor_JMCB_2005,Stock_Watson_Restat_2016}.
    See \citet{AmstadPotterRich} for a comparison to trimmed mean measures of inflation.
    }
First, we find that \emph{trimmed mean and median inflation clearly outperform core inflation} in terms of their root-mean-square error (RMSE) against standard measures of current and future trend inflation.\footnote{
    The RMSE of core inflation against current trend inflation is 21 to 65 percent higher than that of trimmed mean and median inflation, respectively. 
    }  
This finding might seem surprising given communication practices and public attention toward core inflation---the series is often referenced over alternative measures by Federal Open Market Committee (FOMC) officials in speeches \citep{Powell_Speech_2022} and reports to congress \citep{FOMC_Congress_Report_2024}.
However, these results are in line with, and extend to recent high inflation, previous work evaluating core PCE and CPI inflation that also find trimmed mean and median inflation series to be superior \citep*[see, among others,][]{Detmeister_2012_Core_Useful, Ball_et_al_Core_Inflation_2021, Verbrugge_2022_Core_Bad}.\footnote{
    \citet*{Mester2013} find core, median, and trimmed mean inflation to be no better than using headline inflation in linear forecasting models as in \citet{Blinder_Reis_2005}. 
    However, central banks' communication has centered on the raw level of robust inflation series as an indicator for trend inflation and not on their usefulness as inputs for other forecasting models, which motivates our methodology.
    }
Consequently, we focus on the properties of trimmed mean inflation measures in our subsequent analysis, rather than core inflation.

Second, we find \emph{there is no single best robust measure in terms of predictive performance}.
We arrive at this result after constructing alternative trimmed mean inflation series by systematically varying the shares of expenditure with the highest and lowest inflation being excluded, as in \citet{bryan1997efficient}, \citet{Dolmas2005}, \citet{Meyer_Venkatu_2012}, and \citet{Zaman_Cleveland_2013}.
We rank these series based on their forecasting performance, comparing their RMSEs with measures of current and future trend inflation.

Our findings cast a nuanced picture: 
The official median and trimmed mean measures perform nearly as well as the best of the alternative robust measures across periods and trend targets, even though the best trims vary significantly depending on the sample---and whether the current or future inflation trend is targeted.
For instance, the best trim when targeting future trend inflation over the extended 1970 to 2024 sample is much more aggressive, trimming 76 percent of expenditure, than in any subsamples.
By contrast, in the more recent 2000-2024 period, it is optimal to trim only 52 percent of expenditure.\footnote{
    For reference, median inflation trims all but one expenditure category and the official trimmed mean series removes 55 percent of expenditure.
    }
Nevertheless, the RMSEs of the official robust measures and the best of the alternative trimmed mean measures are  statistically indistinguishable in most samples.

Moreover, an important third insight emerges when we look beyond predictive performance: 
\textit{Similarities in predictive performance obscure economically significant differences in the behavior of the series in any given period and across time}. 
The range between the \emph{levels} of the \emph{official} robust inflation series is wide, on average \Paste{range_pre_2020} percentage points before 2020, growing to  \Paste{range_post_2020} percentage points from 2020 onward. 
This finding implies that these measures often provide different signals about current and future inflation in any given month, while tracking the trend equally well only on \emph{average}. 
For example, 12-month median inflation fell from 2.5 to 2.0 percent between March 2020 and February 2021, while trimmed mean inflation was almost flat, and core inflation rose from 1.5 to 1.8 percent. 
\textit{These three series provided different signals about the path of inflation at exactly the time that inflation was heating up.}
The following month, 12-month headline inflation exceeded the FOMC's target of 2 percent and has yet to recover to that level.

This pattern also holds when we broaden the set of robust inflation measures to include \emph{alternative} trimmed mean measures chosen based on their average predictive performance. 
There is no single best series based on predictive performance.
On the contrary, \emph{there is a wide range of trims with statistically the same performance}. 
Moreover, the range of levels for the resulting robust inflation measures remains wide among the measures whose RMSEs against trend inflation are statistically equivalent to the RMSE of the best trims---between \Paste{min_range_eqv_RMSE} and \Paste{max_range_eqv_RMSE} percentage points.

\begin{figure}[tb]
    \centering
    \caption{Range of Robust Measures of Inflation, 2020--2024}
    \includegraphics[width=0.65\textwidth]{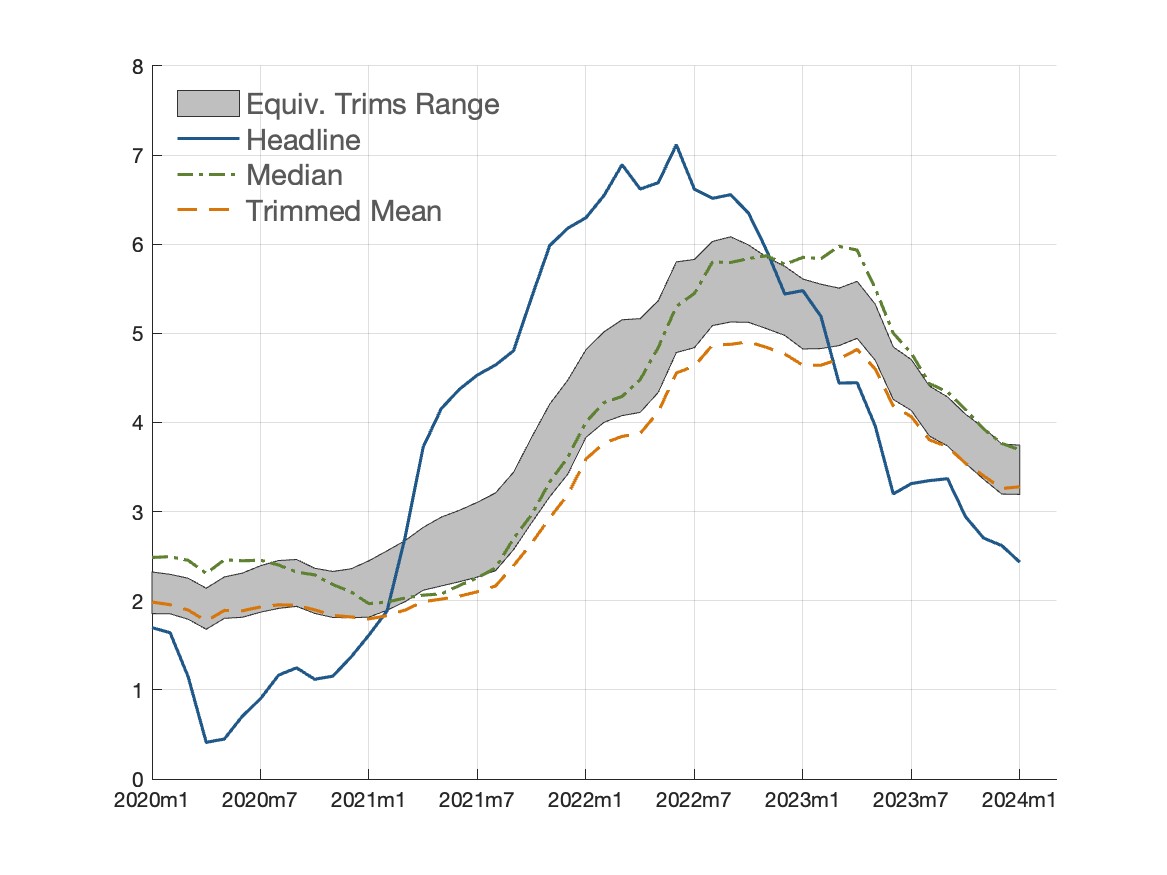}
    \caption*{{\footnotesize\textit{Notes:} 
    The figure shows the range of inflation implied by the set of trimmed mean inflation measures whose root-mean-square error with respect to current trend inflation are statistically equivalent to the best trim at the 5 percent level.
    The figure also includes trimmed mean personal consumption expenditure (PCE) and median PCE as calculated by the authors using the methodologies of \citet{CleMedian} and \citet{DallasTrimmed}, 
    as well as headline PCE inflation.
    }}
    \label{fig:robust_range_2020_2024}
\end{figure}

Nevertheless, \emph{this range is informative about the path of inflation}.
For instance, as we show in Figure \ref{fig:robust_range_2020_2024}, the range of equivalent trims indicated flat inflation from March to November 2020.
By February 2021 its lower and upper bounds had increased by 10 to 20 basis points and proceeded to increase steadily.
The range increased more quickly than median inflation during the first six months of rising inflation after the pandemic, and remained above the official trimmed mean throughout.
The range was also faster to decrease, doing so one month after headline inflation peaked in June 2022---while median and trimmed mean inflation remained high and stable for about ten months before changing their trend. 
This pattern holds throughout the sample. 
In fact, this range (and even its midpoint) provides a better predictor of trend inflation over the whole sample than even the best single trimmed mean inflation series, with an RMSE that is  15 percent smaller than that of the best trim.

Overall, while our results may appear negative, in that \emph{no single} best robust measure emerges, a positive and practically important message also arises:
Communication about inflation trends can be complemented by reporting the \emph{range} of inflation implied by the set of best trimmed mean measures, in addition to the level of any one series. 
This range provides an effective way of communicating information from measures of trend inflation, especially in times of heightened volatility when individual measures conflict with each other.  
Communicating the range of inflation implied by the best trimmed mean measures also aligns with the experimental findings of \citet{Petersen_Kryvtsov_Uncertainty_2023} who show that providing a range of inflation estimates anchors inflation expectations as much as providing a single point estimate, while more effectively reducing the probability assigned by households to very high and very low inflation outcomes.\footnote{
    Communicating inflation and inflation targets affects inflation expectations and monetary policy. 
    See, \citet*{Coibion_Gorodnichenko_Weber_JPE_2022}, \citet{Amy_Handlan_JMP_2022}, and \citet*{Gorodnichenko_et_al_AER_2023} for recent applications and the review by \citet*{Blinder_et_al_JEL_2008}.
    }%

Finally, a fourth insight coming out of our analysis is that \emph{trimmed mean inflation measures are more informative about the state of current trend inflation than about future trend inflation}. 
In fact, the RMSE of the best trimmed mean measure is at least twice as large when predicting future trend inflation (relative to current trend inflation) and the sets of trims with the best forecasting performance are much larger.\footnote{
    This result from our extended sample aligns with the finding that trimmed mean measures are better at nowcasting than forecasting without further adjustments \citep*{Verbrugge2022}.
    }
These results on the performance of individual robust inflation measures speak to their difficulty in capturing changes in inflation trends, made apparent in our extended sample containing several of these episodes in its earlier and later periods.
They suggest that robust inflation measures may be of more practical use for tracking  current than future inflation trends. %

Relative to the literature, our results for alternative trimmed mean measures extend those of \citet{Zaman_Cleveland_2013} and \citet{Meyer_Venkatu_2012} for Consumer Price Index (CPI) to PCE inflation going back to 1960 and update them forward to 2024, including three additional high-inflation episodes. 
We focus on PCE inflation for two main reasons:
First, it is the series targeted by the Federal Reserve when setting monetary policy.
Second, unlike the CPI, the historical PCE series are regularly revised to follow a consistent methodology.

Our analysis also goes beyond studying predictive performance, highlighting implications for the levels of robust inflation series.
The papers cited above examine trimmed mean measures of CPI inflation going up to 2013 and also find wide sets of trims with equivalent forecasting performance.
However, and unlike them, we find these sets  to be asymmetric and to change substantially over time for PCE inflation, even if we restrict the sample to 1970--2013 to cover a similar period.
The sets of best trims skew towards higher upper trims, particularly for the most recent sample covering 2000--2024.
The asymmetry means that the median inflation series is only part of the set of best trims for some samples.
The changing results across targets and samples lead us to conclude that there are no basis for selecting a single series among the PCE trimmed mean inflation measures (implying a range of inflation), unlike \citet{Meyer_Venkatu_2012} who find the median CPI inflation to perform well throughout among other CPI measures and, therefore, argue for its use.

Finally, we also provide a new public good by extending the official trimmed mean and median PCE inflation measures back to 1960 as part of constructing and evaluating robust inflation series.\footnote{
    Our series are available at \url{https://ocamp020.github.io/Robust_Inflation_Series.xlsx}.
    }
These series are currently published starting in 1977, covering only one high inflation episode before 2020; 
a key data limitation.
Consequently, work studying the forecasting properties of robust inflation measures has been constrained to samples in which inflation has been consistently low and stable, missing both the episodes of high inflation in the 1960s and 70s and the recent episode of post-pandemic inflation \citep*[see, among others,][]{Dolmas2005, Rich2007, Mester2013, Meyer_Zaman_2019_BVAR_Median_CPI}.
We therefore complement the literature by analyzing the performance of a wide range of trimmed mean inflation measures over this extended sample.

%% file: Sections_2.0/02m_trimmed_mean.tex
\section{Extending Robust PCE Inflation Measures}

This section details the computation of headline, core, median, and trimmed mean PCE inflation series beginning in 1959 and ending in February 2024, extending the official median and trimmed mean inflation series to cover 1960--77. 
We extend these series replicating the methodologies of the Federal Reserve Banks of Dallas and Cleveland \citep*[see,][]{Bryan1991,bryan1994measuring,Dolmas2005,Carroll2019}.\footnote{
    Currently available official series begin in 1977 to coincide with the introduction of personal computers to the PCE. 
    }
This extension increases the number of months in the sample by almost 40 percent relative to the conventionally available official series, providing us with two additional episodes of rising inflation (1968 and 1973) and covering \Paste{inflation_gt_5pp} months of high inflation (above 5 percent).

Extending the official robust inflation measures to the 1960--77 period is of interest for two reasons:
First, it provides a more consistent view of the patterns for these robust inflation measures in periods of rising and high inflation. 
The 1960--77 period provides us with two additional episodes of rising inflation (1968 and 1973), adding to the four episodes of rising inflation in the post-1977 period. %
There are only three episodes between 1960 and 2024 for which headline PCE inflation is above 5 percent, covering a total of \Paste{inflation_gt_5pp} months, 44 of which fall into the 1960--77 period. %
Second, we can use these series as reference points when we evaluate the predictive performance of a wide set of inflation series in the next section.

In extending the sample, we also document changes in the number of expenditure categories available in PCE data and the frequency of their price updates (see Appendix \ref{agg:replication}). 
We find that prices were not updated regularly for up to 20 percent of expenditure categories prior to 1970, which leads to focus most of our analysis on the period from then onward. 
We also provide a time-consistent set of categories for calculating alternative trimmed mean measures, harmonizing differences in the construction of the official trimmed mean and median inflation series. 

\input{Sections_2.0/02p_data}

\input{Sections_2.0/02q_measures_description}

\input{Sections_2.0/02r_inflation_series}

%% file: Sections_2.0/02p_data.tex
\subsection{PCE Inflation Data}\label{sec:data}

Our analysis is based on the underlying data supplements of the National Income and Product Accounts PCE data release \citep{PCE} between January 1959 and February 2024 to construct robust inflation series.\footnote{ 
    The main input for BEA price indexes is the CPI, but unlike the CPI, the PCE is revised for historical consistency when methods change enabling our consistent backwards extensions. 
    Disaggregated PCE series can be accessed at \url{https://apps.bea.gov/national/Release/XLS/Underlying/Section2All_xls.xlsx}.
    By contrast, extending the median and trimmed mean CPI series is of limited use because the individual CPI component series are not revised when the methodology used to calculate inflation for individual components is changed.
    Changes in methodology can imply significant changes in the volatility of inflation as shown by \citet*{Hazell_Herreno_Nakamura_Steinsson_QJE} and \citet*{Summers_et_al_CPI_2022}.
    } 
The PCE data provide disaggregated price indexes and expenditure weights that cover US consumer spending.

Two main issues accompany the use of these data and the replication of official robust inflation measures. 
First, new expenditure categories are introduced over time, as is the case with expenditure on personal computers, a category introduced in 1977.
The changes in the set of expenditure categories are reflected in differences between the set of categories used in the construction of official trimmed mean inflation and that used for median inflation.
We harmonize the set of expenditure categories by establishing a consistent set of \Paste{ss_cats} series that are available for the entire period or as soon as a new good is introduced.\footnote{ 
    The set of \Paste{ss_cats} categories we use differs from the \Paste{trim_cats} categories used by the Dallas Fed and \Paste{median_cats} categories used by the Cleveland Fed.
    When new expenditure categories are introduced, they often have almost-zero spending, as they represent new goods. 
    In those cases, we assume that the goods they represent were not available before their introduction in the PCE, and we assign to them a retroactive weight of zero.
    The complete list of the series included in each inflation measure can be found in \url{http://dominic-smith.com/data/category_definitions.xlsx}.
    In a previous version of this paper, we established that these differences do not impact the results \citep{Robustv1}. 
    }
Second, the price series for multiple expenditure categories were not updated on a monthly basis before 1970, including owners' equivalent rent, the category with the highest weight (see Figure \ref{fig:nonmissing} in Appendix \ref{agg:replication}). 
This issue is reflected in the levels of the median inflation series, as we show in Figure \ref{fig:robust_measures}.
We therefore focus our analysis in Sections \ref{sec:Evaluation} and \ref{sec:Opt_Trim} on the 1970--2024 period.

%% file: Sections_2.0/02q_measures_description.tex
\subsection{Headline and Core PCE Inflation}

Headline inflation is calculated as a Fisher index of the underlying inflation components at the lowest level of aggregation. 
A Fisher index is the geometric mean of a Laspeyres and a Paasche index, which are calculated respectively as 
\begin{align}
    \Pi_{t}^{L} = \frac{\sum_{i} p_{t}^i q_{t-1}^i }{\sum_{i} p_{t-1}^i q_{t-1}^i }; \quad
    \Pi_{t}^{P} = \frac{\sum_{i} p_{t}^i q_{t}^i }{\sum_{i} p_{t-1}^i q_{t}^i },
\end{align}
where $p_t^i$ and $q_t^i$ are, respectively, the price level and quantity of expenditure category $i$ at time $t$.\footnote{
    Throughout the paper we will focus on inflation over a 12-month period, but plot each series at a monthly frequency. 
    Using inflation over 12 months prevents seasonality from complicating the analysis.
    }
Core PCE inflation is computed in the same way, but excludes all series under food and beverages purchased for off-premises consumption, gasoline and other energy goods, and electricity and gas.
We take headline and core PCE inflation directly from the tables published by the BEA.\footnote{ 
    We use series DPCERG for headline inflation and series DPCCRG  for core inflation \citep{PCE}. 
    In Appendix \ref{app: Housing}, we study the impact of removing housing services from inflation measures. 
    Doing this is similar to proposed measures of ``super-core'' inflation.
    }

\subsection{Trimmed-Mean PCE Inflation Measures}

Trimmed-mean PCE inflation measures select a set of expenditure categories in each month by removing the categories with the lowest inflation rates that represent $\alpha$ percent of expenditure and the categories with the highest inflation rates accounting for $\beta$ percent of expenditure.
Trimmed-mean measures are characterized by these cutoffs and not by the expenditure categories included, which vary every month.
The categories included in a given month are assigned weights using an average of the expenditure on each category at current-period quantities and previous-period quantities, which approximates the weights used in the PCE index formula,
\begin{align}
    \omega_{t}^{i} = \frac{1}{\alpha+\beta}\left(\frac{1}{2}\frac{ p_{t-1}^{i}q_{t-1}^{i} }{ \sum_{i} p_{t-1}^{i}q_{t-1}^{i} } + \frac{1}{2}\frac{ p_{t-1}^{i}q_{t}^{i} }{ \sum_{i} p_{t}^{i}q_{t}^{i} }\right).
\end{align}

The trimmed mean inflation series is the expenditure-weighted mean across the selected categories, where the weights are adjusted to reflect the fact that $\alpha+\beta$ percent of expenditure has been trimmed out. 
Once the monthly gross rates, $\Pi_{t}^{tm,mo}$, are constructed, they are chained to form a yearly inflation index,
\begin{align}
    \Pi_t^{tm} = \prod_{s=0}^{11}\Pi_{t-s}^{tm,mo},\quad  \Pi_{t}^{tm,mo}=\sum_i \omega_t^i \frac{p_{t}^{i}}{p_{t-1}^{i}}.
    \label{eq:Trimmed_Mean_Inflation}
\end{align}
The official trimmed mean inflation measure sets $\alpha=24$ and $\beta=31$, it is published monthly by the Federal Reserve Bank of Dallas \citep{Dolmas2005}. 
It is available from 1977 onward because 1977 is the years personal computers were introduced to the PCE \citep{DallasTrimmed}.

The median inflation series is calculated by trimming out all categories except the one with the median inflation rate in every month (that is, $\alpha=\beta=50$). %
The chained index is
\begin{align}
    \Pi_t^{m} = \prod_{s=0}^{11} \frac{p_{t-s}^{i(m,t-s)}}{p_{t-s-1}^{i(m,t-s)}},
    \label{eq:Median_Inflation}
\end{align}
where $i(m,t-s)$ is the index of the series with the median inflation at time $t-s$.
The official median PCE inflation is published monthly by the Federal Reserve Bank of Cleveland \citep{Bryan1991,bryan1994measuring,Carroll2019} and is also available from 1977 onwards \citep{CleMedian}.

%% file: Sections_2.0/02r_inflation_series.tex
\subsection{Inflation Series}

\begin{figure}[tb]
    \centering
    \caption{Robust Measures of Inflation, 1960--2024}
    \includegraphics[width=0.70\textwidth]{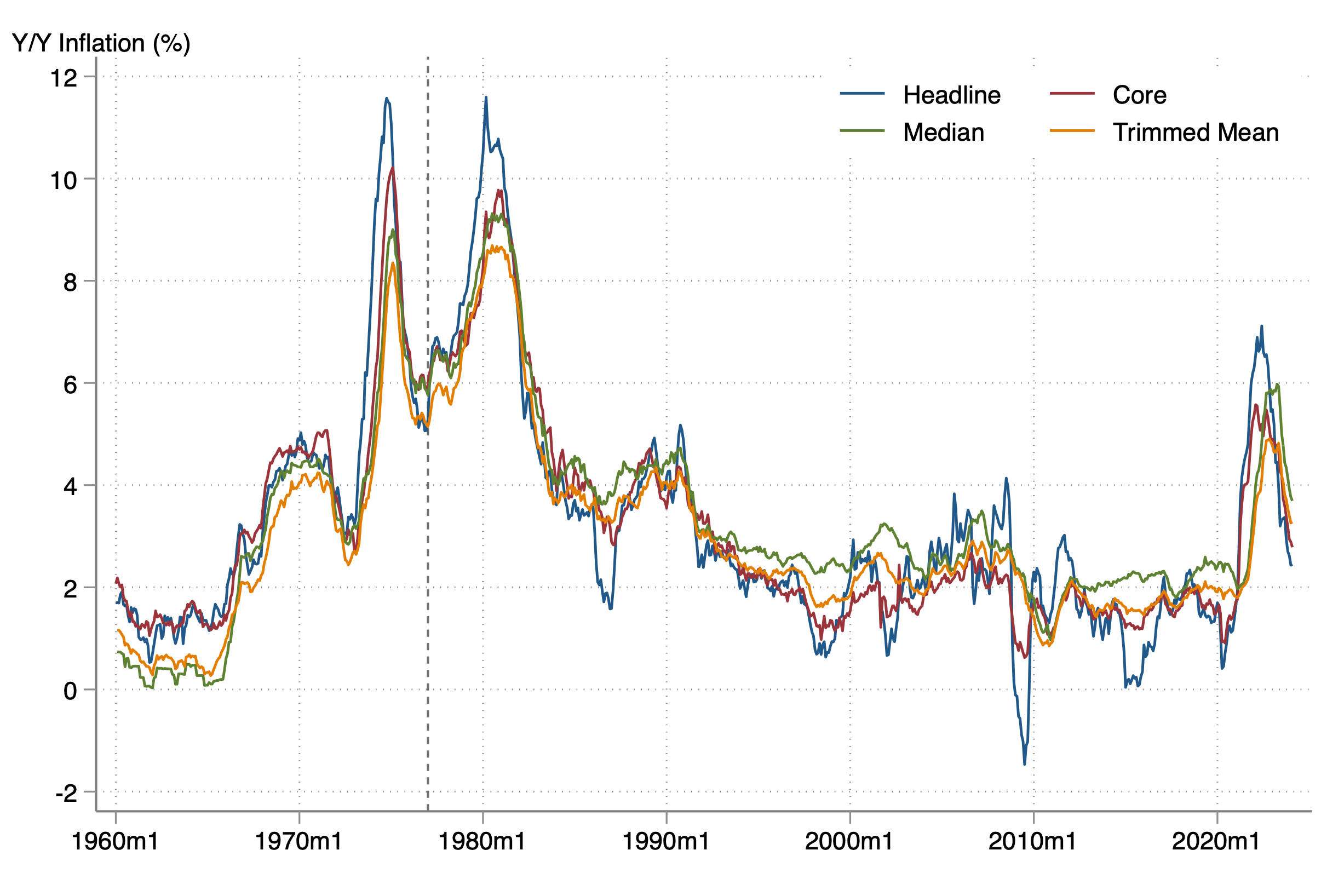}
    \caption*{{\footnotesize\textit{Notes:} 
    The figure shows  trimmed mean personal consumption expenditure (PCE) and median PCE as calculated by the authors using the methodologies of \citet{CleMedian} and \citet{DallasTrimmed}. 
    Appendix \ref{agg:replication} shows that these measures match those produced by the relevant Federal Reserve Banks after 1977. 
    The vertical line in January  1977 indicates that the official trimmed mean and median measures are available starting in 1977.
    Headline and core inflation are taken directly from the PCE data published by the Bureau of Economic Analysis.
    }}
    \label{fig:robust_measures}
\end{figure}

Figure \ref{fig:robust_measures} plots headline, core, trimmed mean, and median PCE inflation. 
We replicate the official trimmed mean and median inflation series after 1977 while also extending them back to 1960 (see Figures \ref{fig:replication_trimmed_mean} and \ref{fig:replication_median} in Appendix \ref{agg:replication}).
These two extended robust inflation measures track the behavior of headline inflation despite significant disagreement between them, which translates into similar predictive performance as we show in the next section. 
The average range between the levels of the robust inflation series is \Paste{range_infl_measures} percentage points. 
These differences between the series increase at the end of the sample, with the average range increasing to \Paste{mad_2021_and_2022} percentage points between 2021 and 2024 (see Figure \ref{fig:robust_measures_min_max} in Appendix \ref{app:Additional_Figures}).\footnote{
    There are other disagreements. 
    For instance, the sign of the change (increasing or decreasing) of core PCE inflation matches that of headline PCE in \Paste{sign_match_core_pce} the months in our sample; the values for median and trimmed mean PCE inflation are \Paste{sign_match_median_pce} and \Paste{sign_match_trimmed_mean} months, respectively.
    }
As the next section makes clear, this pattern of a similar predictive performance (over time) paired with economically significant differences in month-to-month levels is common across a wide range of trimmed mean measures.

\begin{figure}[tb]
    \centering
    \caption{Range of Underlying Inflation, 1960--2024}
    \includegraphics[width=0.70\textwidth]{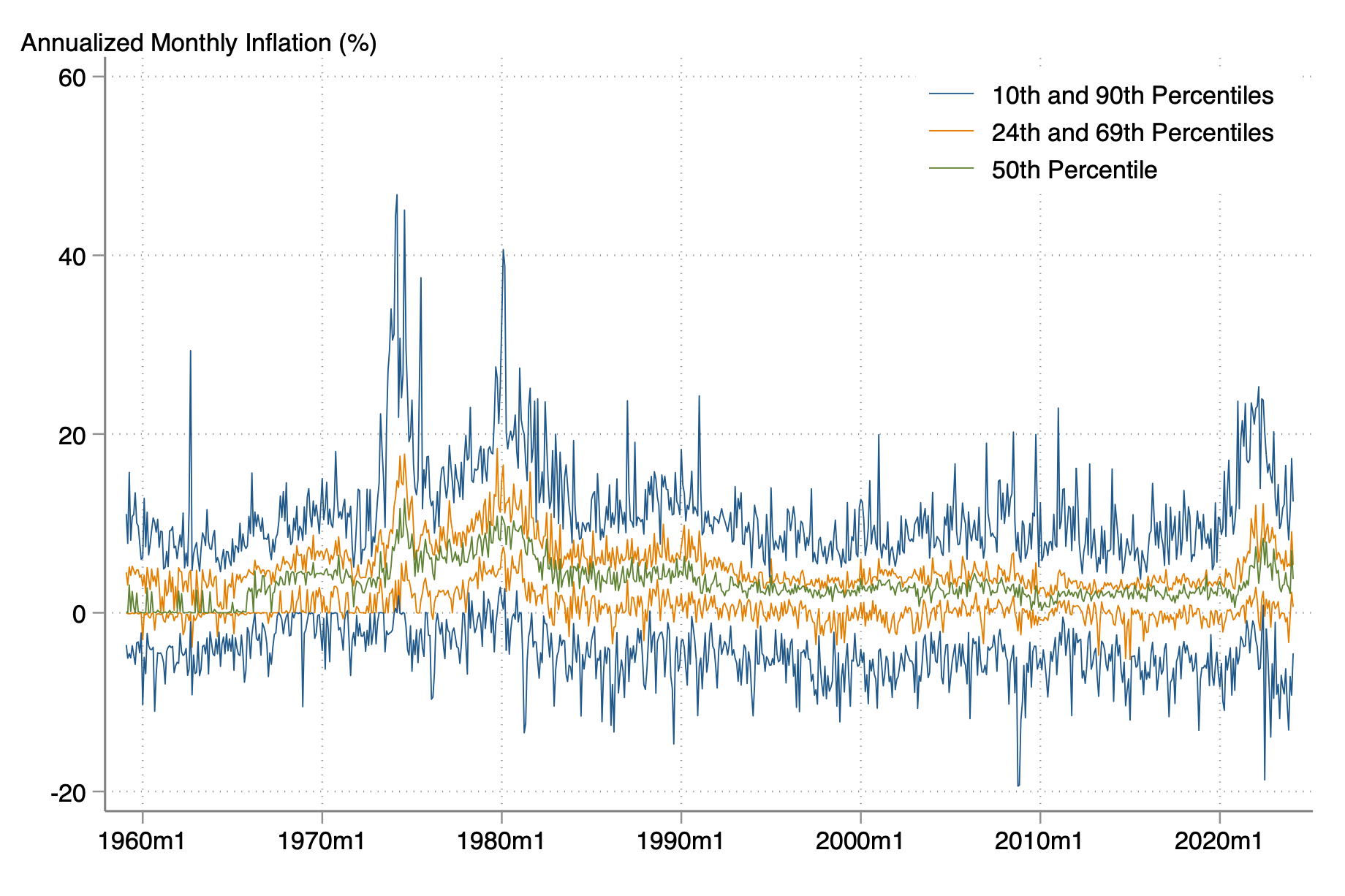}
    \caption*{{\footnotesize\textit{Notes:} 
    The figure shows the authors' calculations of the range of  inflation series used for different inflation measures from 1960 to 2024.
    The lines correspond to the 10th and 90th percentiles of the cross section of monthly inflation rates in the 177 series considered for the trimmed mean measure, the 24th and 69th percentiles of the PCE inflation series that correspond to the range used for trimmed mean inflation, and the median inflation series.
    Percentiles are weighted using the average real expenditure on each category in months $t$ and $t-1$.
    }}
    \label{fig:robust_measures_pct}
\end{figure}

The differences between the inflation measures come from the underlying set of expenditure categories used to compute them. 
Trimming categories narrows the range of inflation rates used to construct robust inflation measures. 
Figure \ref{fig:robust_measures_pct} illustrates this by plotting the 10th, 24th, 50th, 69th, and 90th percentiles of month-to-month inflation across expenditure categories.
The 24th and 69th percentiles correspond to the trims of the official trimmed mean inflation series, and the 50th percentile to the trim of the official mean inflation series.

The distribution of inflation rates across categories is remarkably wide, with a range between the 10th and 90th percentiles of \Paste{range_90_10} percentage points on average, which makes trimming consequential in many periods, particularly during those with high headline inflation. 
The interquartile range (the range between the 25th and 75th percentiles of the distribution) is \Paste{range_75_25} percentage points on average, of similar magnitude to the range between the levels of the three robust inflation measures in Figure \ref{fig:robust_measures}.

The set of categories which are included in the calculations of the trimmed mean and median inflation measures varies over time, as it depends on the contemporaneous distribution of inflation across expenditure categories. 
However, there are certain regularities.
The categories most commonly excluded are food and fuel categories (for example, eggs, vegetables, gasoline), which are some of the categories excluded from core PCE. 
This set is much larger for median inflation.
In fact, 62 of the \Paste{ss_cats} categories we consider never coincide with the median inflation category, while every category is included in the trimmed mean at some point.
Table \ref{tab:Excluded_Categories} lists the categories most commonly excluded and included in the construction of median and trimmed mean inflation.

%% file: Sections_2.0/03m_Evaluation.tex
\section{Evaluating Official Robust Inflation Measures}\label{sec:Evaluation}

How well do the official robust inflation measures match current and future inflation trends across multiple samples? 
Our analysis shows that trimmed mean and median inflation outperform core inflation across all samples and objectives. 
Moreover, the performance of these two series is similar across the samples we consider, with the official trimmed mean series slightly outperforming median inflation in terms of capturing current- and future-trend inflation.
However, the similar performance of the series obscures significant differences in their monthly levels, which our analysis lays out in full in Section \ref{sec:Opt_Trim}.

\subsection{Evaluation Criteria}

To evaluate the performance of inflation series, we construct ex-post measures of current and future trend inflation, which aim to smooth out the transitory components of inflation.
Then we compare each inflation measure to these target trend measures and calculate the root mean squared error (RMSE).
This is the same criteria used by \citet{Dolmas2005} to select the trimming cutoffs of the official trimmed mean inflation series.
We consider a long sample (1970--2024), an old sample (1970--1989), and a recent sample (starting in 2000).

Our main measure of current trend inflation is a 36-month centered moving average of headline inflation.
Thus, it includes data from 18 months before to 18 months after the current period.
This measure was proposed by \citet{bryan1997efficient} and has been used as a benchmark since. 
It provides an ex-post proxy for what underlying inflation was at a point in time.

Our main measure of future trend inflation is constructed as a 12-month forward moving average of headline inflation that starts one year ahead of the current period, which corresponds to the two-years-ahead annual inflation. 
Thus, it includes data between 12 and 24 months in the future. 
As \citet{Meyer_Venkatu_2012} highlight, future inflation is particularly relevant as decision makers, such as central bankers, are forward-looking and make decisions based on the expected behavior of inflation.
\citet{Zaman_Cleveland_2013} and \citet{Meyer_Venkatu_2012} use the annualized CPI inflation rate over the next 36 months as the main target in their analysis (similar to our alternative forward measure described below) but find similar results when using the 24-months-ahead annualized rate of CPI inflation.

We also construct an alternative measure of current trend inflation using the trend component of \citeauthor{Christiano_Fitzgerald_2003}'s (\citeyear{Christiano_Fitzgerald_2003}) band-pass filter (removing frequencies below 39 months) and an alternative measure of future trend inflation constructed as the 24-month forward-looking average rate of inflation as in \citealp{Dolmas2005}.\footnote{
    The trend component of the band pass filter removes high frequency movement in inflation. \citet{Dolmas2005} found that removing frequencies below 29 months maximizes the correlation between the resulting trend series and the Federal Funds rate target set by the Federal Open Market Committee. 
    }
We plot all these series in Figure \ref{fig:trend_ts} of Appendix \ref{app: Alternative_Trend_Measures}, where we also provide the results of our evaluation of the official series against the alternative trend inflation measures.

Given a target $\bar{\pi}$ (for current or future trend inflation), we evaluate how well an inflation measure $\pi^i$, $i\in\{\text{core, trimmed mean, median}\}$, tracks it over a given sample. 
We do so by calculating the RMSE for each candidate robust measure $i$ as\footnote{
    We use the level of inflation  $\pi^i$ without transforming it. 
    Transformations such as linear regressions improve the fit of robust inflation measures such as median inflation by removing a consistent bias in certain measures \citep{Rich2021}; however, these measures are most often communicated and used without such transformations.
    } 
\begin{align}
    \text{RMSE}^i = \sqrt{\frac{1}{T} \sum_t (\pi^i_{t} - \bar{\pi}_t)^2}. \label{eq: RMSE_Def}
\end{align}

\subsection{Performance of Official Measures}

A clear result emerges from the comparison of the official robust inflation measures (core, trimmed mean, and median) against current and future trend inflation:
Trimmed-mean and median inflation outperform core inflation in all samples regardless of whether we compare them against current or future inflation trends, as Table \ref{tab:prediction} shows. 
For instance, the RMSE of trimmed mean and median inflation is between 20 and 25 percent lower than that of core inflation when targeting current trend inflation in the 1970--2024 sample.\footnote{
    We also compare the performance of robust inflation measures to the performance of headline PCE inflation. %
    Headline inflation corresponds to the no-trimming limit. 
    The RMSE that the official trimmed mean and median inflation series produce is between 50 and 75 percent of the RMSE of headline inflation, verifying that trimming does improve the series' predictive performance.
    } 
Furthermore, these differences are always statistically significant, as we verify in the last column of Table \ref{tab:prediction} where we report the p-value of the \citet{Diebold_Mariano_1995} test under the null hypothesis of no difference in the prediction error between core inflation and the official trimmed mean and median inflation series.

Although these results may be somewhat surprising given the attention core inflation receives in policy discussions and news coverage, they are in line with \citet{Detmeister_2012_Core_Useful}, \citet{ Ball_et_al_Core_Inflation_2021}, \citet{Verbrugge_2022_Core_Bad}, and other studies that also find that core inflation is outperformed by trimmed mean measures using other targets and time periods.\footnote{
    \citet{Mester2013} also find that there is no advantage to core inflation over alternative robust measures in the context of linear prediction models. 
    Unlike us, they do not find trimmed mean and median inflation to be significantly better.
    Their analysis differs from ours in its scope (going only until 2013) and method, as we use the level of the robust inflation measures as predictors of trend inflation in line with the practice of policy makers and previous studies. 
    }
The focus on core inflation started during the Greenspan era as chair of the FOMC \citep[see,][]{Blinder_Reis_2005} and the FOMC started publishing statistics on core PCE inflation next to headline PCE inflation in 2007 \citep{Mester2013}.
The FOMC still makes explicit mention of core inflation (over other measures) in official reports to congress \citep{FOMC_Congress_Report_2024} and in speeches by its chair \citep{Powell_Speech_2022}.
This practice is not shared by other central banks.
For instance, the \citet{BoC} and the \citet{Riksbank} reference versions of the trimmed mean and median inflation measures as their ``preferred'' measures of ``underlying'' inflation.

An additional result arises when comparing robust inflation measures:
The official trimmed mean inflation series slightly outperforms the official median inflation series in capturing the behavior of current and future trend inflation, except when targeting current trend inflation in the restricted sample (1970--89), when the pattern reverses. %
However, the RMSEs produced by the two official measures are not statistically different in most cases (see the second-to-last column of Table \ref{tab:prediction}).  %
The differences between the performances of trimmed mean and median inflation are higher when predicting current trend inflation than when predicting the future trend, but the null hypothesis is only strongly rejected for the recent sample (starting in 2000).
These same patterns hold when looking at alternative trend inflation measures as we show in Table \ref{tab:prediction_All} of Appendix \ref{app: Alternative_Trend_Measures}.

\begin{table}[tb]
    \caption{Predicting Performance of Official Measures}
    \begin{center}
    \begin{threeparttable}
        \begin{tabular}{ll S S S S S}
        \hline \hline
        \multirow{2}{*}{Target} & \multirow{2}{*}{Sample} &  \multicolumn{3}{c}{PCE Inflation Measure RMSE} & \multicolumn{2}{c}{DM Test: $\Pr\left(z>\left|\text{DM}\right|\right)$}\\
         \cline{3-5}\cline{6-7}
         &  & {Core} & {Trimmed Mean} & {Median}  & {TM vs Med.} & {Core vs Other} \\
        \hline 
                         & {1970-2024} & 1.47 & 1.11 & 1.18 & 0.015 & 0.000 \\
        {Current Trend}  & {1970-1989} & 1.84 & 1.62 & 1.52 & 0.088 & 0.001 \\
                         & {2000-2024} & 1.34 & 0.81 & 1.02 & 0.000 & 0.004 \\
        \hline
                         & {1970-2024} & 2.45 & 2.14 & 2.17 & 0.197 & 0.000 \\
        {Future Trend}   & {1970-1989} & 3.35 & 3.02 & 3.02 & 0.903 & 0.000 \\
                         & {2000-2024} & 1.99 & 1.66 & 1.71 & 0.161 & 0.019 \\
        \hline
    \end{tabular}
    \begin{tablenotes}
        \item {\footnotesize \textit{Notes:} 
        The table presents the predictive performance of personal-consumption-expenditure (PCE) inflation measures with respect to different trend inflation targets for different samples.
        Performance is measured by the series' root-mean-square error (RMSE) with respect to trend inflation.
        The table reports the RMSEs for core, trimmed mean, and median inflation. 
        The last two columns report the p-value of the \citet{Diebold_Mariano_1995} test for the difference between the RMSEs of trimmed mean and median inflation and the difference of the core inflation series and the best of the trimmed mean and the median inflation series. 
        }
    \end{tablenotes}
    \end{threeparttable}
    \end{center}
    \label{tab:prediction} 
\end{table}

However, this last result obscures a key insight from our analysis: 
Despite the official trimmed mean and median inflation series' similar prediction errors, their levels differ substantially in most months. %
The average range between the levels of these two series is \Paste{range_infl_long} percentage points over our sample and grows larger by the end of the sample (see Figure \ref{fig:robust_measures_min_max} in Appendix \ref{app:Additional_Figures}).
These differences can result in meaningful discrepancies in the signals that these series provide.
As we showed in Figure \ref{fig:robust_range_2020_2024} in the Introduction, 12-month median inflation was decreasing and trimmed mean inflation was almost flat at the same time that headline inflation was rising above the FOMC's target in the first months of 2021.

This key insight also arises from analyzing alternative robust measures as we do below; namely, relying on the average prediction performance of the series obscures underlying differences in their predictions for any given month.
Further, we see no systematic way to determine which series is providing the best signal at any point in time.

%% file: Sections_2.0/04m_Additional_Trimmed_Mean.tex
\section{Optimal Trimmed-Mean Measures}\label{sec:Opt_Trim}

The analysis in this section systematically varies the trim cutoffs used to construct trimmed mean measures by considering all integer combinations of trims for $\alpha,\beta\in[0,50]$, in a similar way to \citet{Zaman_Cleveland_2013} and \citet{Meyer_Venkatu_2012}.
The resulting set of measures includes the official trimmed mean, median, and headline inflation series as special cases. 
As before, we evaluate their performance based on their RMSEs against current and future trend inflation and contrast their predictive performances with differences in the levels of trend inflation implied by the different measures. 
Appendix \ref{app: Alternative_Trimmed_Mean_Measures} extends the results to alternative measures of trend inflation.

\subsection{Optimal trimming cutoffs}
There are large differences between the optimal trimming cutoffs and the trims of the official measures ($\alpha=24$ and $\beta=36$, or $\alpha=\beta=50$).
The optimal cutoffs are asymmetric, trimming more from the top than from the bottom $(\beta^\star>\alpha^\star)$.
More importantly, they vary significantly across targets and when comparing the most recent sample (2000--24) to the longer and older samples. 
Table \ref{tab:best_trim} presents the results.

However, our results also show that this large variation in optimal trims does not translate into a significantly lower RMSE.
In fact, the official measures perform almost as well as those with the optimal trimming cutoffs. 
The RMSE of the optimal trims is at most 4 percent lower, and the differences are never statistically significant at the 5 percent level.
The last column of Table \ref{tab:best_trim} reports the p-value of the \citet{Diebold_Mariano_1995} test when comparing the predictive performance of the best trim with the better of the two official robust inflation measures (see Table \ref{tab:prediction}).

\begin{table}[tbh!]
    \caption{Best Trims for Trimmed-Mean Inflation}
    \begin{center}
    \begin{threeparttable}
    \begin{tabular}{ll SS S S S}
        \hline \hline
        \multirow{2}{*}{Target} & \multirow{2}{*}{Sample} &  \multicolumn{3}{c}{Best Trims} & \multicolumn{1}{c}{Official Trims} & {DM Test}\\
         \cline{3-5}
         &  &  {Lower} & {Upper} & {RMSE} & {min(RMSE)} & {$\Pr\left(z>\left|\text{DM}\right|\right)$} \\
        \hline 
                         & {1970-2024} & 17 & 19 & 1.08 & 1.11 & 0.065 \\
        {Current Trend}  & {1970-1989} & 18 & 16 & 1.46 & 1.52 & 0.235 \\
                         & {2000-2024} & 18 & 23 & 0.80 & 0.81 & 0.403 \\
        \hline
                         & {1970-2024} & 35 & 41 & 2.11 & 2.14 & 0.213 \\
        {Future Trend}   & {1970-1989} & 15 & 17 & 2.91 & 3.02 & 0.412 \\
                         & {2000-2024} & 24 & 28 & 1.63 & 1.66 & 0.288 \\
        \hline
    \end{tabular}
    \begin{tablenotes}
        \item {\footnotesize \textit{Notes:} 
        The table reports the best trim as determined by the predictive performance across trims for different targets of trend inflation and different samples.   
        The table also reports the root-mean-square error (RMSE) for the best trims and the lower RMSE of the official trimmed and median inflation series. 
        The last column reports the p-value of the \citet{Diebold_Mariano_1995} test for the difference between these RMSEs. %
        }
    \end{tablenotes}
    \end{threeparttable}
    \end{center}
    \label{tab:best_trim} 
\end{table}

\begin{figure}
    \begin{centering}
        
    \caption{RMSE across Trims}\label{fig: RMSE_All}
    
    \begin{subfigure}[t]{0.48\textwidth}
    \caption{Current Trend: 1970-2024}
    \includegraphics[width=1.0\textwidth]{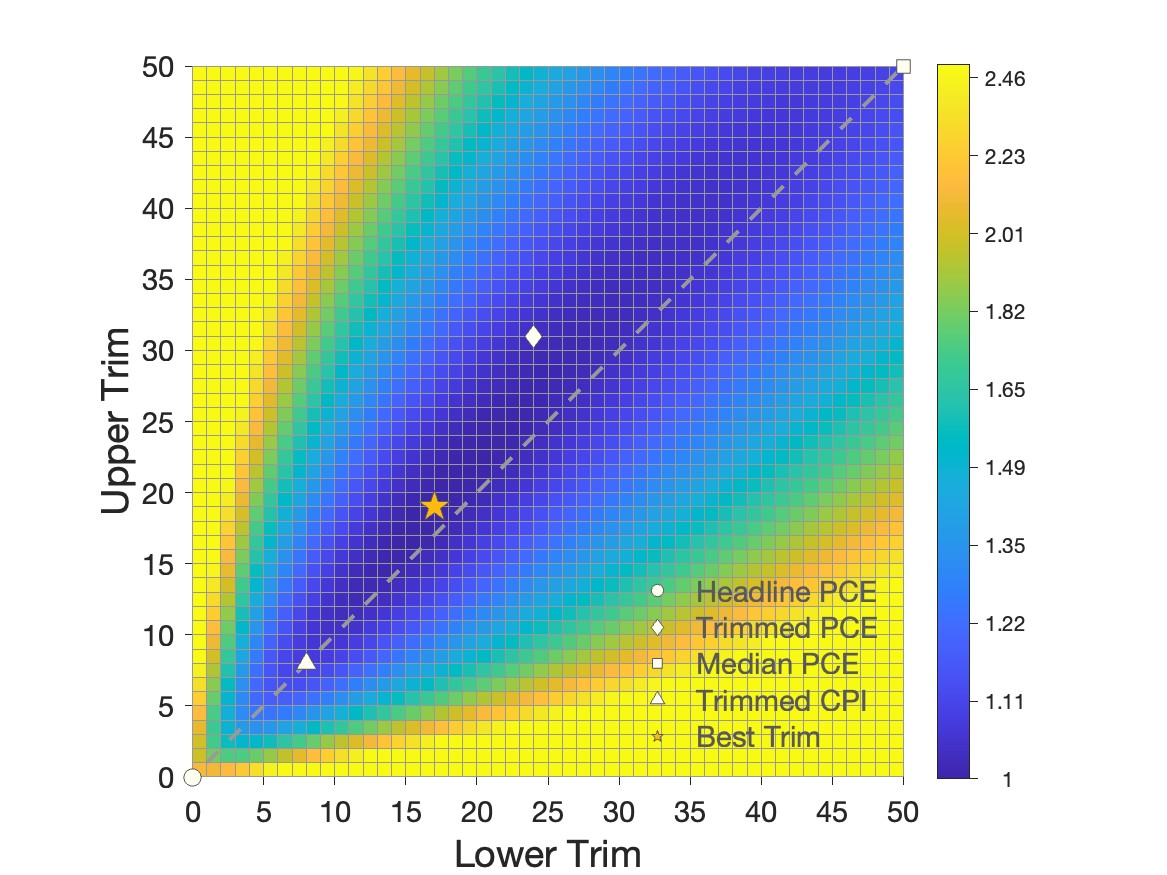} 
    \end{subfigure}
    ~
    \begin{subfigure}[t]{0.48\textwidth}
    \caption{Future Trend: 1970-2024}
    \includegraphics[width=1.0\textwidth]{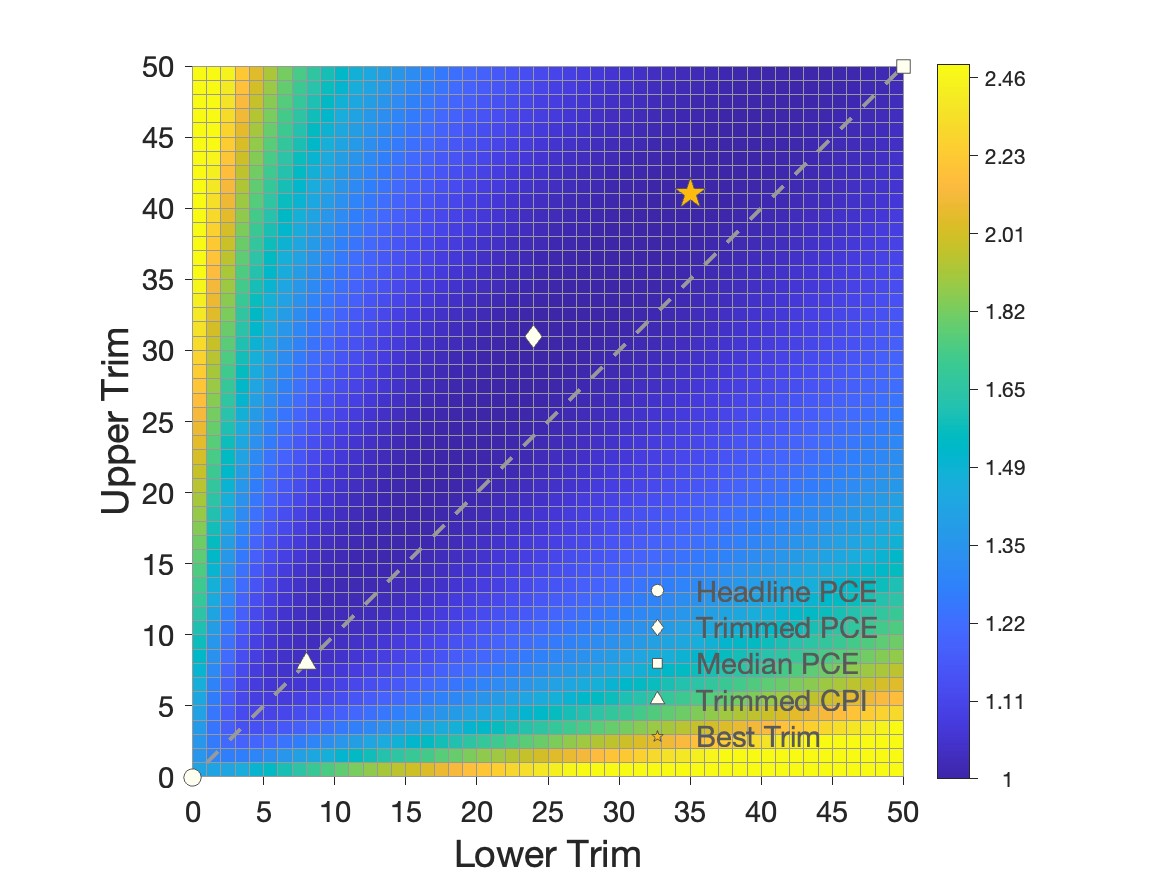} 
    \end{subfigure}
    
    \begin{subfigure}[t]{0.48\textwidth}
    \caption{Current Trend: 1970-1989}
    \includegraphics[width=1.0\textwidth]{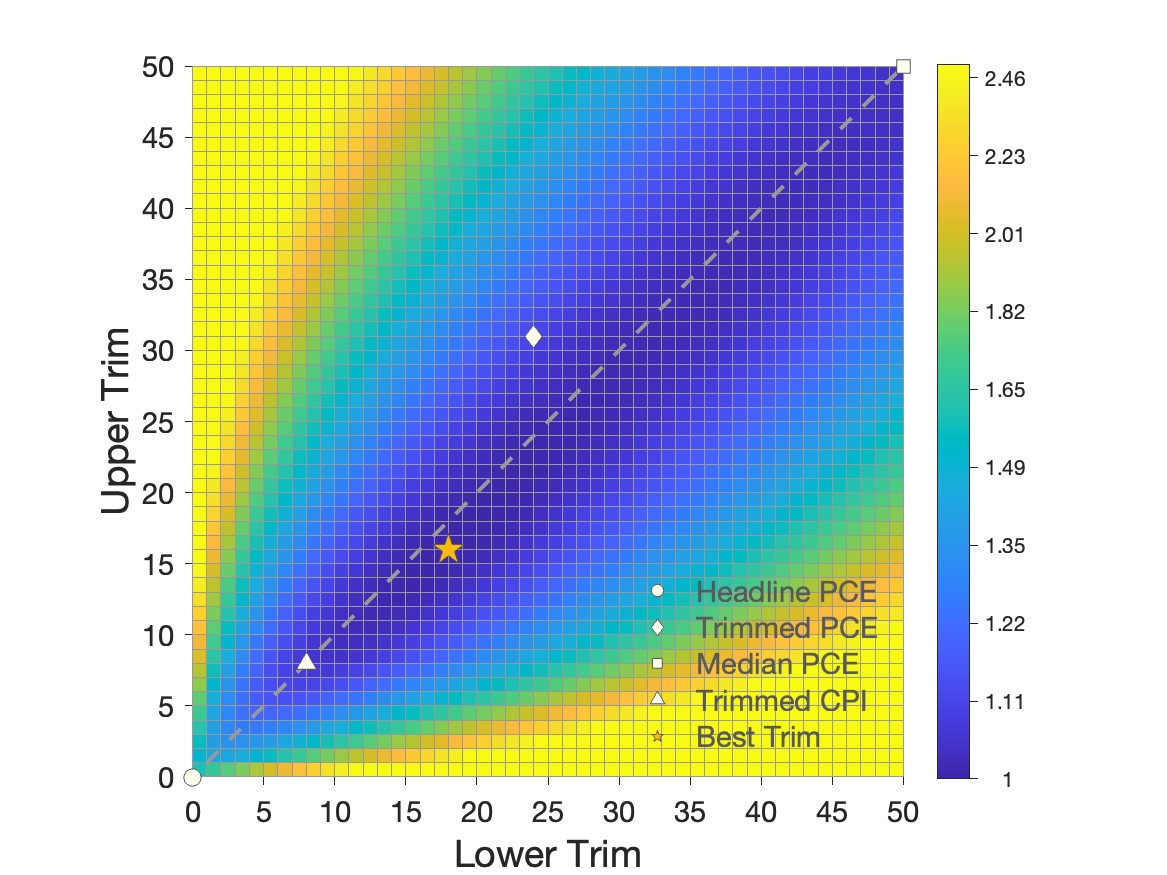} 
    \end{subfigure}
    ~
    \begin{subfigure}[t]{0.48\textwidth}
    \caption{Future Trend: 1970-1989}
    \includegraphics[width=1.0\textwidth]{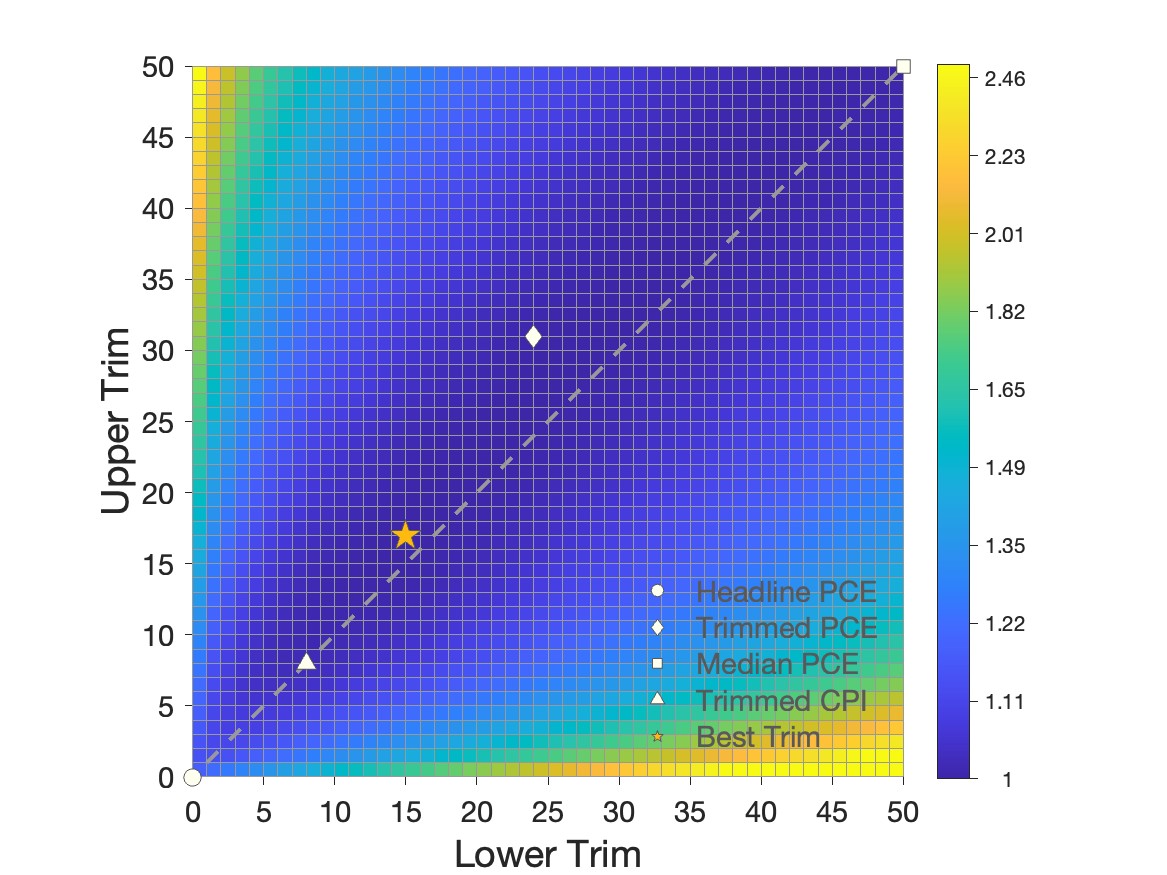}
    \end{subfigure}

    \begin{subfigure}[t]{0.48\textwidth}
    \caption{Current Trend: 2000-2024}
    \includegraphics[width=1.0\textwidth]{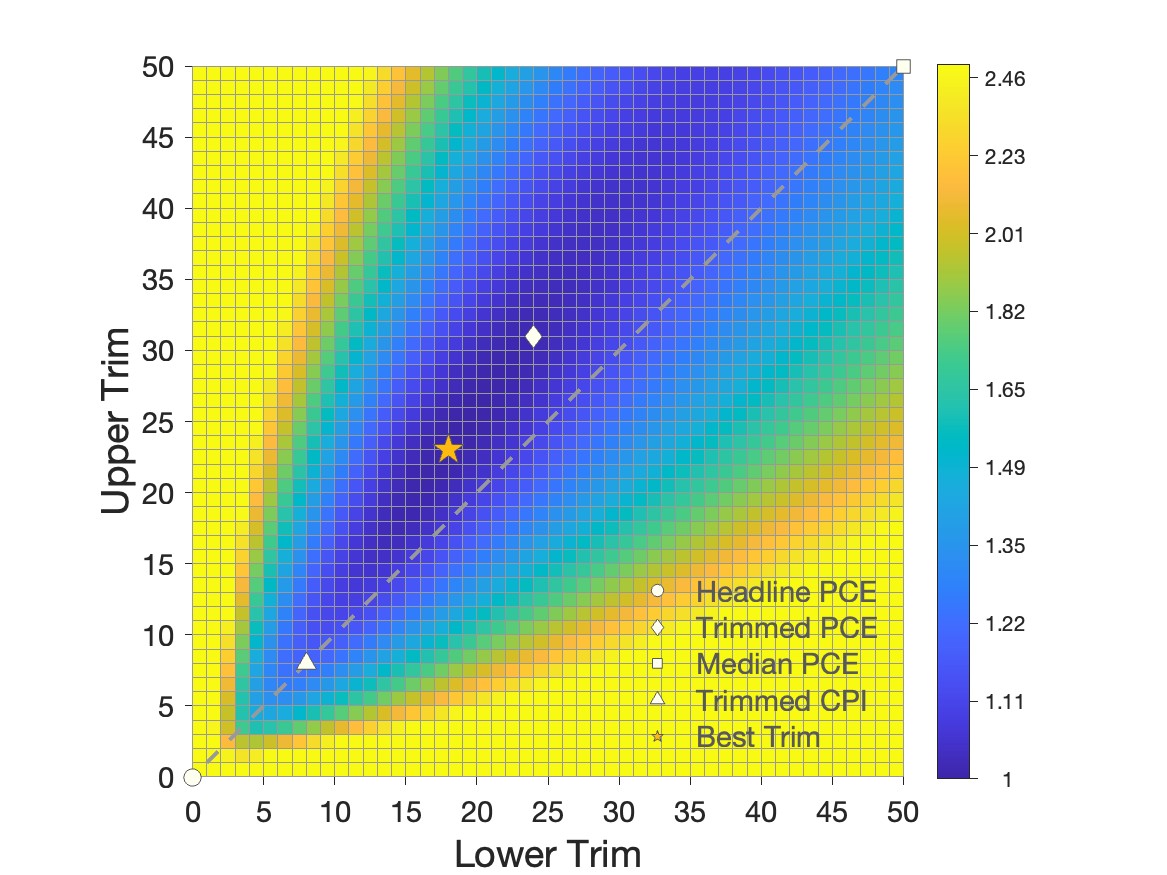}
    \end{subfigure}
    ~
    \begin{subfigure}[t]{0.48\textwidth}
    \caption{Future Trend: 2000-2024}
    \includegraphics[width=1.0\textwidth]{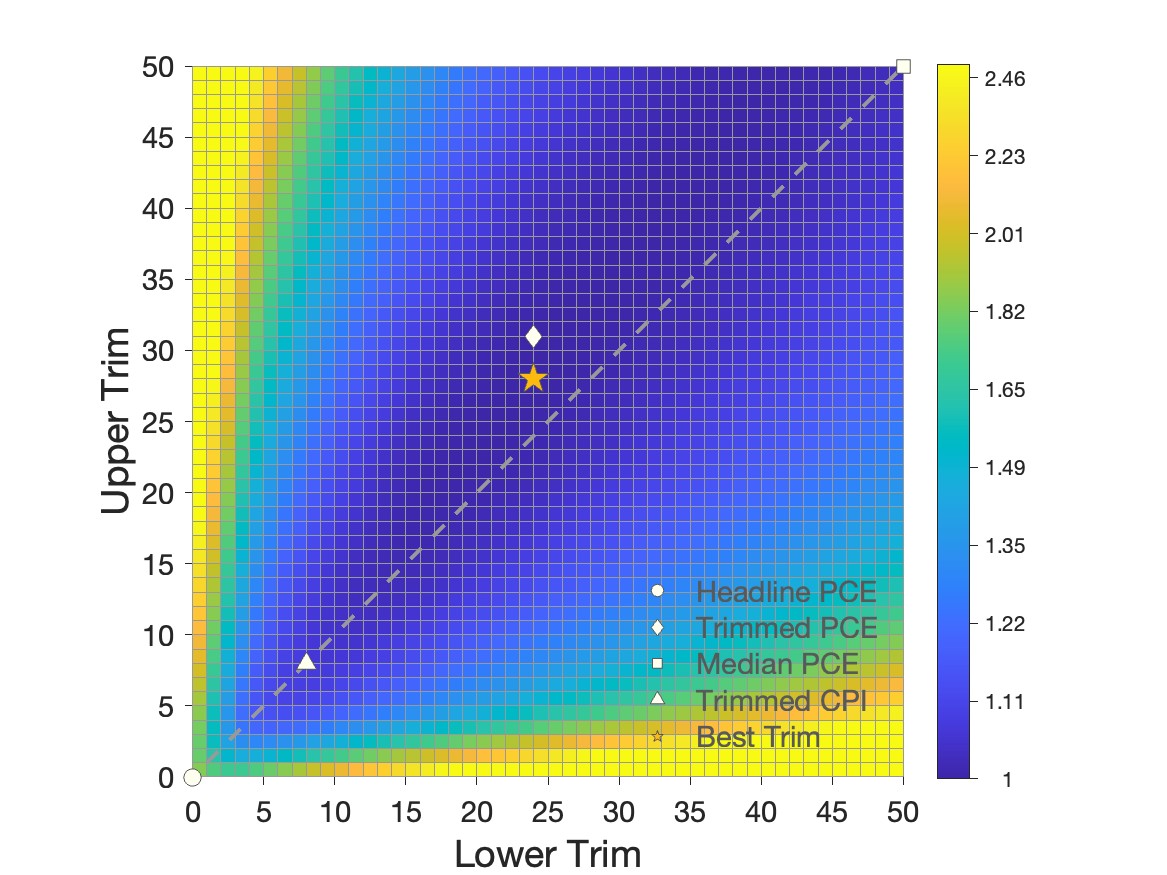}
    \end{subfigure}

    \end{centering}
    
    \vspace{-0.2cm}
    \singlespacing
    \footnotesize{}\textit{Notes:}
    The figures show heat maps of the root-mean-square error (RMSE) when targeting current and future trend inflation with different trimmed mean inflation measures.
    To ensure comparability across plots, the RMSEs are reported relative to the RMSE of the best trim reported in Table \ref{tab:best_trim}.
\end{figure}

\begin{figure}
    \begin{centering}
        
    \caption{Statistical Difference of RMSE across Trims}\label{fig: DM_Test}
    
    \begin{subfigure}[t]{0.48\textwidth}
    \caption{Current Trend: 1970--2024}
    \includegraphics[width=1.0\textwidth]{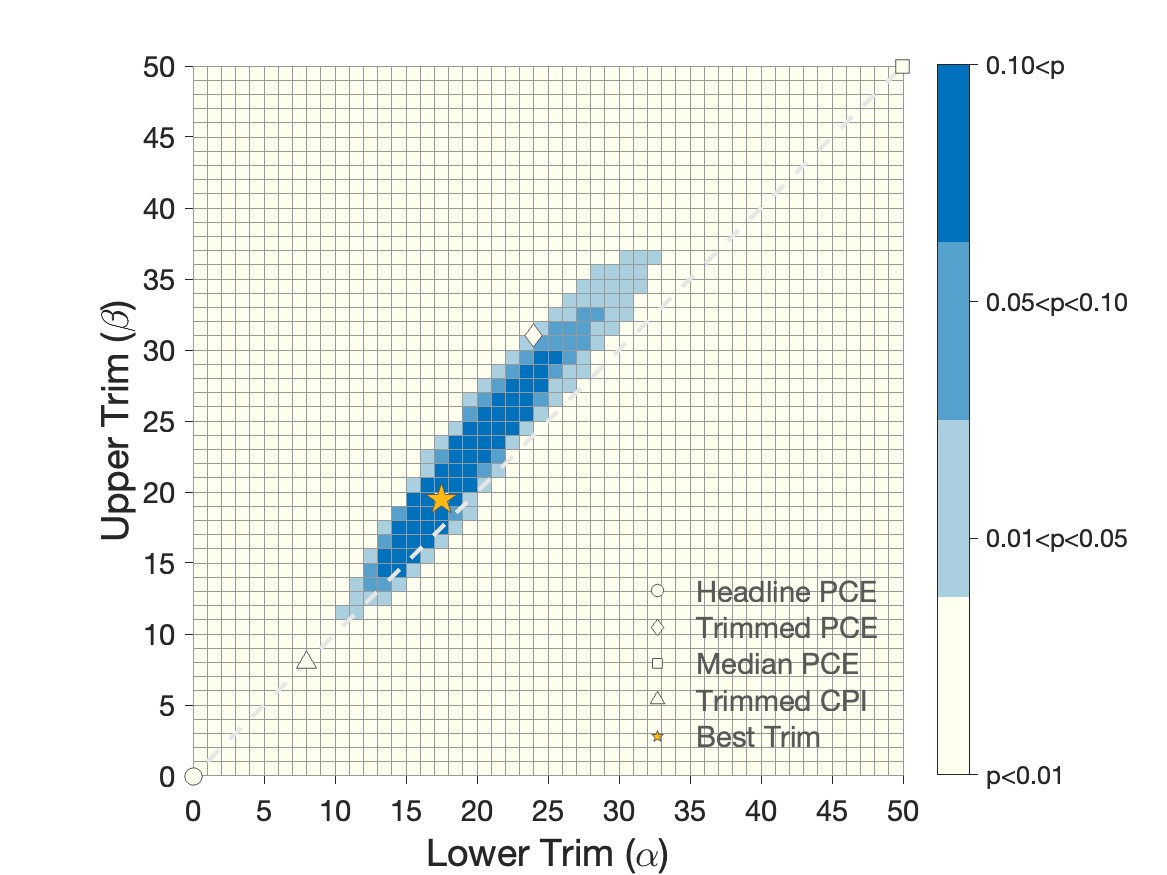}
    \end{subfigure}
    ~
    \begin{subfigure}[t]{0.48\textwidth}
    \caption{Future Trend: 1970--2024}
    \includegraphics[width=1.0\textwidth]{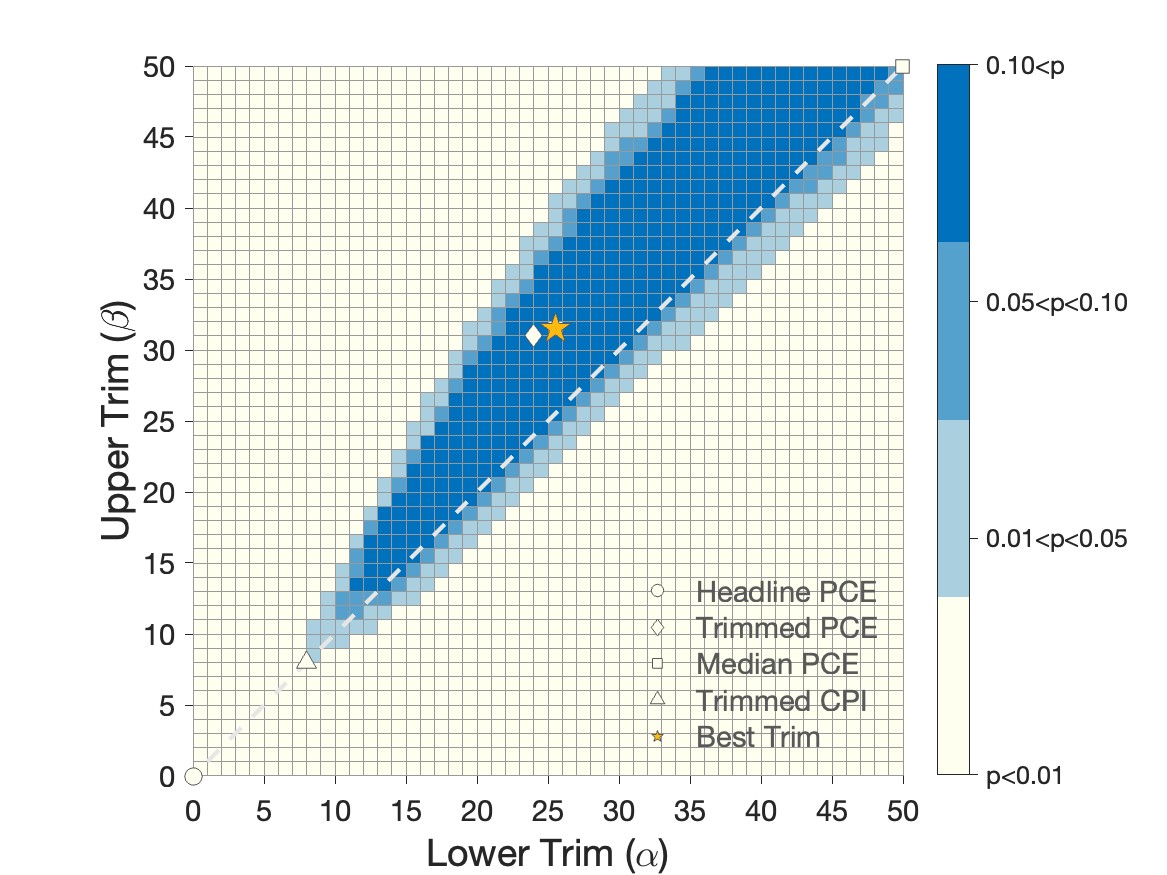}
    \end{subfigure}
    
    \begin{subfigure}[t]{0.48\textwidth}
    \caption{Current Trend: 1970--89}
    \includegraphics[width=1.0\textwidth]{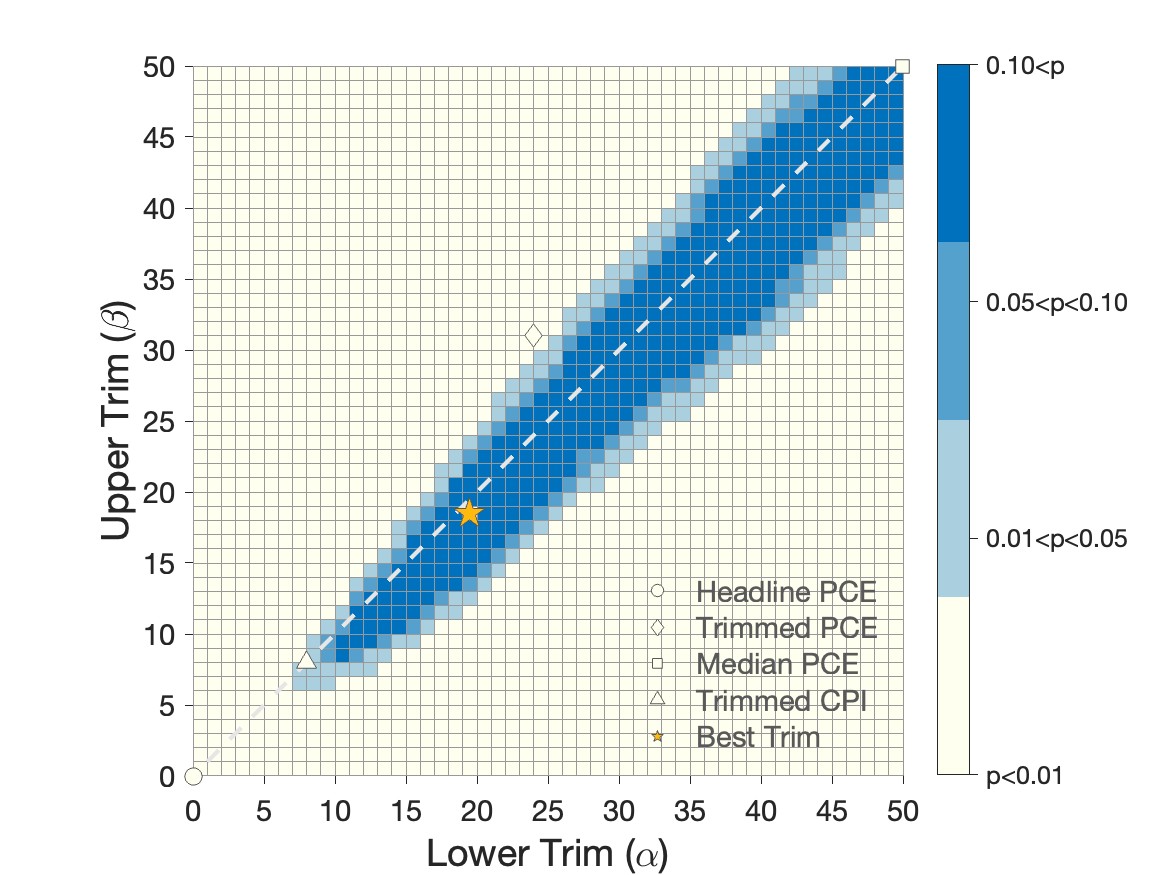}
    \end{subfigure}
    ~
    \begin{subfigure}[t]{0.48\textwidth}
    \caption{Future Trend: 1970--89}
    \includegraphics[width=1.0\textwidth]{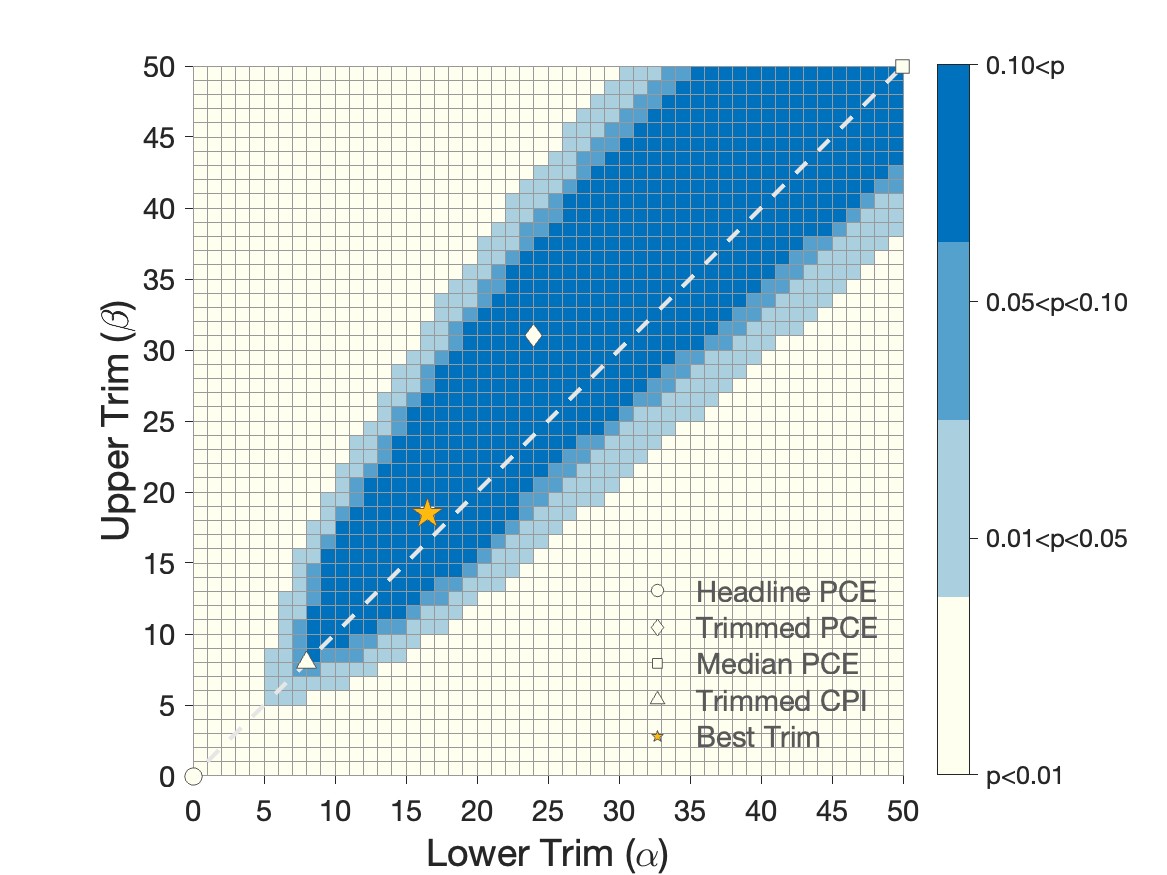}
    \end{subfigure}

    \begin{subfigure}[t]{0.48\textwidth}
    \caption{Current Trend: 2000--2024}
    \includegraphics[width=1.0\textwidth]{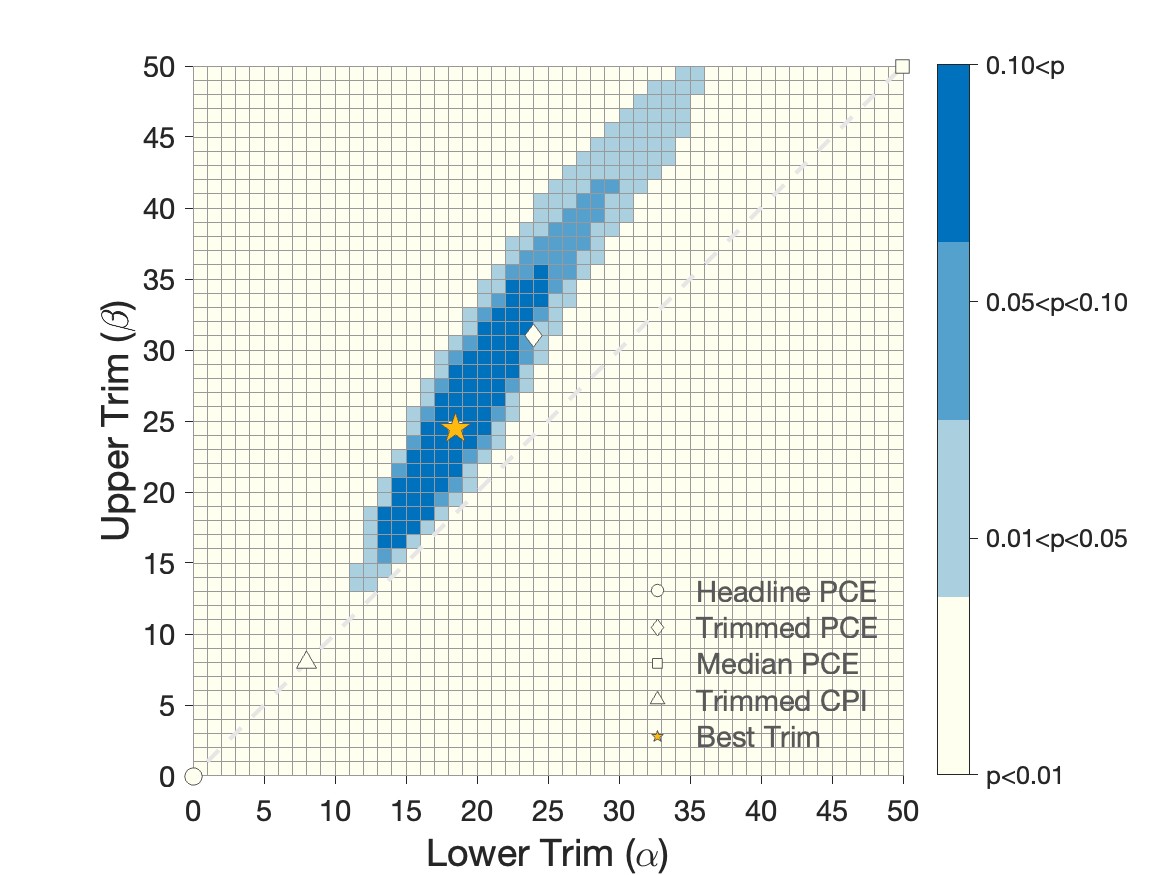}
    \end{subfigure}
    ~
    \begin{subfigure}[t]{0.48\textwidth}
    \caption{Future Trend: 2000--2024}
    \includegraphics[width=1.0\textwidth]{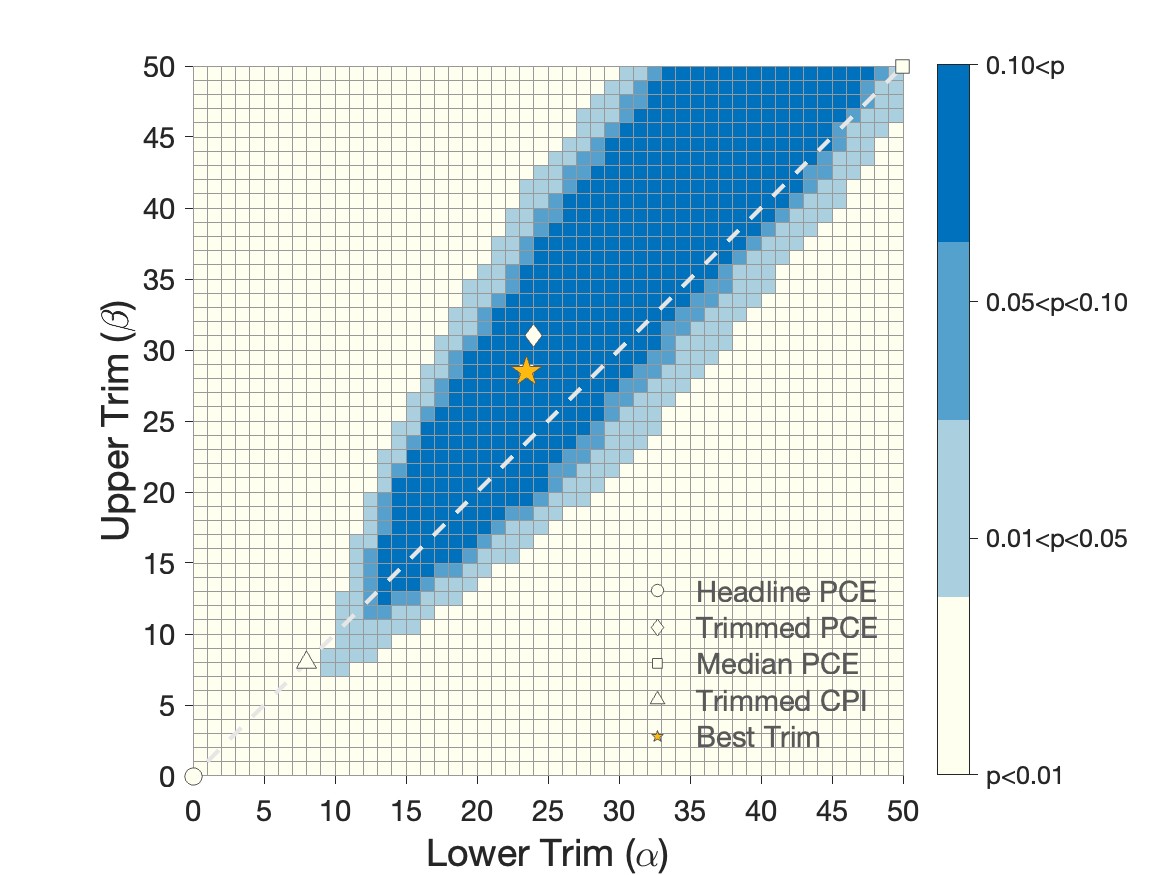}
    \end{subfigure}

    \end{centering}
    
    \vspace{-0.2cm}
    \singlespacing
    \textit{\footnotesize{}Notes:}\footnotesize{} 
    The figures group trims according to the outcome of the \citet{Diebold_Mariano_1995}  test, which compares the root-mean-square error (RMSE) implied by the trims with the RMSE of the best trim as presented in Table \ref{tab:best_trim}. 
    The trims are grouped based on the p-value of the test. 
    The darkest region consists of trims whose RMSE is statistically equivalent to the lowest RMSE across all trims.

\end{figure}

\paragraph{No single optimal trim}
The similarity in the performance of the official robust inflation measures and the optimal trimmed measures is part of a larger pattern:
For every target and sample, a wide set of trims has a similar forecasting performance.
Moreover, there is a set of \textit{near-optimal} trims whose RMSE is statistically indistinguishable from the RMSE of the best trim.
These near-optimal trims are asymmetric---once again trimming more of the high-inflation categories, $\beta>\alpha$---and their sets are much larger (and asymmetric) when targeting future trend inflation.
Figures \ref{fig: RMSE_All} and \ref{fig: DM_Test} present these results by plotting the RMSEs of all trim combinations for the two target trends and the three samples (relative to the lowest RMSE across all trim combinations) and the sets of near-optimal trims for different cutoffs of the Diebold-Mariano test, respectively.
Figures \ref{fig: RMSE_Alternative} and \ref{fig: DM_Test_Alternative} of Appendix \ref{app: Alternative_Trimmed_Mean_Measures} present the results for our alternative measures of trend inflation.

This pattern is explained by the effect of trimming on the underlying distribution of monthly inflation rates. 
As trimming cutoffs increase, the range of inflation rates taken into account is reduced, affecting the variability and the mean prediction of the resulting trimmed mean series. 
Take, for instance, an increase in the upper trim that eliminates more series with high inflation. 
This reduces the average and standard deviation of the prediction implied by the trimmed mean series.
However, the reduction in variability is empirically smaller than the reduction in the average prediction, actually making the trimmed mean series \emph{relatively} more variable as the upper trim increases.
In fact, the coefficient of variation (the ratio of the standard deviation to the mean) increases rapidly with the trimming cutoff, moving from 0.6 to 0.8.\footnote{ 
    Figure  \ref{fig: Trimmed_Mean_Properties} makes the effect of trimming precise by plotting the coefficient of variation and the RMSE for trimmed mean series that target current trend inflation.
    Near the optimal trim, the RMSE does not change much, but the slope of the coefficient of variation is positive.
    }  
As a result, the RMSE of the resulting series does not change much, while similar RMSEs can be obtained from different trim combinations that imply different average predictions.

Our results are consistent with previous work that compared different trimmed mean inflation series. 
See, for example, \citet*{Zaman_Cleveland_2013} and \citet{Meyer_Venkatu_2012} who evaluate trimmed mean CPI measures up until 2013 and also find wide sets of trims to have equivalent forecasting performance.
However, they find these sets to be formed by symmetrical trims, unlike what we find in our extended analysis of PCE inflation series.
In particular, they find median CPI inflation (corresponding to symmetric trims of $\alpha=\beta=50$) to be among the set of near-optimal trims. 
Our results show asymmetrical trims to be superior in the context of PCE inflation, particularly for the most recent sample starting in 2000.
This asymmetry makes it so that the median PCE inflation measure is actually not part of the set of near-optimal trims except for the older sample covering 1970--1989.

\subsection{Performance beyond prediction errors}

A naive reading of the results above might suggest that a wide range of trims is functionally equivalent, with the only difference given by the choice of trim points when targeting the behavior of current or future trend inflation. 
However, the small differences in the RMSEs of different trims hide significant variation in the prediction levels of the implied trimmed mean series.

\begin{figure}
    
    \begin{centering}
        
    \caption{Prediction Range across Best Trims across Time}\label{fig: Prediction_Level}
    
    \begin{subfigure}[t]{0.48\textwidth}
    \caption{Centered Trend}\label{fig: Prediction_Current}
    \includegraphics[width=1.0\textwidth]{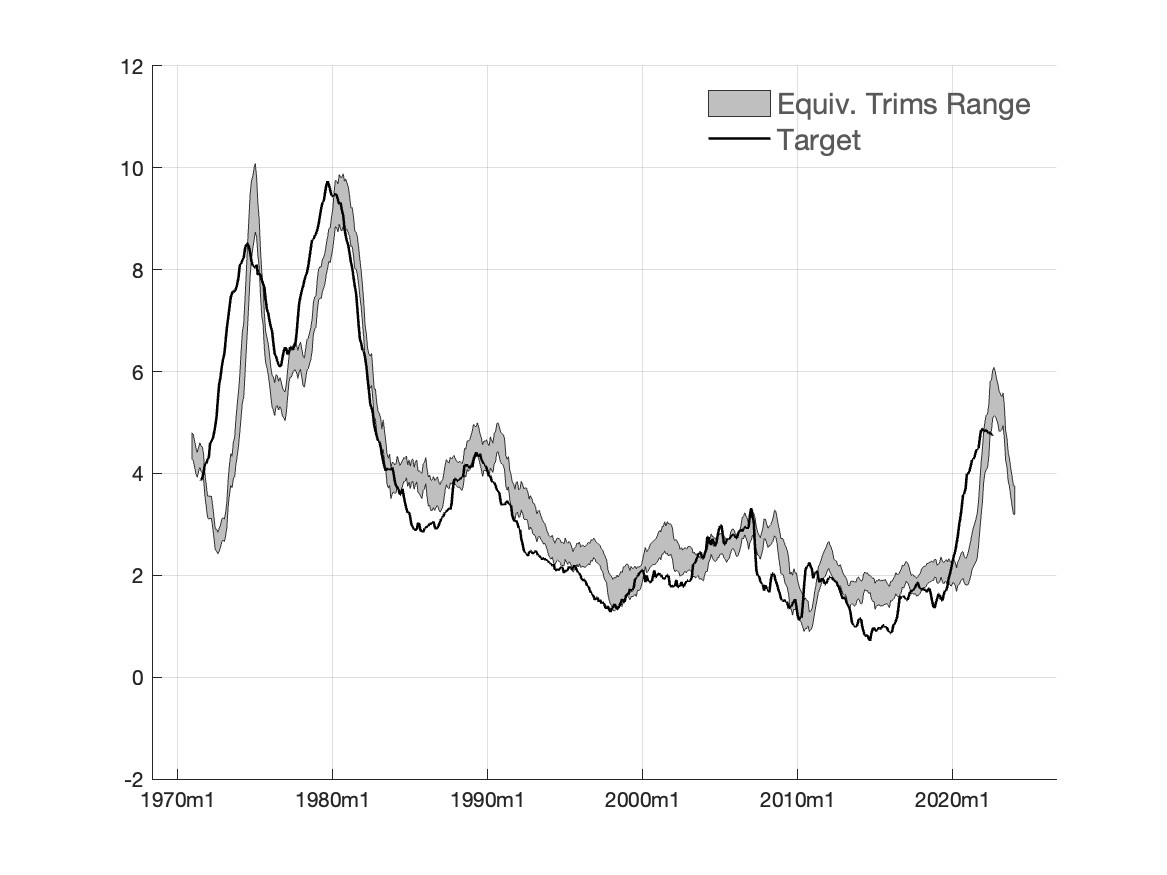}
    \end{subfigure}
    ~
    \begin{subfigure}[t]{0.48\textwidth}
    \caption{Future Trend}\label{fig: Prediction_Level_Future}
    \includegraphics[width=1.0\textwidth]{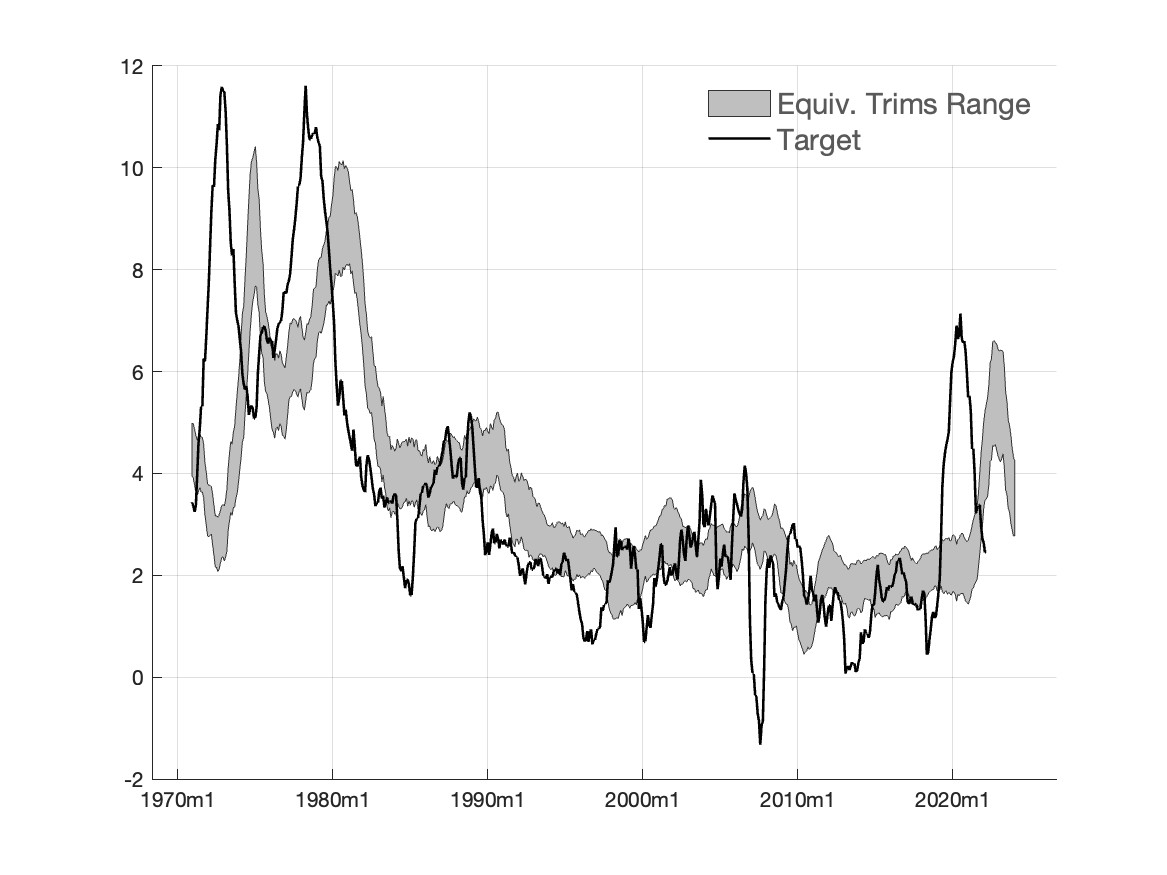}
    \end{subfigure}
    
    \end{centering}
    
    \vspace{-0.2cm}
    \singlespacing
    \footnotesize{}\textit{Notes:}
    The figures plot the range of predictions for the current and future trends between 1970 and 2024 of the set of trimmed mean measures of inflation that are statistically equivalent at the 5 percent level according to the \citet{Diebold_Mariano_1995} test of the difference of their root-mean-square errors (RMSEs) with respect to the RMSE of the best trim for each target (see Table \ref{tab:best_trim}). 
    The start- and end-dates for each figure reflect the data requirements for the ex-post measures of trend inflation.
    
\end{figure}

Figure \ref{fig: Prediction_Level} plots the range of inflation implied by the sets of near-optimal trims---those whose prediction error is closest to the error of the best trim---for the 1970--2024 sample when targeting current and future trend inflation.\footnote{
    Figure \ref{fig: Prediction_Range} plots the distance between the lowest and highest predictions for the current and future trend for the set of near-optimal trims and the best 50 and 100 trims according to their RMSE. 
    Figure \ref{fig: Inf_Range_by_Trims} complements this information by presenting the average range of inflation implied by each trim combination.
    }
The average range is \Paste{min_range_eqv_RMSE} percentage points when targeting current trend inflation and \Paste{max_range_eqv_RMSE} percentage points when targeting future trend inflation.
This variation in prediction levels makes it even more complicated to select a single optimal series, as there are other near-optimal series with different implications for trend inflation in any given month.

Nevertheless, the behavior of the range---taken as a whole---is informative about the behavior of trend inflation. 
Consider, for instance, the range for trend inflation leading to February 2024 implied by the set of near-optimal trims described above.
The range started to decline after August 2022 (one month after headline inflation peaked), while median and trimmed mean inflation only started to decrease in May 2023, when they started tracking the upper and lower edges of the range; see Figure \ref{fig:robust_range_2020_2024}. 
The final predictions for current trend inflation in February 2024 are between 3.05 and 3.55 percent, a range of 50 basis points.\footnote{
    Figure \ref{fig: Prediction_HM} plots the inflation predictions among the near-optimal trims in February 2024. 
    Figure \ref{fig: robust_range_examples} plots the range implied by the set of near-optimal trims in the 1971--1977 and 2008--2012 periods.
    }

Focusing on the \emph{range of inflation levels} implied by the best trimmed mean measures can therefore provide a useful indicator of trend inflation. 
Indeed, we verify next that the range (and even its midpoint) provides better predictors of trend inflation over the whole sample than even the best single trimmed mean inflation measure. 
The range implied by the best trimmed mean inflation measures also provides a transparent way to capture and communicate uncertainty over the behavior of inflation, especially when accompanied by selected robust inflation series (such as median or trimmed mean inflation). 
We observe that the range is widest at times when headline inflation is high or is undergoing rapid changes (see Figure \ref{fig: Prediction_Range}). 
In this way, the range provides additional and useful information for policy makers and the public at large.\footnote{
    \citet{Petersen_Kryvtsov_Uncertainty_2023} find no detrimental effects of communicating ranges over point estimates of inflation. 
    Instead, ranges are more informative for households with extreme values of expected inflation.
    }

\subsection{Back to prediction: The range of near-optimal trims}
Taken together, our results indicate that focusing on a set of trimmed mean inflation measures, instead of selecting a single measure, may provide complementary information on the behavior of trend inflation to that in any one inflation measure.
The range of inflation generated by the set of near-optimal trims described above is a natural candidate for this because they are all equally good on average at tracking trend inflation, and because the information they provide can be effectively communicated through their range of inflation predictions, as Figure \ref{fig: Prediction_Level} illustrates.
Communication can be streamlined by focusing on the range of inflation generated by the set of best 50 trims, which has similar properties to the one generated by the set of equivalent trims (which contains 74 series), while being less complex and in this way easier to communicate.

Moreover, the range of inflation implied by the set of near-optimal trims is a better predictor of the behavior of trend inflation than the point estimate of even the best trim combination.
To gauge the range's predictive performance we compute its RMSE as the minimum distance between the value of the target trend measure and the range,
\begin{align}
    \text{RMSE}^{\text{range}} = \sqrt{\frac{1}{T} \sum_t \left( \max\left\{\; \bar{\pi}_t-\max_{i\in\text{range}}\{{\pi^i_{t}}\} \;,\; \min_{i\in\text{range}}\{{\pi^i_{t}}\} - \bar{\pi}_t \;,\; 0 \;\right\} \right)^2}, 
    \label{eq: RMSE_Range}
\end{align}
so that the error is zero in a month if the trend falls within the range and the distance to the nearest endpoint if the trend falls outside. 
For instance, the range in Figure \ref{fig: Prediction_Current} delivers a RMSE of 0.73 when tracking current trend inflation, well below the RMSE of the best trimmed mean series (1.08, Table \ref{tab:best_trim}).

While a lower RMSE is to be expected when using a range of inflation instead of a point estimate, the gains in predictive performance do not come entirely from the width of the range. 
Two simple additional calculations help to show this.
First, there is information in the range beyond its width. 
Using the midpoint of the range as a point estimate for trend inflation and computing the RMSE as in equation \eqref{eq: RMSE_Def} delivers a RMSE of 0.91 over the whole sample. 
This is almost half of the difference between the RMSE of the range prediction and the point estimate of the best trim.
Second, not all ranges are equally informative.
Consider a range around the official trimmed mean series of the same width as the range in Figure \ref{fig: Prediction_Current}.
This new range delivers a RMSE of 0.90 (computed as in \ref{eq: RMSE_Range}), virtually the same as using only the point estimate from the midpoint of the near-optimal trims' range.

\paragraph{Current vs future inflation}
Finally, our analysis also shows that measures of trimmed mean inflation have a hard time tracking changes in future trend inflation, instead lagging its movements, as Figure \ref{fig: Prediction_Level} makes clear. 
This is particularly evident in the two instances of high inflation in the pre-1977 sample---that we added to the official series---and at the end of the sample, when inflation rises again.
In fact, the predicted range for future trend inflation does not start increasing until inflation has already peaked.
By contrast, the range of trimmed mean inflation measures is more informative about the behavior of current trend inflation. %
These results suggest that trimmed mean measures are best used to understand current inflation, with other methods such as those in \citet{Verbrugge2022} better suited for predicting future inflation.

%% file: Sections_2.0/05m_Conclusions.tex
\section{Conclusion}

Economists often use robust inflation series such as core inflation or median inflation to communicate with the public and gauge the behavior of trend inflation. 
Our evaluation of the performance of the official series for various targets and across various samples shows them to be robust across time and comparable with the performance of the best trimmed mean inflation measure, which selects trimming cutoffs to minimize prediction error.

Among the official robust measures, trimmed mean and median inflation clearly outperform core inflation, constituting a somewhat surprising result given policy practices and public attention to core inflation. 
However, a more consequential result concerning the choice of optimal robust measures emerges from our analysis. 
Results based on the average predictive performance of the series obscure an underlying pattern of trimmed mean measures. A wide range of measures have the same predictive performance, but they produce substantially different predictions in any given month.

We conclude that following a set of trimmed mean measures rather than a single series may provide additional insight about the behavior of inflation.
This information can be effectively communicated by reporting the range of predictions from the set of measures with the best predictive performance, as we do in Figures \ref{fig:robust_range_2020_2024} and \ref{fig: Prediction_Level}.
This range is more informative about trend inflation than any single series among the ones we consider, so that even its midpoint outperforms the best trimmed mean inflation measure in tracking inflation.
In this way, this range informs us about the behavior of trend inflation while indicating uncertainty in a way that is easy to communicate to the public.
It can also provide a more credible signal to households with more uncertainty about inflation, anchoring their inflation expectations as shown in \citet{Petersen_Kryvtsov_Uncertainty_2023}.

While we also explored alternative trimmed mean measures not reported here and found similar results,\footnote{
    In particular, we constructed trimmed mean measures excluding housing, which is the single largest expenditure category and  one of the series most commonly included in the official trimmed mean and median inflation measures (see Table \ref{tab:Excluded_Categories}). 
    We found that housing does not play a large role in the behavior of trimmed mean inflation measures. 
    All of our results are preserved when excluding housing from the set of expenditure categories, as we show in Appendix \ref{app: Housing}.
    } 
future work may consider a twist to our analysis. 
Such work may consider evaluating the predictive power of robust measures of inflation for particular \emph{alternative targets}---for example, current or future inflation rates relevant for different sub-groups of the US population. 
This differential evaluation may provide valuable new insights into robustness in the context of heterogeneous effects of monetary policy.

%% file: Appendix/a01m_Replication.tex
\section{Replication of Trimmed-Mean and Median PCE Inflation Series}\label{agg:replication}

\begin{figure}[tbh]
    \centering
    \caption{Replication of Trimmed-Mean PCE Inflation, 1960--2024}
    \includegraphics[scale=0.70]{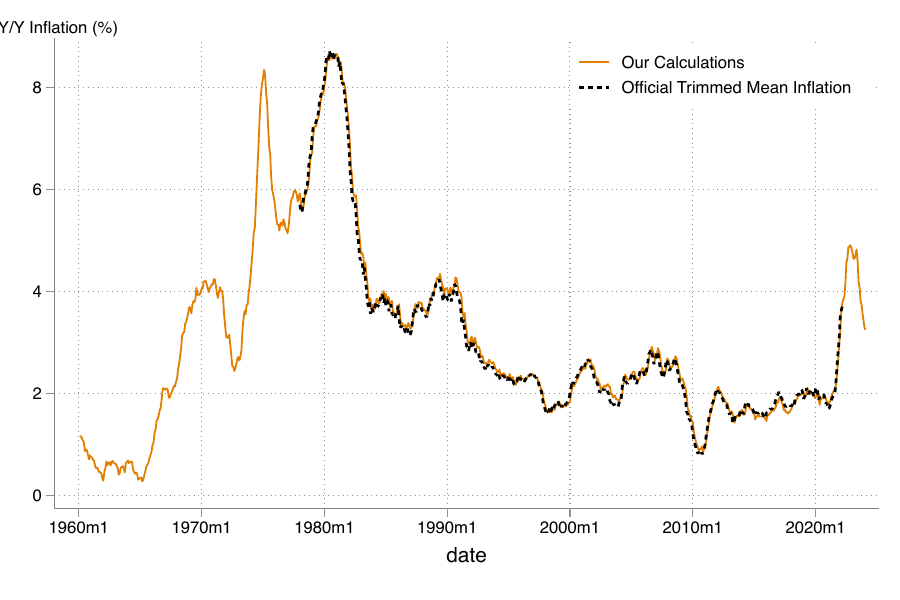}
    \caption*{{\footnotesize\textit{Notes:} The figure shows the authors' calculation of trimmed mean PCE inflation along with the official trimmed mean PCE series.} 
    
    }
    \label{fig:replication_trimmed_mean}
\end{figure}

\begin{figure}[bh!]
    \centering
    \caption{Replication of Median PCE Inflation, 1960--2024}
    \includegraphics[scale=0.70]{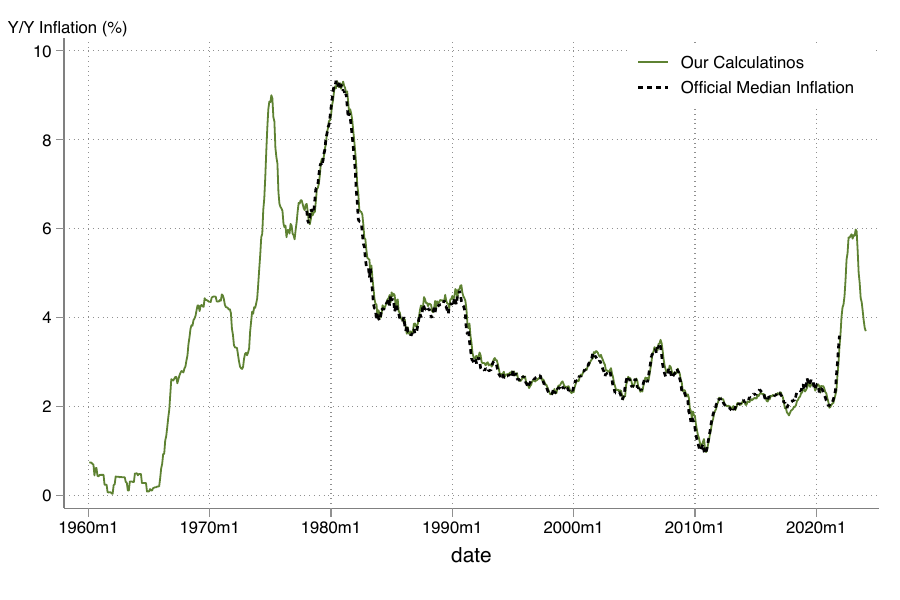}
    \caption*{{\footnotesize\textit{Notes:} The figure shows the authors' calculation of median PCE inflation along with the official median PCE series.
    
    }}
    \label{fig:replication_median}
\end{figure}

\begin{figure}
    \centering
    \caption{Number of Series with No Monthly Price Changes}
    \begin{subfigure}[b]{0.95\textwidth}
    \caption{Trimmed Mean}
    \includegraphics[width=\textwidth]{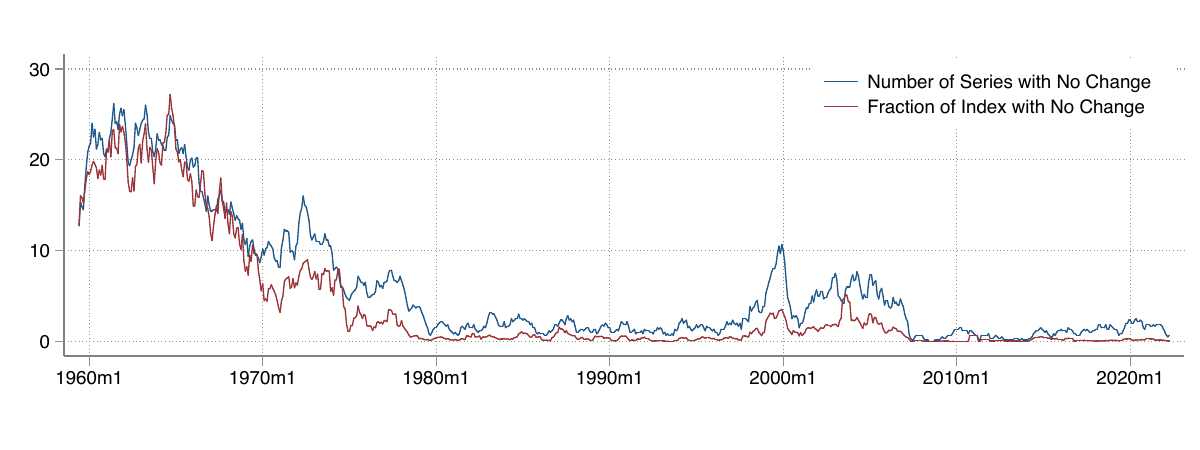}
    \end{subfigure}
    \begin{subfigure}[b]{0.95\textwidth}
    \caption{Median}
    \includegraphics[width=\textwidth]{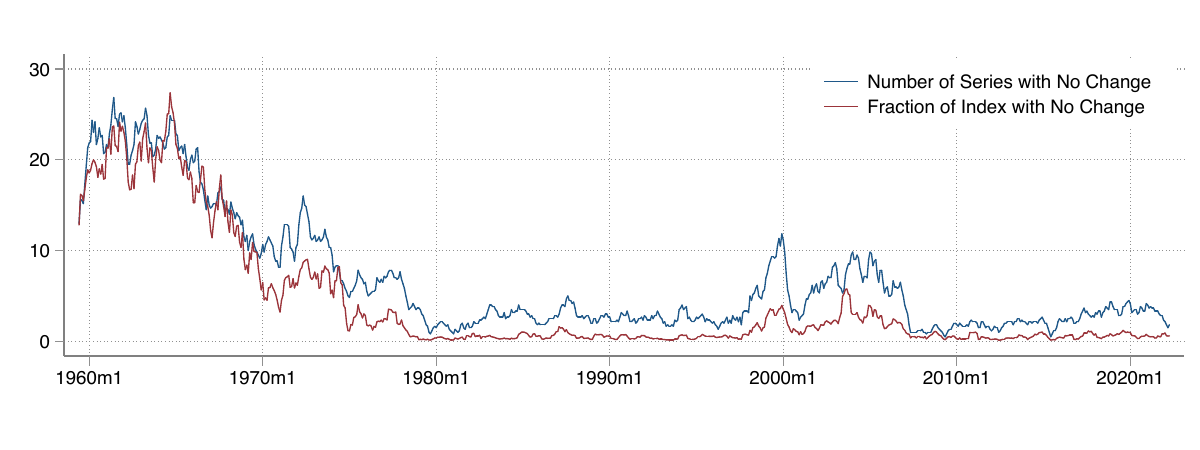}
    \end{subfigure}
    \begin{subfigure}[b]{0.95\textwidth}
    \caption{Time Consistent}
    \includegraphics[width=\textwidth]{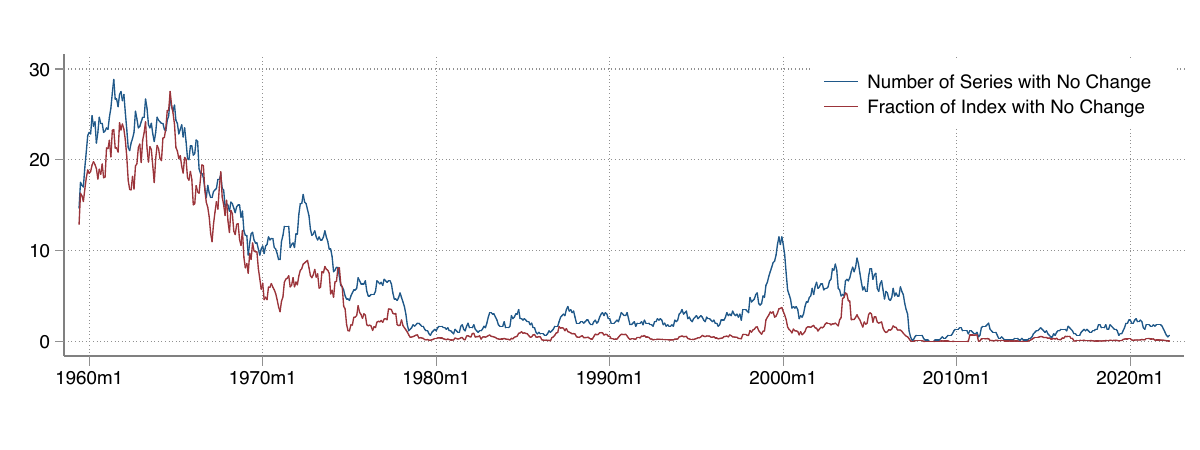}
    \end{subfigure}
    \caption*{{\footnotesize\textit{Notes:} The lines plot the number of series and fraction of total expenditure (in percent) with no monthly price change for each of the three sets of series used in the paper.}} 
    \label{fig:flat}
\end{figure}

\begin{figure}
    \centering
    \caption{Number of Series with Positive Expenditure}
    \begin{subfigure}[b]{0.95\textwidth}
    \caption{Trimmed Mean}
    \includegraphics[width=\textwidth]{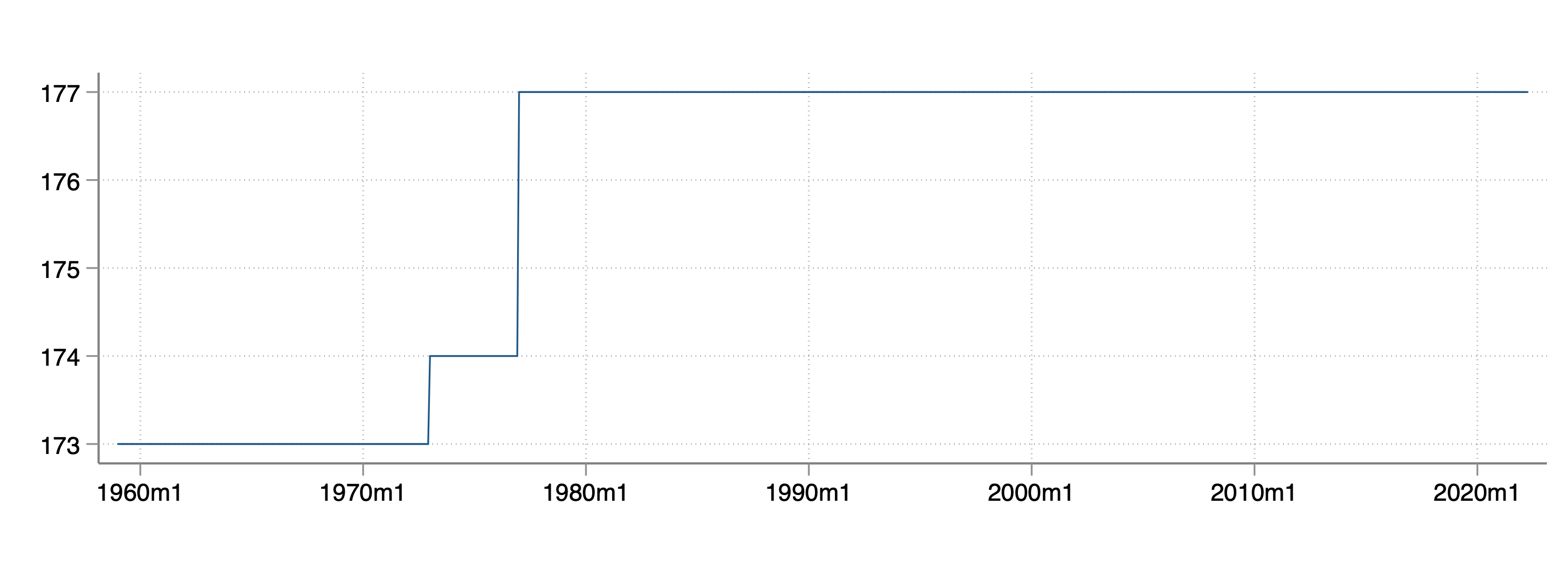}
    \end{subfigure}
    \begin{subfigure}[b]{0.95\textwidth}
    \caption{Median}
    \includegraphics[width=\textwidth]{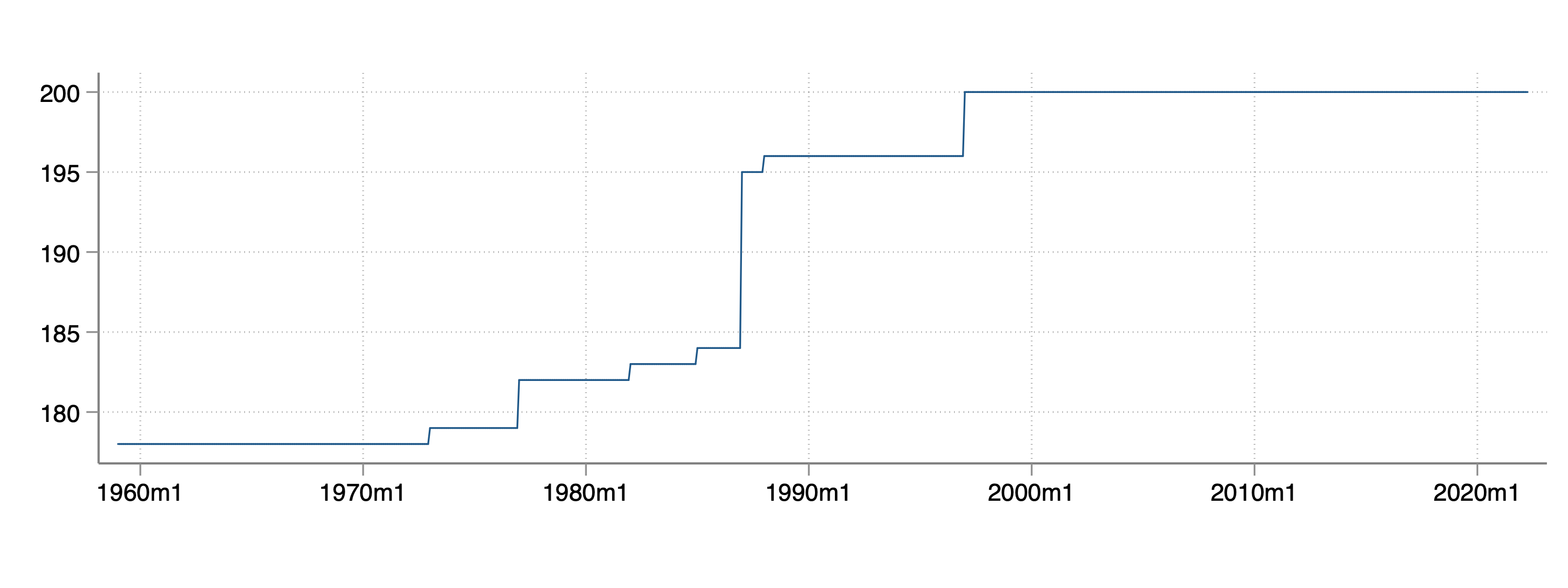}
    \end{subfigure}
    \begin{subfigure}[b]{0.95\textwidth}
    \caption{Time Consistent}
    \includegraphics[width=\textwidth]{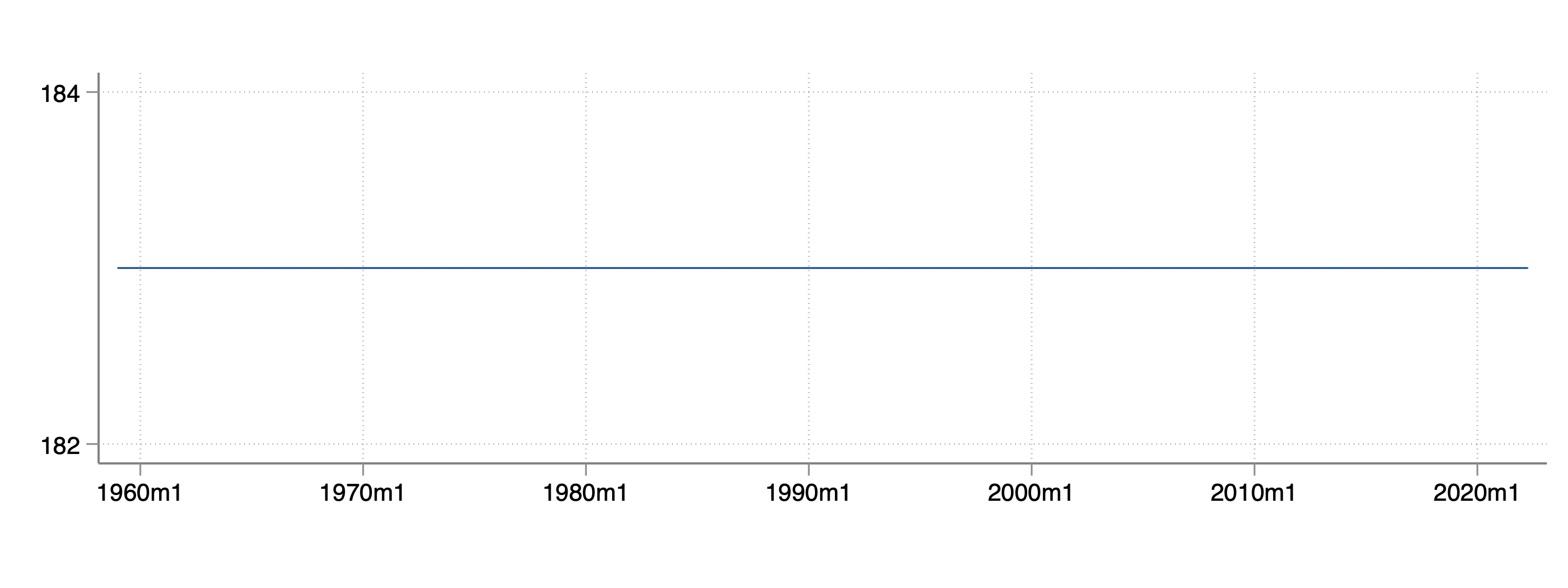}
    \end{subfigure}
    \caption*{{\footnotesize\textit{Notes:} The lines plot the number of series with positive expenditure in the PCE series over time from the set of series considered by the trimmed mean inflation, median inflation, and the time consistent set constructed in the paper.}} 
    \label{fig:nonmissing}
\end{figure}

%% file: Appendix/a02m_Add_Figures.tex
\section{Additional Results}\label{app:Additional_Figures}

\subsection{Time Series Properties}

\begin{figure}[tbh]
    \centering
    \caption{Range of Robust Inflation Measures, 1960--2024}
    \includegraphics[scale=0.7]{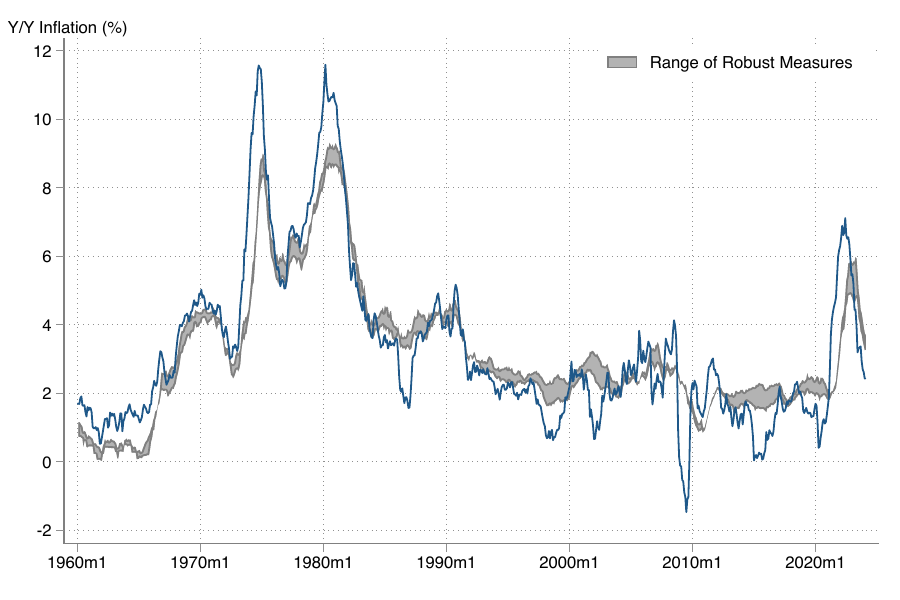}
    \caption*{{\footnotesize\textit{Notes:} 
    The figure shows the authors' calculations of the range of robust inflation measures (core inflation, median inflation, and trimmed mean inflation) from 1960 to 2024.
    We report year-over-year inflation to smooth out variations in monthly inflation.
    The range is shown in the shaded area. 
    The blue line corresponds to headline inflation.
    }}
    \label{fig:robust_measures_min_max}
\end{figure}

The level of agreement between the series is also captured by the range of values they cover, shown in Figure \ref{fig:robust_measures_min_max} along with the level of headline PCE inflation. 
The range is \Paste{range_infl_measures} percentage points on average over the whole sample, \Paste{range_infl_low} percentage points when inflation is less than 2.5 percent, and \Paste{range_infl_high} percentage points when inflation is above 5 percent. 
Thus, the range values covered by robust inflation measures is disproportionately wider when inflation is low  than when inflation is high.
This again shows that there is substantially more agreement between the signals provided by the different inflation measures when inflation is high.

\input{Tables/d31m_summary_by_period}

\begin{figure}[tbh]
    \centering
    \caption{Time-Series Variability of Measures of Inflation, 1960--2024}
    \includegraphics[scale=0.75]{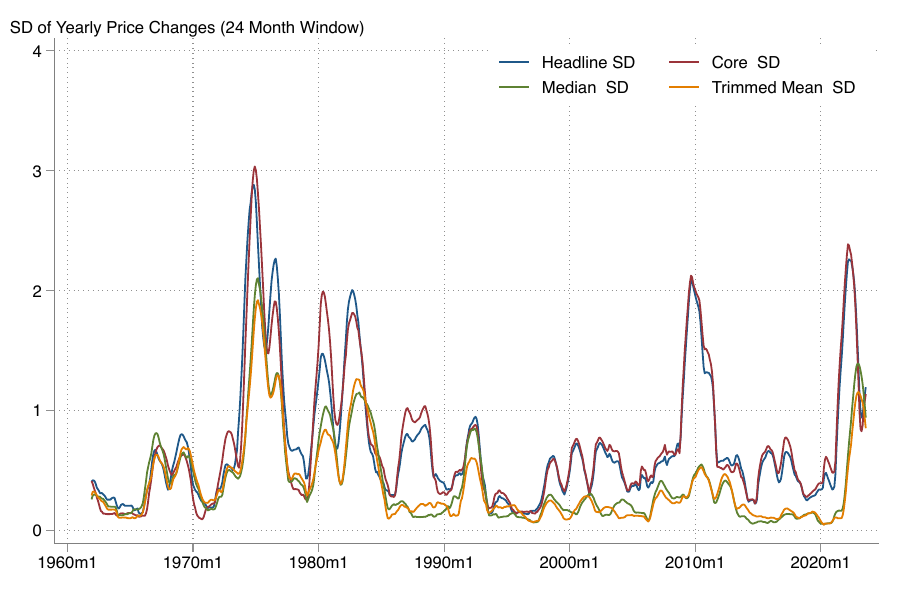}
    \caption*{{\footnotesize\textit{Notes:} 
    The figure shows the authors' calculations of the standard deviations of headline inflation, core inflation, median inflation, and trimmed mean inflation for a rolling window of 24 months.
    }}
    \label{fig:robust_measures_sd}
\end{figure}

The variability of the robust inflation measures is also higher during low-inflation episodes despite these measures being constructed to be less responsive to transitory movements in inflation. 
Even though the robust inflation measures are overall less volatile than headline inflation, this pattern does not hold throughout the whole sample. 
Table \ref{tab:Mean_StD_Inflation} reports the mean, standard deviation, and coefficient of variation of the four inflation series for different samples that depend on the level of headline inflation, Figure \ref{fig:robust_measures_sd} plots the time series of the standard deviations.

When inflation is below 2.5 percent, median and trimmed mean inflation are more volatile than headline inflation, and when inflation is between 2.5 and 5 percent all three robust inflation measures are more variable than headline inflation.
Moreover, the coefficient of variation is highest when headline inflation is below 2.5 percent. 
The robust inflation measures also change their ranking in terms of how volatile they are. 
Core inflation is the most volatile in the complete sample, but median inflation is more volatile when inflation is low (below 2.5 percent) and trimmed mean inflation is more volatile when inflation is high (above 5 percent).

\begin{table}[tb]
    \caption{Most Commonly Excluded and Included Expenditure Categories}
    \begin{center}
    \begin{threeparttable}
    \begin{tabular}{llll}
        \hline \hline
        & Median & Trimmed Mean & Middle 80\%  \\
        &        &              & (10, 10) Trim \\
               \multicolumn{4}{c}{Most Commonly Excluded}\\
        \hline 
    1 &   & Eggs & Eggs \\
    2 & 66 series are & Food on farms & Vegetables \\
    3 & never median & Vegetables    & Food on farms \\
    4 &   & Fruit         & Fuel Oil \\
    5 &                         & Gasoline      & Gasoline         \\
        \hline
               \multicolumn{4}{c}{Most Commonly Included}\\ \hline
    1 & Owner-occ homes & Owner-occ homes & Owner-occ homes \\
    2 & Other purchased meals & Other purchased meals & Other purch meals \\
    3 & Tenant-occ homes & Casino gambling & Tenant-occ homes \\
    4 & Nonprofit hospitals & Owner-occ mobile homes  & Casino gambling \\
    5 & Physician services & Tenant-occ homes  & Lotteries \\
    \hline 
    \end{tabular}
    \begin{tablenotes}
        \item {\footnotesize \textit{Notes:} 
        The table reports the five expenditure categories most commonly excluded and the five most commonly included when computing median and trimmed mean inflation  as well as those excluded and included when trimming the middle 90 percent of expenditure, setting trims to $\alpha=\beta=10$, with a consistent set of inflation categories.
        All the results are from the authors' calculations of the series reported in Figure \ref{fig:robust_measures}.
        In the case of median inflation (first column), all categories but one are included in a given month, so we report the number of series that are never included. 
        A series is considered "included" if any of the weight of the series is used in the calculation.
        
        }
    \end{tablenotes}
    \end{threeparttable}
    \end{center}
    \label{tab:Excluded_Categories} %
\end{table}

\input{Appendix/a02o_Official_Trimmed_Mean}

\input{Appendix/a02p_Opt_Trimmed_Mean}

\input{Appendix/a02q_Housing}

%% file: Tables/d31m_summary_by_period.tex
\renewcommand{\tabcolsep}{15pt}
\begin{table}[tbh]
    \caption{Summary Statistics: Inflation Measures}
    \begin{center}
    \begin{threeparttable}
    \begin{tabular}{l cccc}
        \hline \hline
         & \multicolumn{4}{c}{Inflation Measures} \\
         & \multirow{2}{*}{Headline} & \multirow{2}{*}{Core} & \multirow{2}{*}{Median} & Trimmed \\
         &     &      &        & Mean    \\
        \cline{2-5}
         & \multicolumn{4}{c}{Full Sample (748 months)}\\
        \cline{2-5}
        Mean        & 3.27  &  3.21  &  3.33  &  2.96 \\
        Std. Dev.   & 2.42  &  2.13  &  2.01  &  1.86 \\
        Coeff. Var. & 0.74  &  0.66  &  0.60  &  0.63 \\
        \cline{2-5}
         & \multicolumn{4}{c}{$\pi<2.5\%$ (373 months)}\\
        \cline{2-5}
        Mean        & 1.55  &  1.73  &  2.01  &  1.72 \\
        Std. Dev.   & 0.67  &  0.53  &  0.95  &  0.70 \\
        Coeff. Var. & 0.43  &  0.31  &  0.47  &  0.41 \\
        \cline{2-5}
         & \multicolumn{4}{c}{$2.5\%\leq\pi<5\%$ (252 months)}\\
        \cline{2-5}
        Mean        & 3.61  &  3.51  &  3.55  &  3.17 \\
        Std. Dev.   & 0.71  &  1.06  &  0.84  &  0.77 \\
        Coeff. Var. & 0.20  &  0.30  &  0.24  &  0.24 \\
        \cline{2-5}
         & \multicolumn{4}{c}{$5\%\leq\pi$ (123 months)}\\
        \cline{2-5}
        Mean        & 7.76  &  7.09  &  6.85  &  6.31 \\
        Std. Dev.   & 2.00  &  1.59  &  1.60  &  1.57 \\
        Coeff. Var. & 0.26  &  0.22  &  0.23  &  0.25 \\
        \hline
    \end{tabular}
    \begin{tablenotes}
        \item {\footnotesize \textit{Notes:} The numbers are mean, standard deviation, and coefficient of variation of the different inflation measures for different samples determined by the level of PCE inflation. All numbers are in percentage points.}
    \end{tablenotes}
    \end{threeparttable}
    \end{center}
    \label{tab:Mean_StD_Inflation}
\end{table} 
\renewcommand{\tabcolsep}{10pt}

%% file: Appendix/a02o_Official_Trimmed_Mean.tex
\FloatBarrier
\subsection{Alternative Measures of Trend Inflation}\label{app: Alternative_Trend_Measures}

\begin{figure}[bh!]
    \centering
    \caption{Time Series of Trend Inflation}
    \begin{subfigure}[b]{\textwidth}
    \centering 
    \caption{Measures of Current Trend Inflation}
    \includegraphics[scale=0.7]{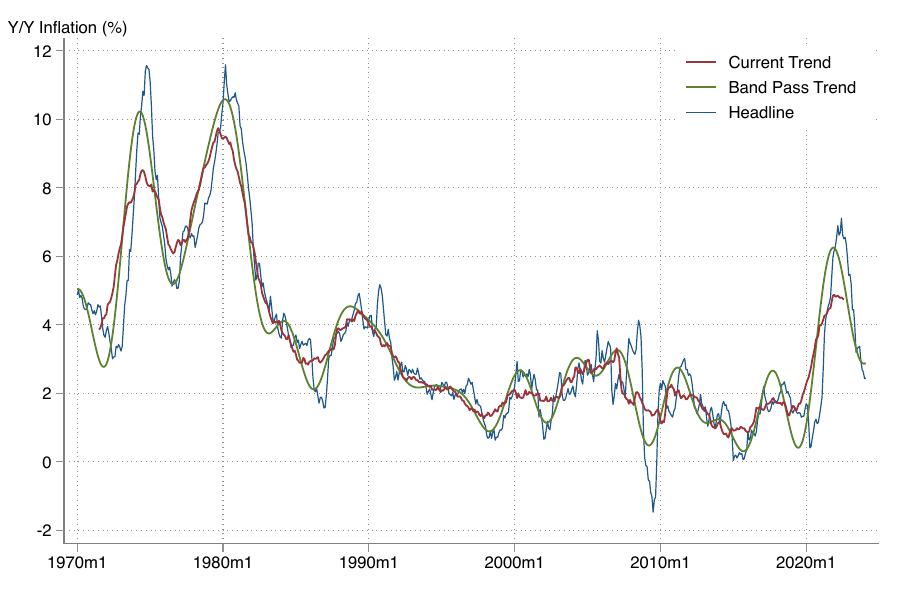}
    
    \end{subfigure}
    
    \begin{subfigure}[b]{\textwidth}
    \centering 
    \caption{Measures of Future Trend Inflation}
    \includegraphics[scale=0.7]{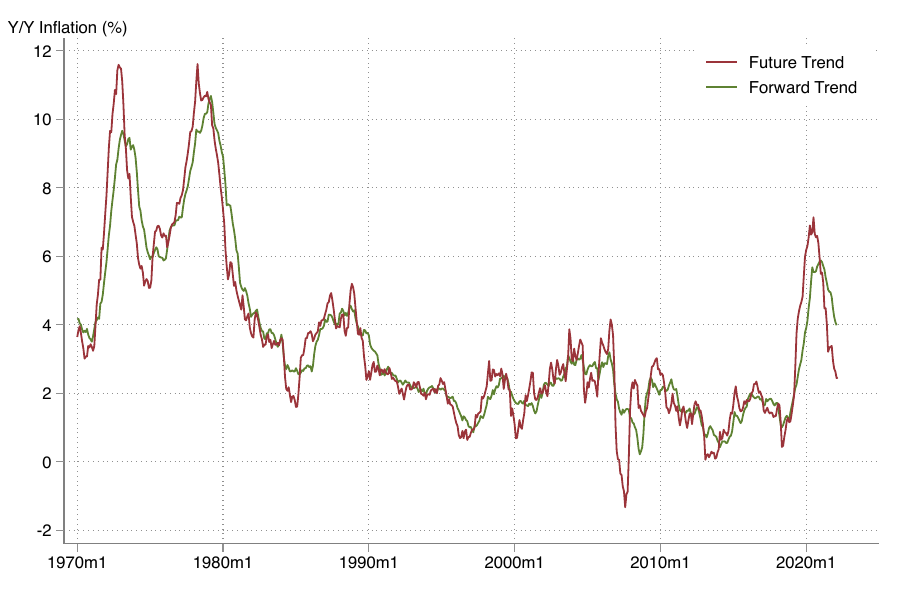}
    \end{subfigure}
    
    \caption*{{\footnotesize\textit{Notes:} 
    The figures show the authors' calculations of two measures of trend inflation. 
    The first panel presents two measures of current trend inflation, a 36-month centered inflation trend (current trend), and a band-pass filter trend (band-pass trend) as described in Section \ref{sec:Evaluation}, together with the series of year-on-year headline PCE inflation.
    The second panel presents two measures of future trend inflation, a 12-month forward moving average of headline inflation with data between 12 and 24 months ahead (future trend), and a 24-month forward moving average (forward trend).
    }}
    \label{fig:trend_ts}
\end{figure}

\begin{table}[tbh!]
    \caption{Ranking of Various Methods of Calculating Robust Measures}
    \begin{center}
    \begin{threeparttable}
       \begin{tabular}{ll S S S S S}
        \hline \hline
        \multirow{2}{*}{Target} & \multirow{2}{*}{Sample} &  \multicolumn{3}{c}{PCE Inflation Measure RMSE} & \multicolumn{2}{c}{DM Test: $\Pr\left(z>\left|\text{DM}\right|\right)$}\\
         \cline{3-5}\cline{6-7}
         &  & {Core} & {Trimmed Mean} & {Median}  & {TM vs Med.} & {Core vs Other} \\
        \hline 
                    & {1970-2024} & 1.47 &1.11 & 1.18 & 0.015 & 0.000 \\
        {Current}   & {1970-1989} & 1.84 &1.62 & 1.52 & 0.088 & 0.001 \\
        {Trend}     & {2000-2024} & 1.34 &0.81 & 1.02 & 0.000 & 0.004 \\
        \hline
                    & {1970-2024} & 1.50 &1.27 & 1.34 & 0.011 & 0.001 \\
        {Band-Pass} & {1970-1989} & 1.79 &1.66 & 1.57 & 0.042 & 0.011 \\
        {Trend}     & {2000-2024} & 1.35 &1.07 & 1.27 & 0.000 & 0.040 \\
        \hline
                    & {1970-2024} & 2.45 &2.14 & 2.17 & 0.197 & 0.000 \\
        {Future}    & {1970-1989} & 3.35 &3.02 & 3.02 & 0.903 & 0.000 \\
        {Trend}     & {2000-2024} & 1.99 &1.66 & 1.71 & 0.161 & 0.019 \\
        \hline
                    & {1970-2024} & 1.96 &1.67 & 1.71 & 0.104 & 0.000 \\
        {Forward}   & {1970-1989} & 2.68 &2.39 & 2.36 & 0.614 & 0.001 \\
        {Trend}     & {2000-2024} & 1.56 &1.26 & 1.36 & 0.001 & 0.052 \\
        \hline
    \end{tabular}
    \begin{tablenotes}
        \item {\footnotesize \textit{Notes:} 
        The table presents the predictive performance of different PCE inflation measures with respect to different trend inflation targets in different samples.
        The performance is measured with the series' root-mean-square error (RMSE) with respect to trend inflation.
        The table reports the RMSEs for core, trimmed mean, and median inflation. 
        The last two columns report the p-value of the \citet{Diebold_Mariano_1995} test for the difference between the RMSEs of the trimmed mean and median inflation series and the difference of the core inflation series and the best of the trimmed mean and the median inflation series. 
        }
    \end{tablenotes}
    \end{threeparttable}
    \end{center}
    \label{tab:prediction_All}
\end{table}

%% file: Appendix/a02p_Opt_Trimmed_Mean.tex
\clearpage
\newpage
\FloatBarrier
\subsection{Alternative Trimmed Mean Measures of Inflation}\label{app: Alternative_Trimmed_Mean_Measures}

\begin{table}[tbh!]
    \caption{Best Trims for Trimmed-Mean Inflation}
    \begin{center}
    \begin{threeparttable}
    \begin{tabular}{ll SSS S S}
        \hline \hline
        \multirow{2}{*}{Target} & \multirow{2}{*}{Sample} &  \multicolumn{3}{c}{Best Trims} & {Official Trims} & {DM Test} \\
         \cline{3-5}
         &  &  {Lower} & {Upper} & {RMSE} & {min(RMSE)} & {$\Pr\left(z>\left|\text{DM}\right|\right)$} \\
        \hline 
                    & {1970-2024} & 17 & 19 & 1.08 & 1.11 & 0.065 \\
        {Current}   & {1970-1989} & 18 & 16 & 1.46 & 1.52 & 0.235 \\
        {Trend}     & {2000-2024} & 18 & 23 & 0.80 & 0.81 & 0.403 \\
        \hline
                    & {1970-2024} & 11 & 12 & 1.15 & 1.27 & 0.000 \\
        {Band-Pass} & {1970-1989} & 10 &  9 & 1.39 & 1.57 & 0.007 \\
        {Trend}     & {2000-2024} & 13 & 16 & 1.00 & 1.07 & 0.019 \\
        \hline
                    & {1970-2024} & 35 & 41 & 2.11 & 2.14 & 0.213 \\
        {Future}    & {1970-1989} & 15 & 17 & 2.91 & 3.02 & 0.412 \\
        {Trend}     & {2000-2024} & 24 & 28 & 1.63 & 1.66 & 0.288 \\
        \hline
                    & {1970-2024} & 14 & 16 & 1.62 & 1.67 & 0.053 \\
        {Forward}   & {1970-1989} & 13 & 13 & 2.28 & 2.36 & 0.108 \\
        {Trend}     & {2000-2024} & 16 & 19 & 1.22 & 1.26 & 0.084 \\
        \hline
    \end{tabular}
    \begin{tablenotes}
        \item {\footnotesize \textit{Notes:} 
        The table reports the best trims for different targets of trend inflation over different samples as determined by the predictive performance across trims. 
        The root-mean-square error (RMSE) of the best trim is also reported along with the lowest RMSE of the official trimmed and median inflation series. 
        The last column reports the p-value of the \citet{Diebold_Mariano_1995} test for the difference between the RMSEs of the best trim and the lowest of the official series. 
        }
    \end{tablenotes}
    \end{threeparttable}
    \end{center}
    \label{tab:best_trim_All}
\end{table} 

\FloatBarrier

\begin{figure}
    \begin{centering}
        
    \caption{RMSE across Trims}\label{fig: RMSE_Alternative}
    
    \begin{subfigure}[t]{0.48\textwidth}
    \caption{Band-Pass Trend: 1970--2024}
    \includegraphics[width=1.0\textwidth]{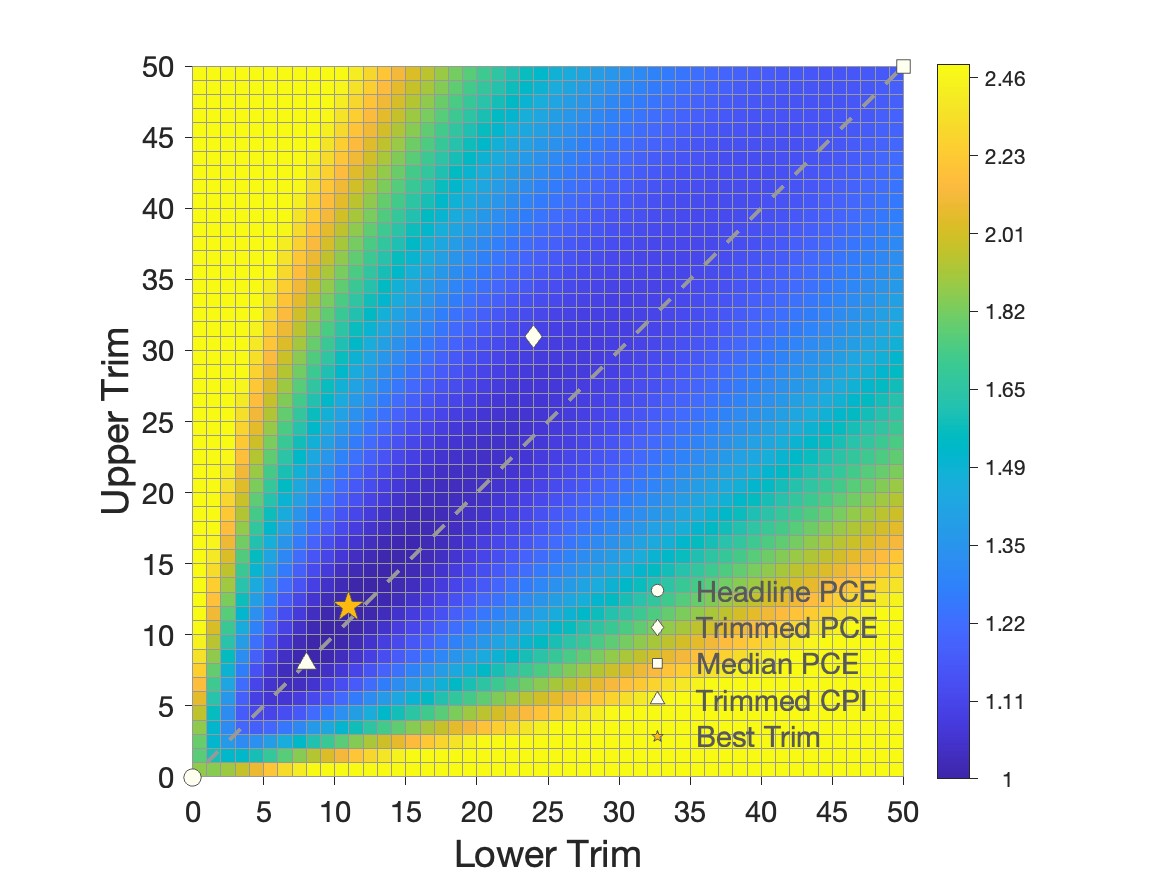}
    \end{subfigure}
    ~
    \begin{subfigure}[t]{0.48\textwidth}
    \caption{Forward Trend: 1970--2024}
    \includegraphics[width=1.0\textwidth]{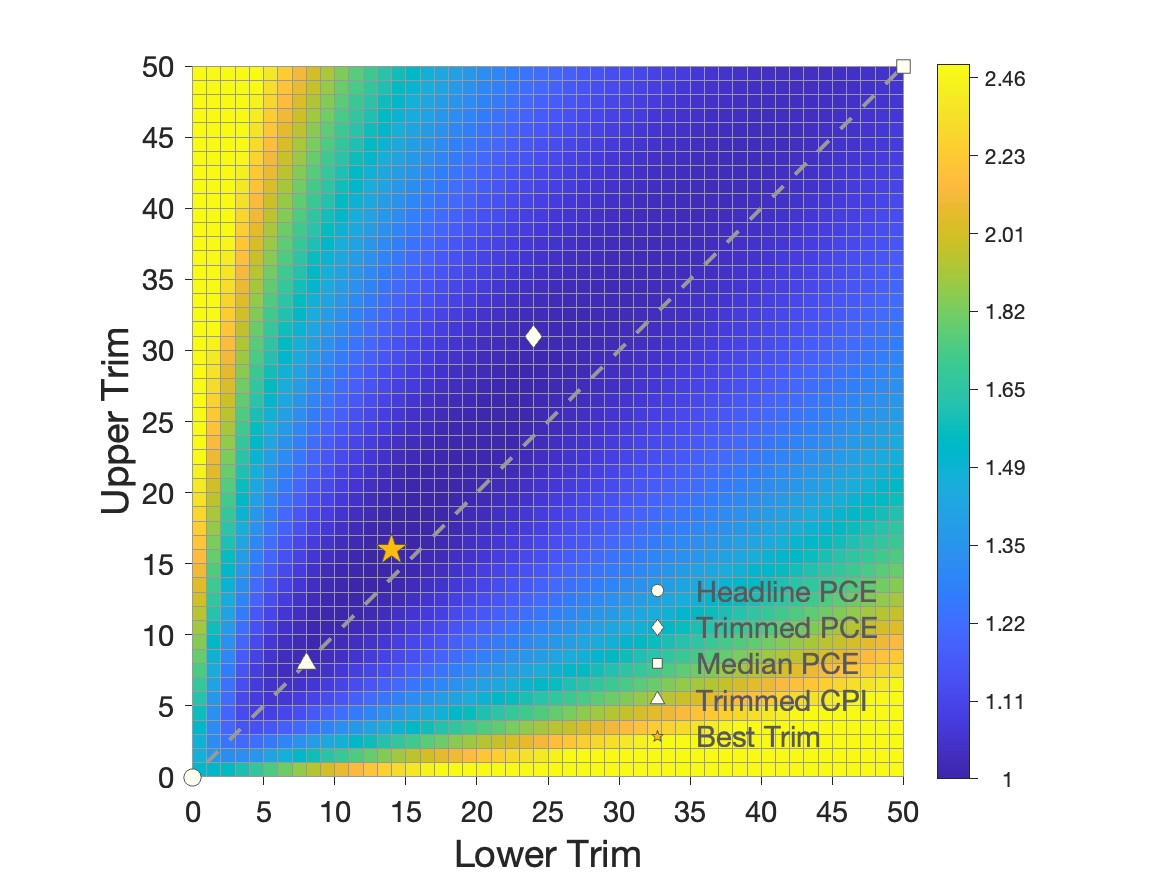}
    \end{subfigure}
    
    \begin{subfigure}[t]{0.48\textwidth}
    \caption{Band Pass Trend: 1970--89}
    \includegraphics[width=1.0\textwidth]{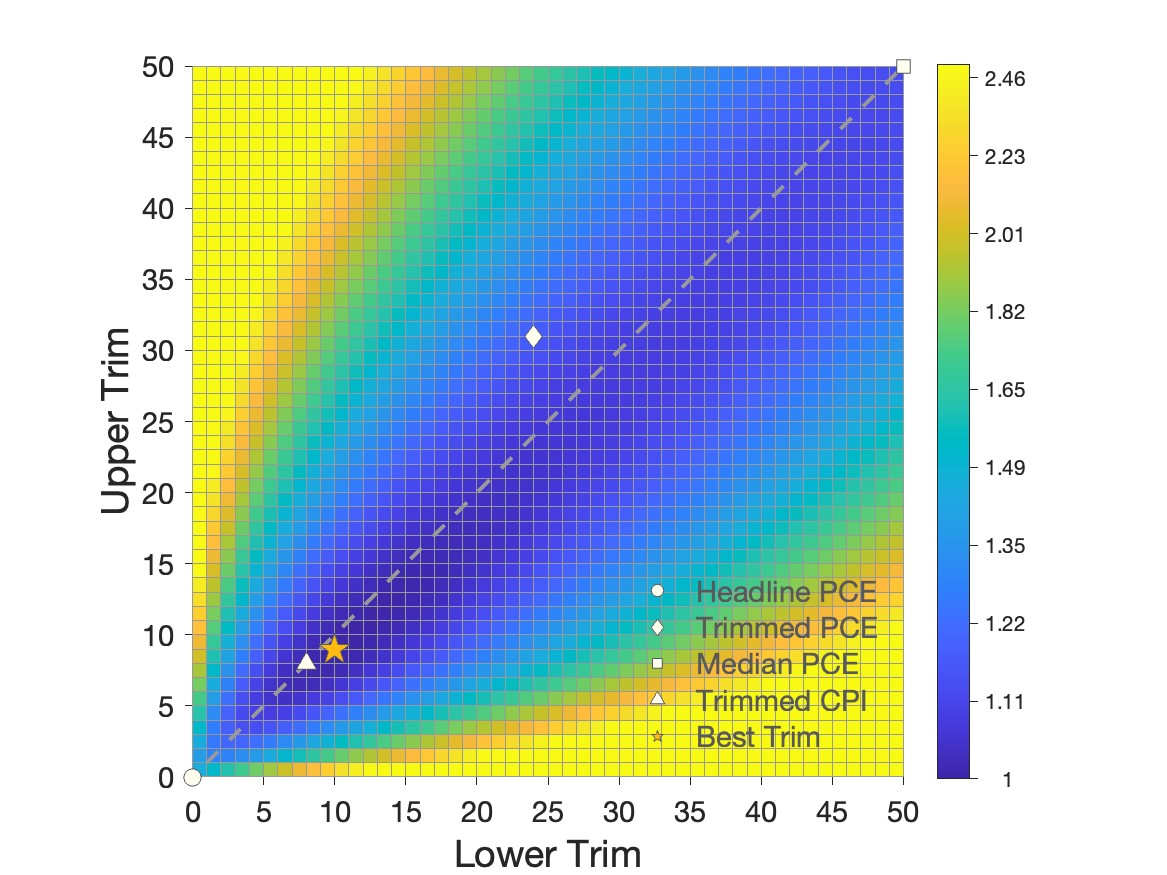}
    \end{subfigure}
    ~
    \begin{subfigure}[t]{0.48\textwidth}
    \caption{Forward Trend: 1970--89}
    \includegraphics[width=1.0\textwidth]{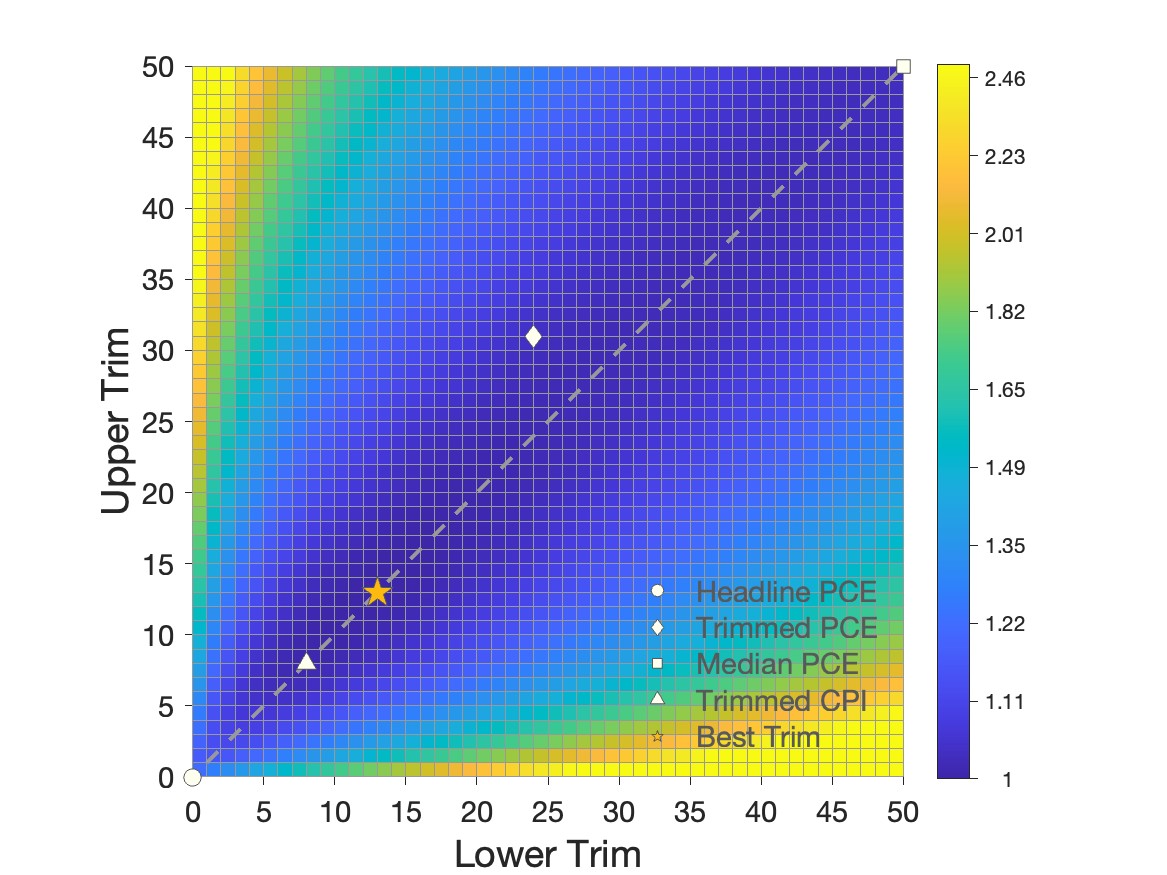}
    \end{subfigure}

    \begin{subfigure}[t]{0.48\textwidth}
    \caption{Band-Pass Trend: 2000--2024}
    \includegraphics[width=1.0\textwidth]{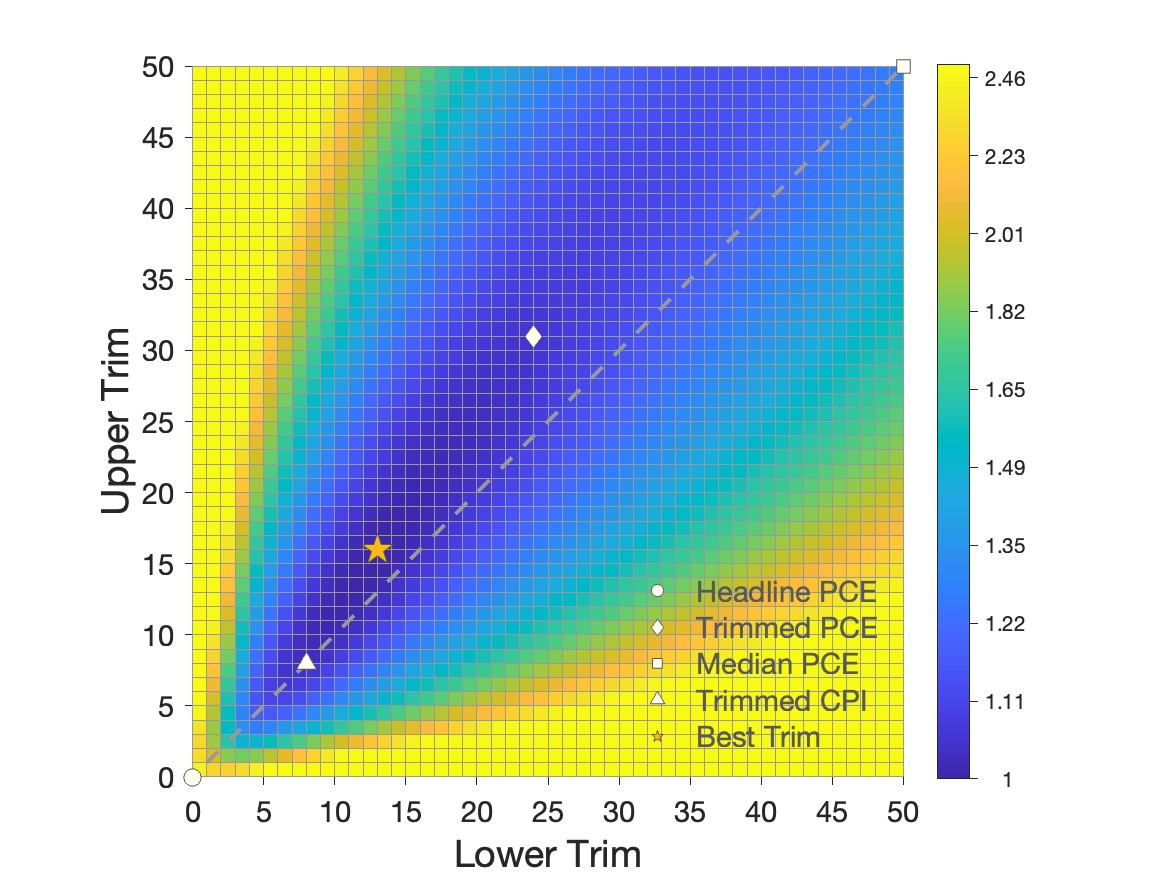}
    \end{subfigure}
    ~
    \begin{subfigure}[t]{0.48\textwidth}
    \caption{Forward Trend: 2000--2024}
    \includegraphics[width=1.0\textwidth]{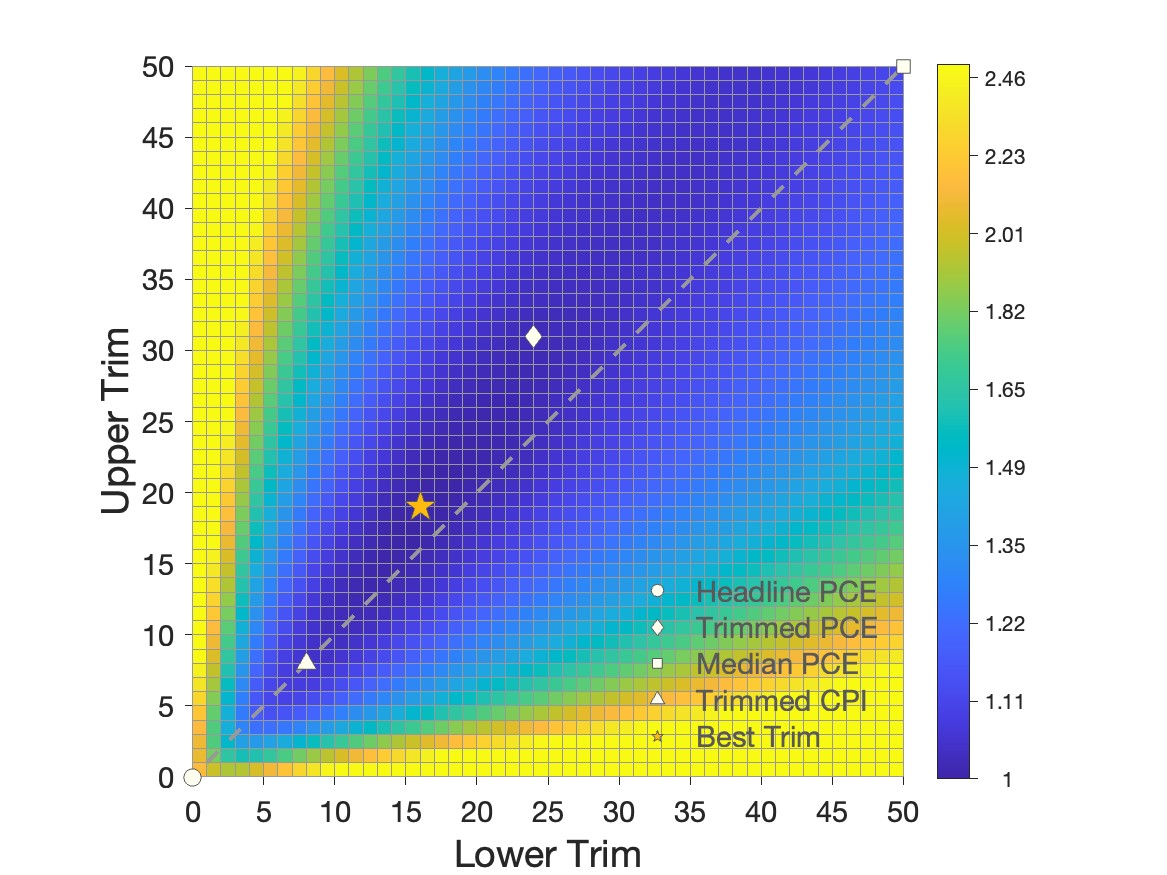}
    \end{subfigure}

    \end{centering}
    
    \vspace{-0.2cm}
    \singlespacing
    \textbf{\textit{\footnotesize{}Note:}}\textit{\footnotesize{} 
    The figures show heat maps of the root-mean-square error (RMSE) when targeting band-pass and forward trend inflation with different trimmed mean inflation measures.
    To ensure comparability across plots, the RMSE numbers are reported relative to the RMSE of the best trim reported in Table \ref{tab:best_trim_All}.
    }
    
\end{figure}

\begin{figure}
    \begin{centering}
        
    \caption{Statistical Difference of RMSE across Trims - Alternative Trends}\label{fig: DM_Test_Alternative}
    
    \begin{subfigure}[t]{0.48\textwidth}
    \caption{Band-Pass Trend: 1970--2024}
    \includegraphics[width=1.0\textwidth]{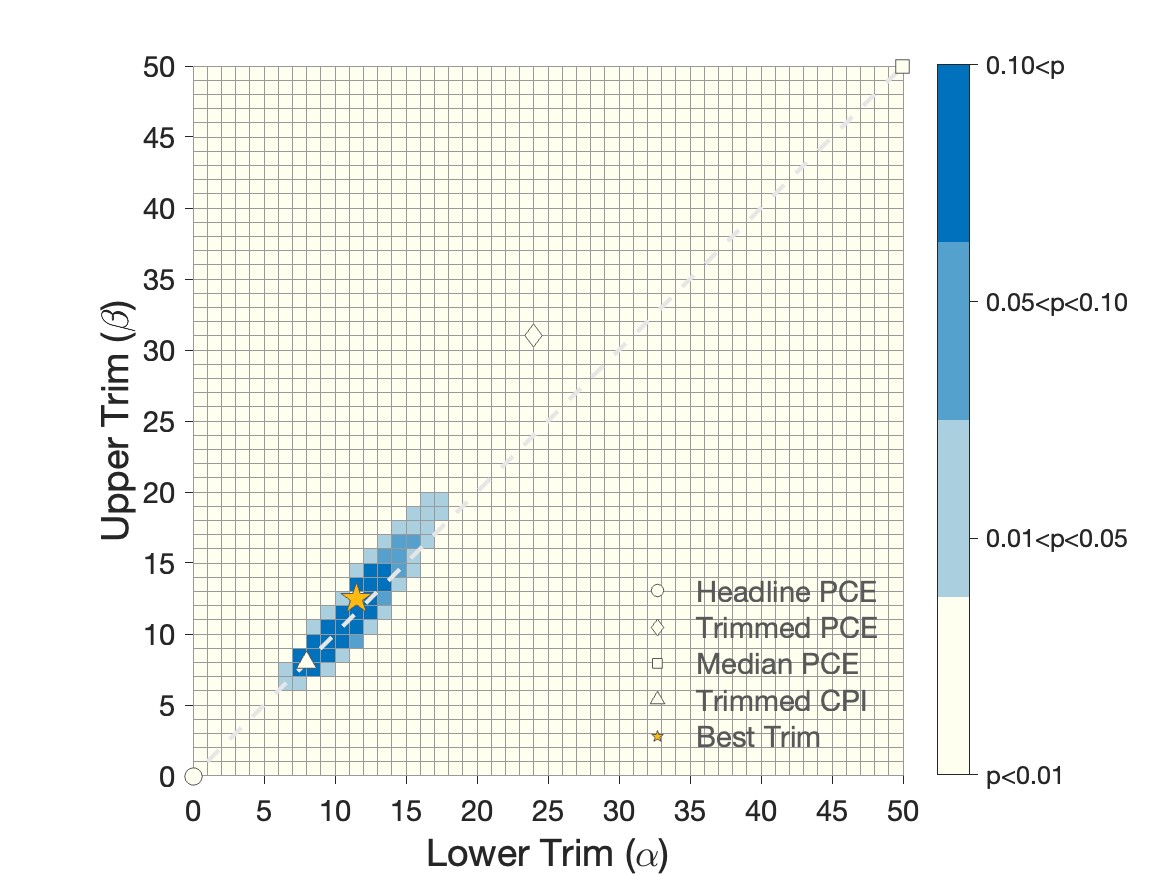}
    \end{subfigure}
    ~
    \begin{subfigure}[t]{0.48\textwidth}
    \caption{Forward Trend: 1970--2024}
    \includegraphics[width=1.0\textwidth]{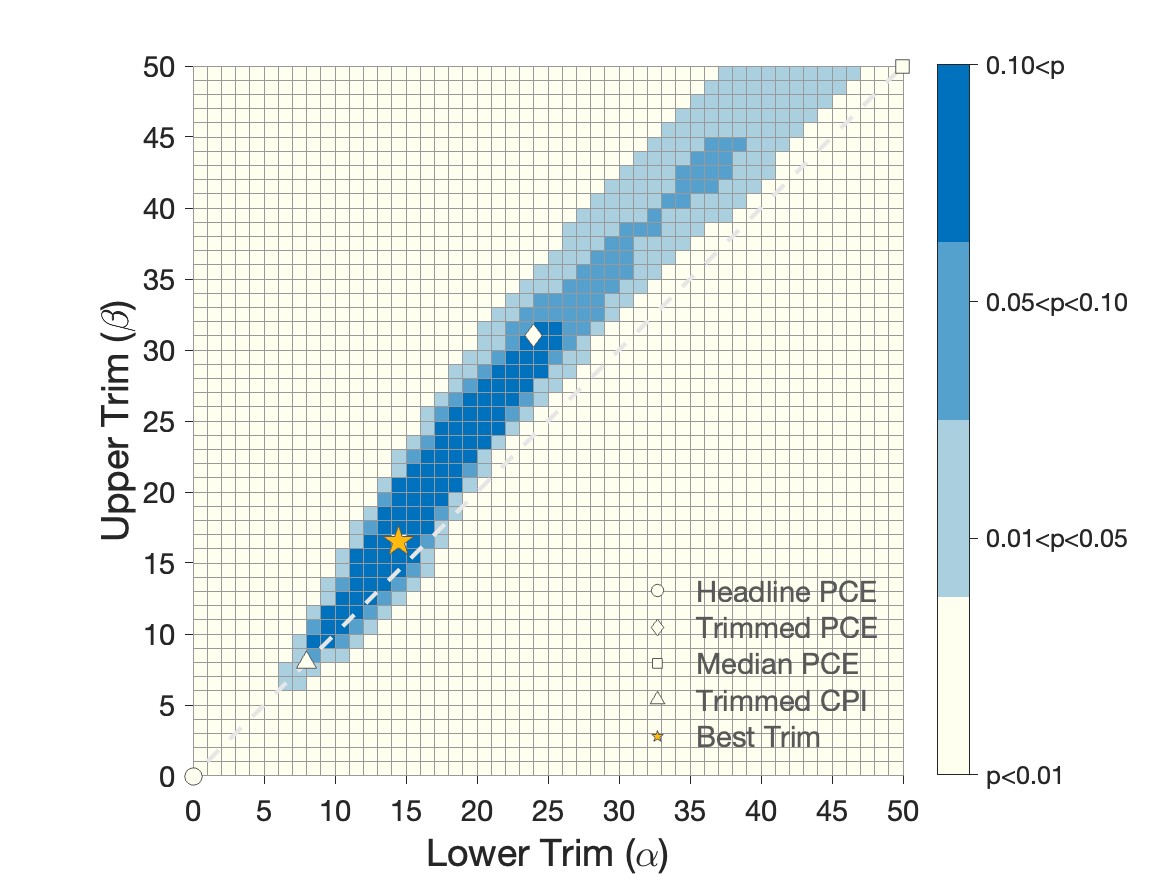}
    \end{subfigure}
    
    \begin{subfigure}[t]{0.48\textwidth}
    \caption{Band-Pass Trend: 1970--89}
    \includegraphics[width=1.0\textwidth]{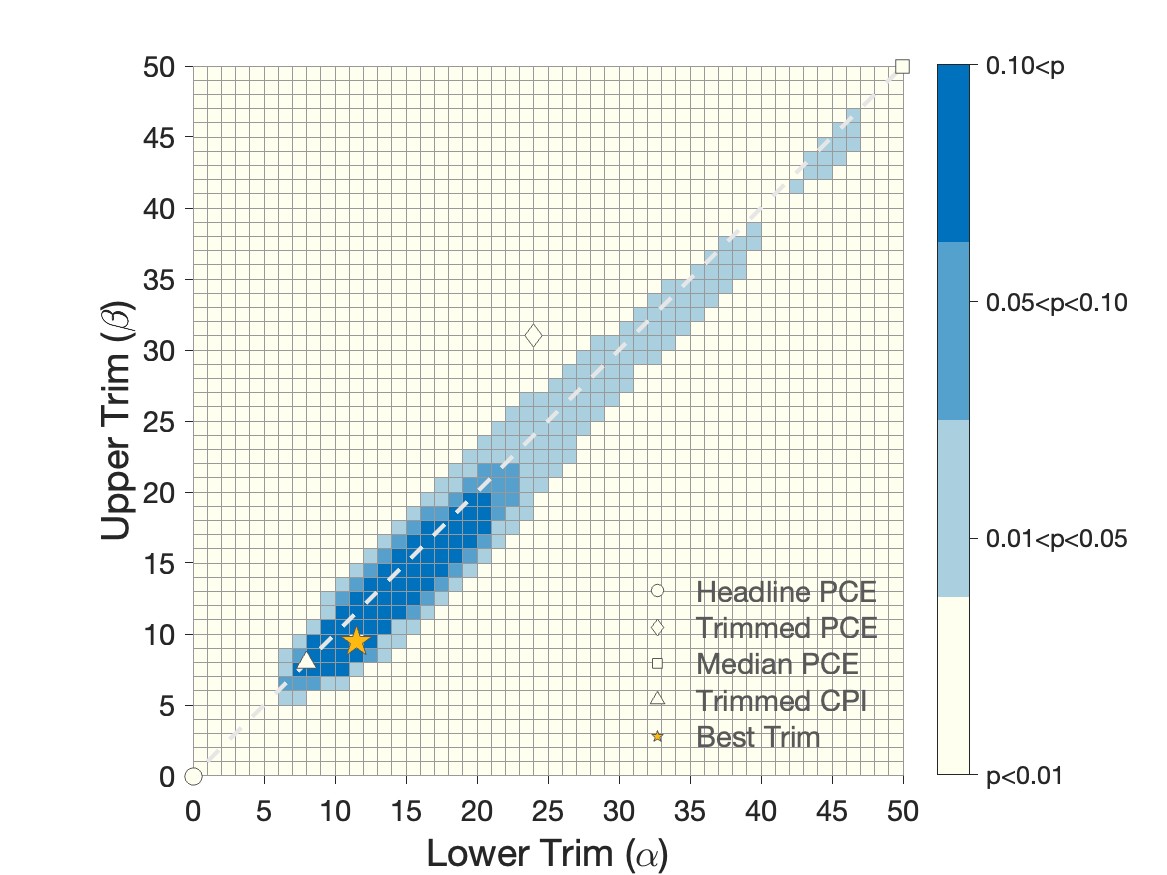}
    \end{subfigure}
    ~
    \begin{subfigure}[t]{0.48\textwidth}
    \caption{Forward Trend: 1970--89}
    \includegraphics[width=1.0\textwidth]{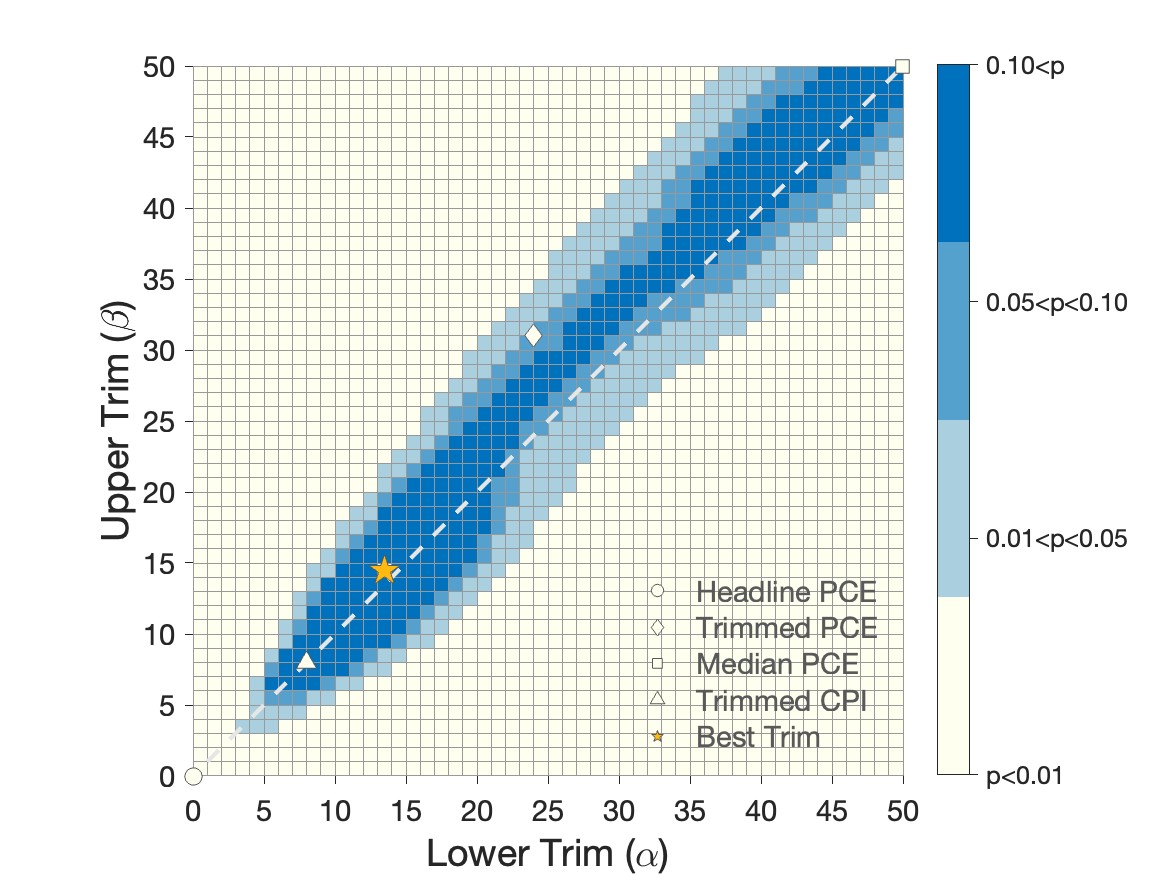}
    \end{subfigure}

    \begin{subfigure}[t]{0.48\textwidth}
    \caption{Band-Pass Trend: 2000--2024}
    \includegraphics[width=1.0\textwidth]{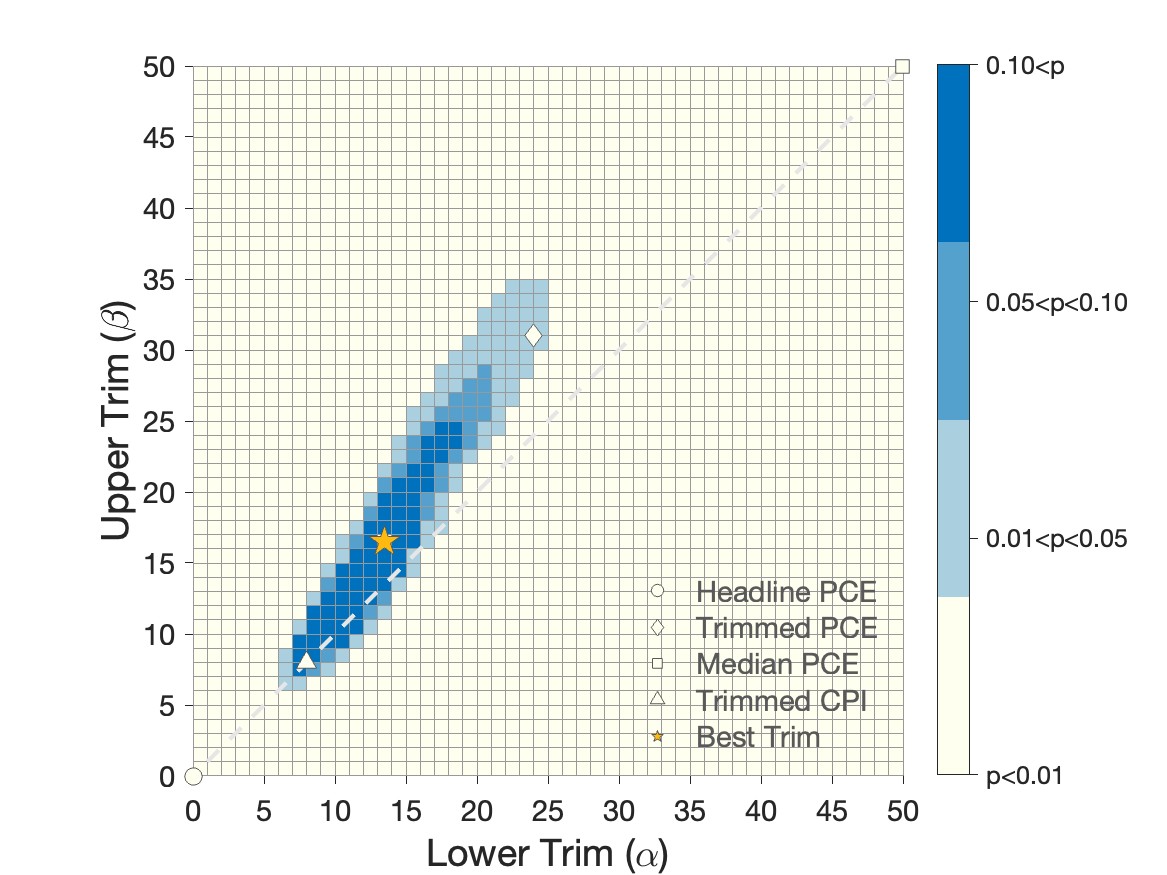}
    \end{subfigure}
    ~
    \begin{subfigure}[t]{0.48\textwidth}
    \caption{Forward Trend: 2000--2024}
    \includegraphics[width=1.0\textwidth]{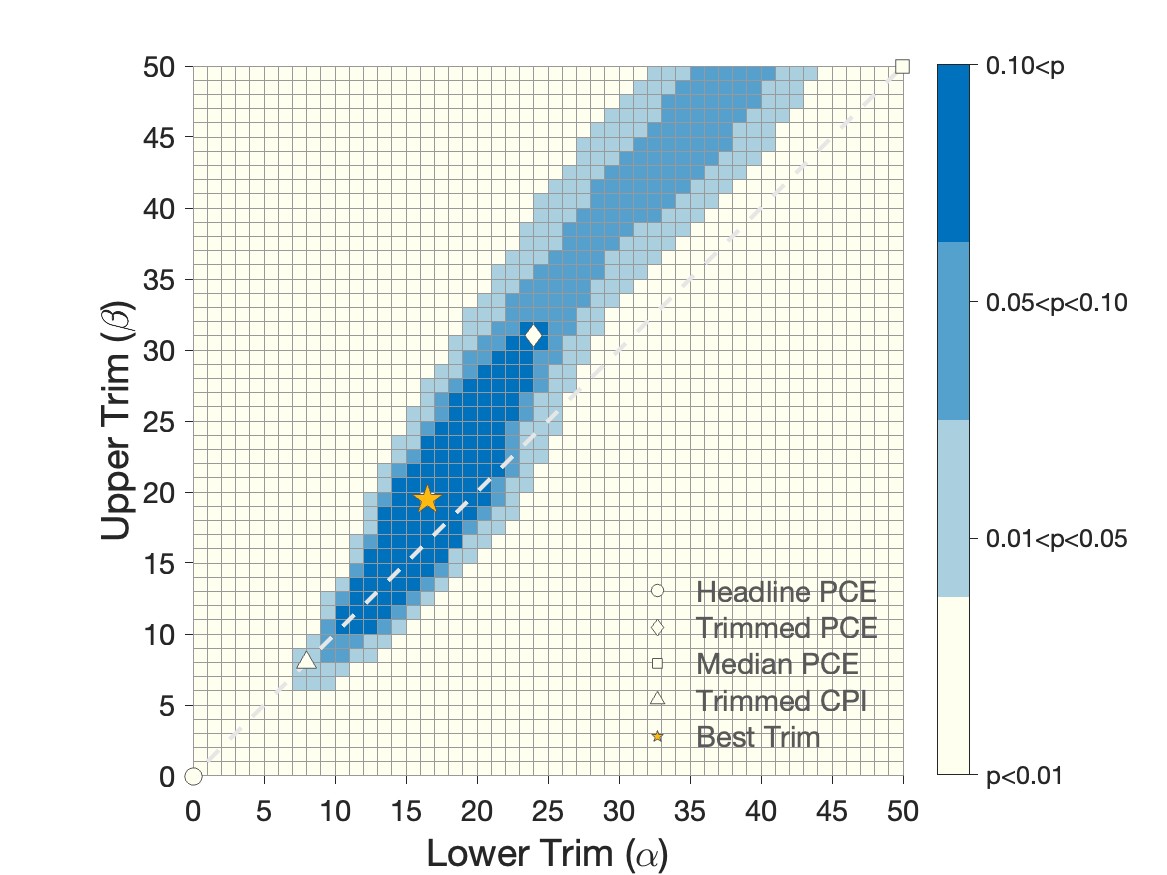}
    \end{subfigure}

    \end{centering}
    
    \vspace{-0.2cm}
    \singlespacing
    \textit{\footnotesize{}Note:}\footnotesize{} 
    The figures group trims according to the outcome of the \citet{Diebold_Mariano_1995}  test, which compares the root-mean-square error (RMSE) implied by the trims with the RMSE of the best trim as presented in Table \ref{tab:best_trim_All}. 
    The trims are grouped based on the p-value of the test. 
    The darkest region consists of trims whose RMSE is statistically equivalent to the lowest RMSE across all trims.

\end{figure}

\begin{figure}
    
    \begin{centering}
        
    \caption{The Behavior of Trimmed Mean Measures}\label{fig: Trimmed_Mean_Properties}
    
    \begin{subfigure}[t]{0.48\textwidth}
    \caption{Coefficient of Variation}
    \includegraphics[width=0.9\textwidth]{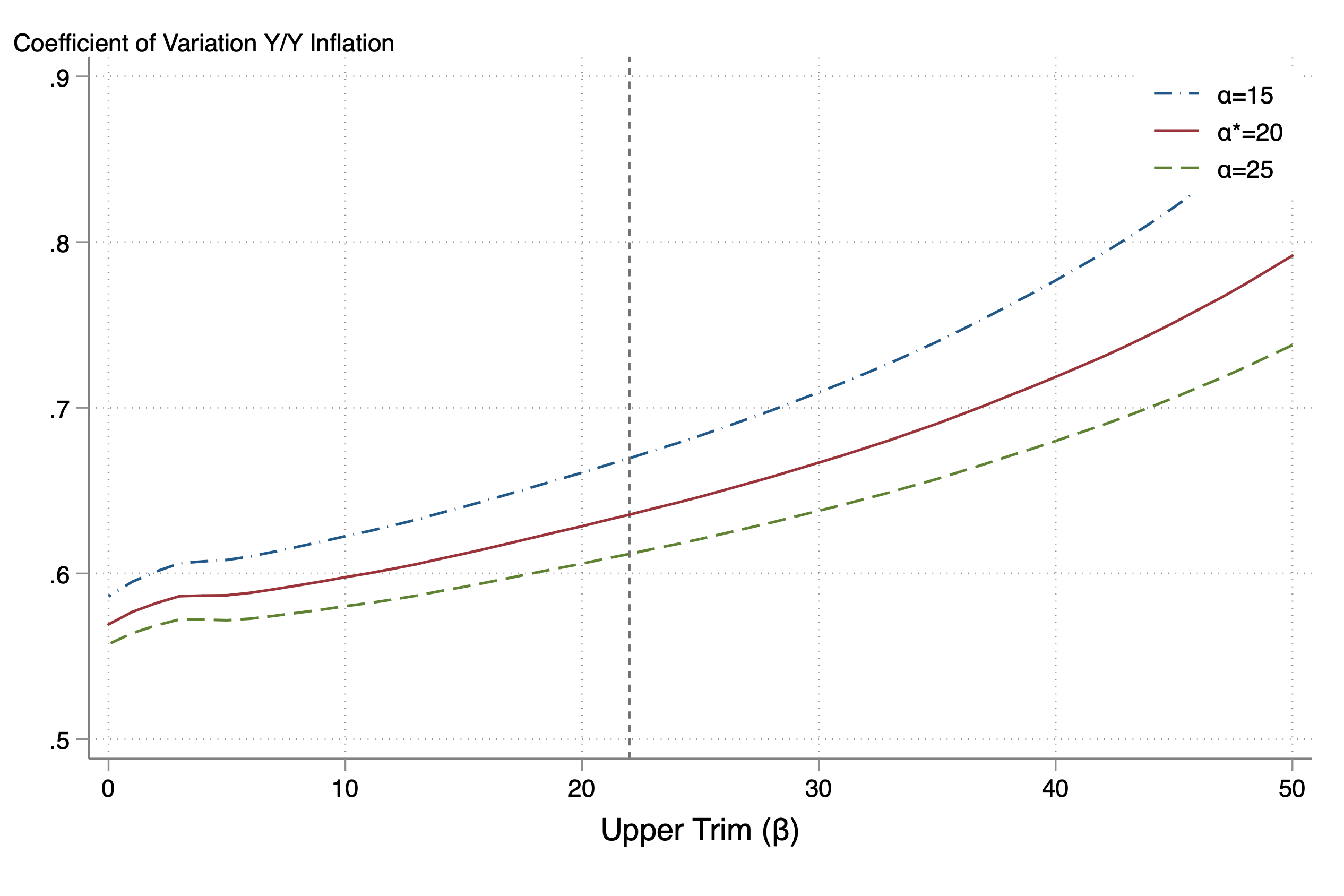}
    \end{subfigure}
    ~
    \begin{subfigure}[t]{0.48\textwidth}
    \caption{RMSE - Current Trend}
    \includegraphics[width=0.9\textwidth]{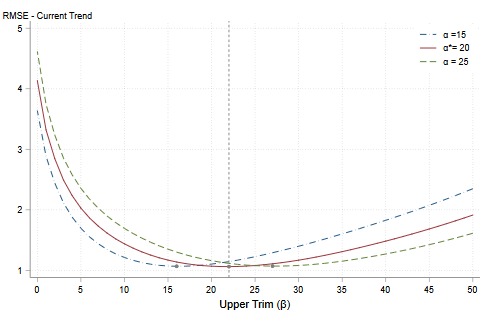}
    \end{subfigure}
    
    \end{centering}
    
    \vspace{-0.2cm}
    \singlespacing
    \footnotesize{}\textit{Notes:}
    The left panel shows the coefficient of variation for different upper trims for trimmed mean measures with three different levels of lower trim.  
    The coefficient of variation is the ratio between the standard deviation and the mean of thee series over the 1970--2024 sample.
    The right panel shows the RMSE of the series when targeting current trend inflation. 
    The dotted vertical line signals the optimal upper trim and the $*$ indicates the optimal lower trim.

\end{figure}

\begin{figure}
    
    \begin{centering}
        
    \caption{Prediction Range across Best Trims across Time}\label{fig: Prediction_Range}
    
    \begin{subfigure}[t]{0.48\textwidth}
    \caption{Centered Trend}
    \includegraphics[width=1.0\textwidth]{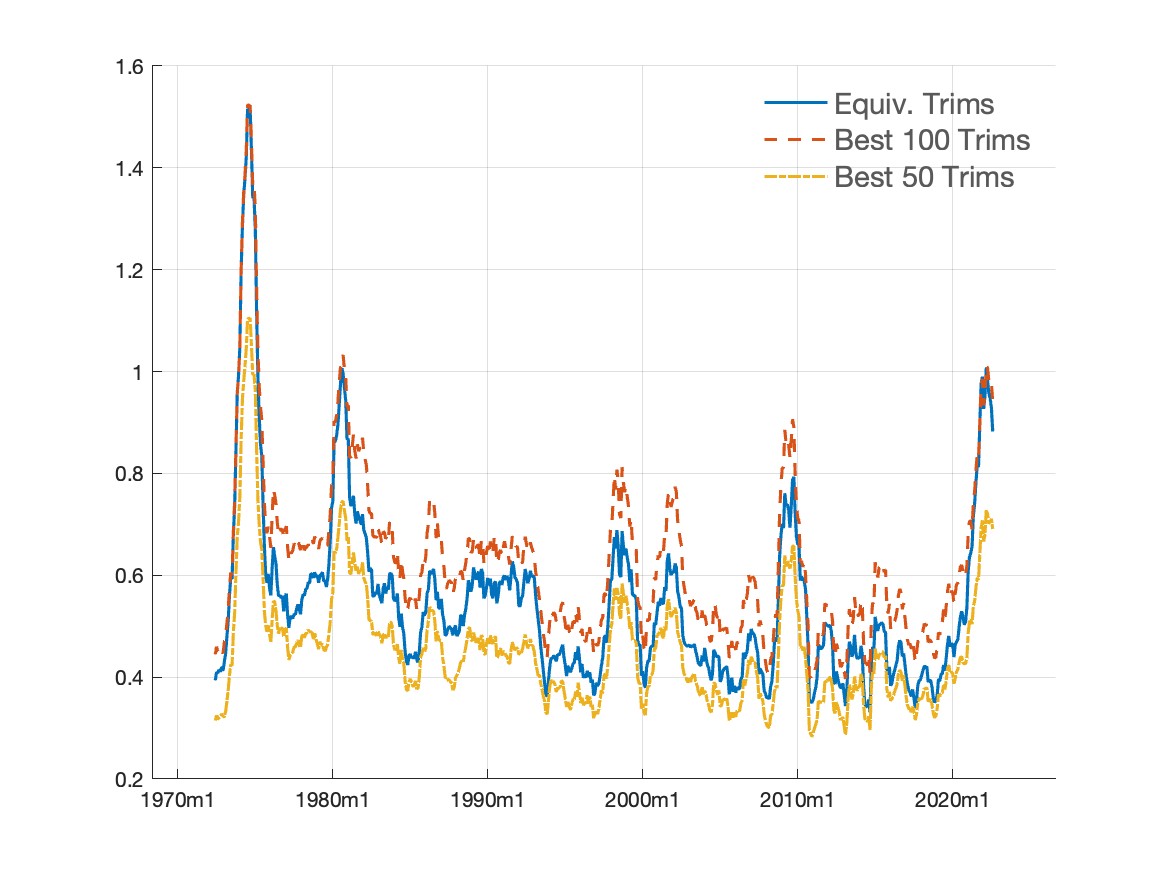}
    \end{subfigure}
    ~
    \begin{subfigure}[t]{0.48\textwidth}
    \caption{Future Trend}
    \includegraphics[width=1.0\textwidth]{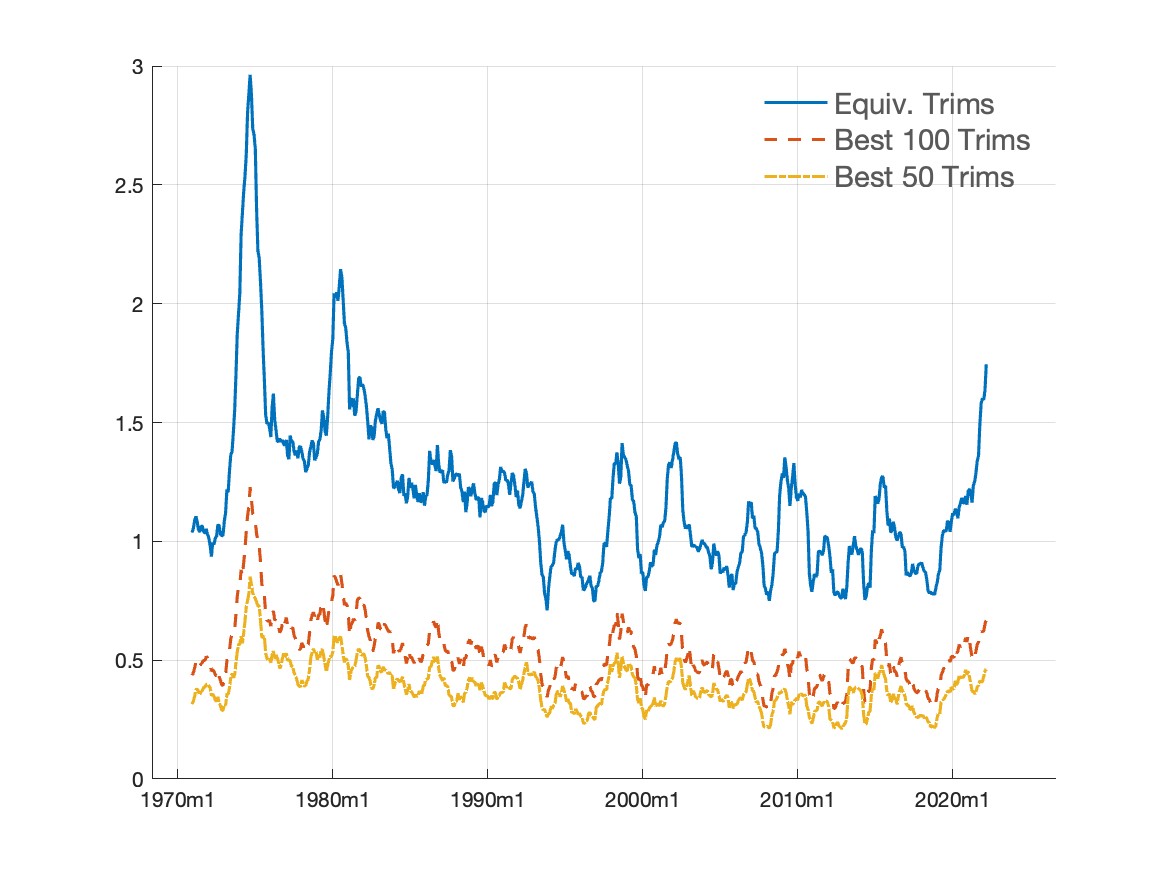}
    \end{subfigure}
    
    \end{centering}
    
    \vspace{-0.2cm}
    \singlespacing
    \footnotesize{}\textit{Notes:}
    The figures plot the range between the lowest and highest predictions for the current and future trends between 1970 and 2024 given a set of trimmed mean measures of inflation. 
    There are three ranges for each inflation target: first, the range implied by considering the trims that are statistically equivalent at the 5 percent level according to the \citet{Diebold_Mariano_1995} test of the difference of their root-mean-square errors (RMSEs) with respect to the RMSE of the best trim (see Table \ref{tab:best_trim}); second, the range implied by considering the best 100 trims as ranked by their RMSE; third, the range implied by considering the best 50 trims as ranked by their RMSE.

\end{figure}

\begin{figure}
    \begin{centering}
    
    \caption{Average Range of Inflation by Trims}\label{fig: Inf_Range_by_Trims}
    
    \includegraphics[scale=0.25]{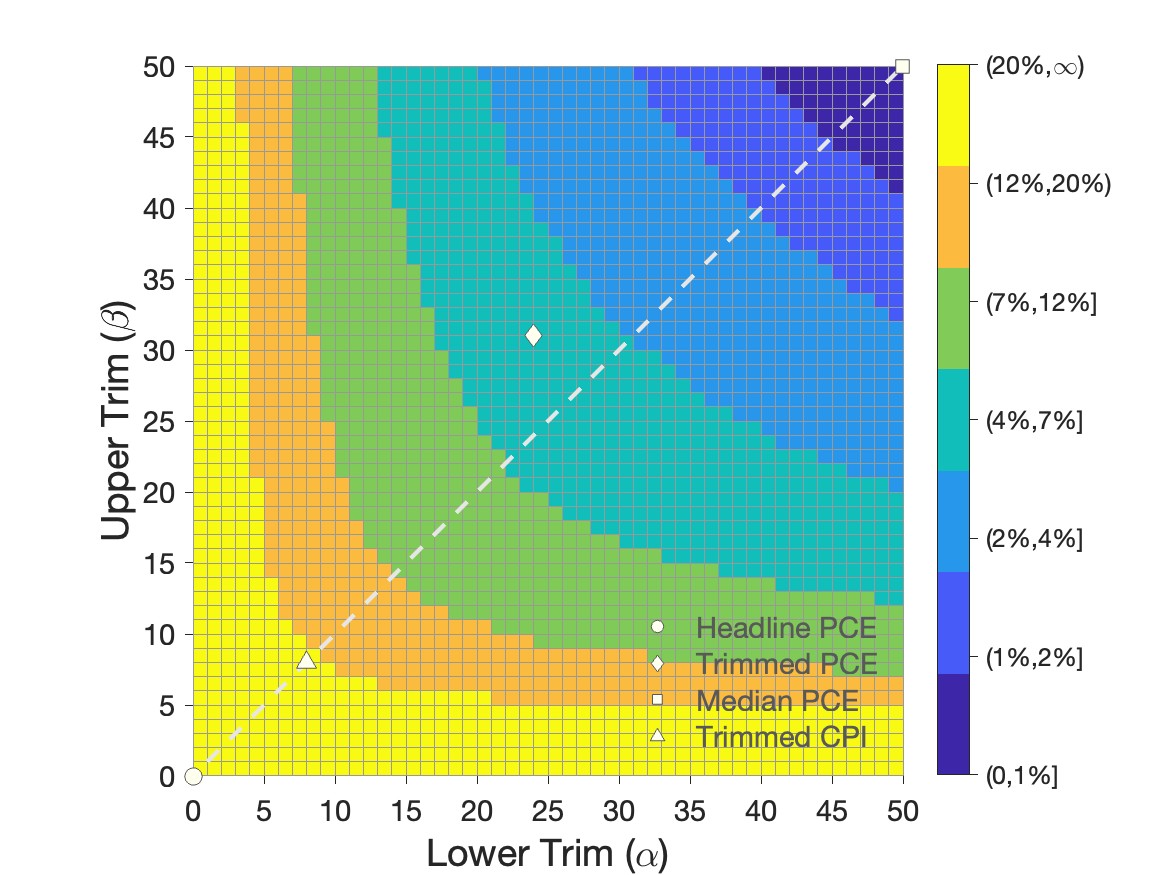}
    
    \end{centering}
    
    \vspace{-0.2cm}
    \singlespacing
    \textit{\footnotesize{}Note:}\footnotesize{} 
    The figure shows the average range of inflation rates across individual expenditure categories implied by each trim combination, $\pi_{1-\beta}-\pi_{\alpha}$.

\end{figure}

\begin{figure}
    
    \begin{centering}
        
    \caption{Range of Robust Measures of Inflation}\label{fig: robust_range_examples}
    
    \begin{subfigure}[t]{0.48\textwidth}
    \caption{1971--1976}\label{fig:robust_range_1970s}
    \includegraphics[width=1.0\textwidth]{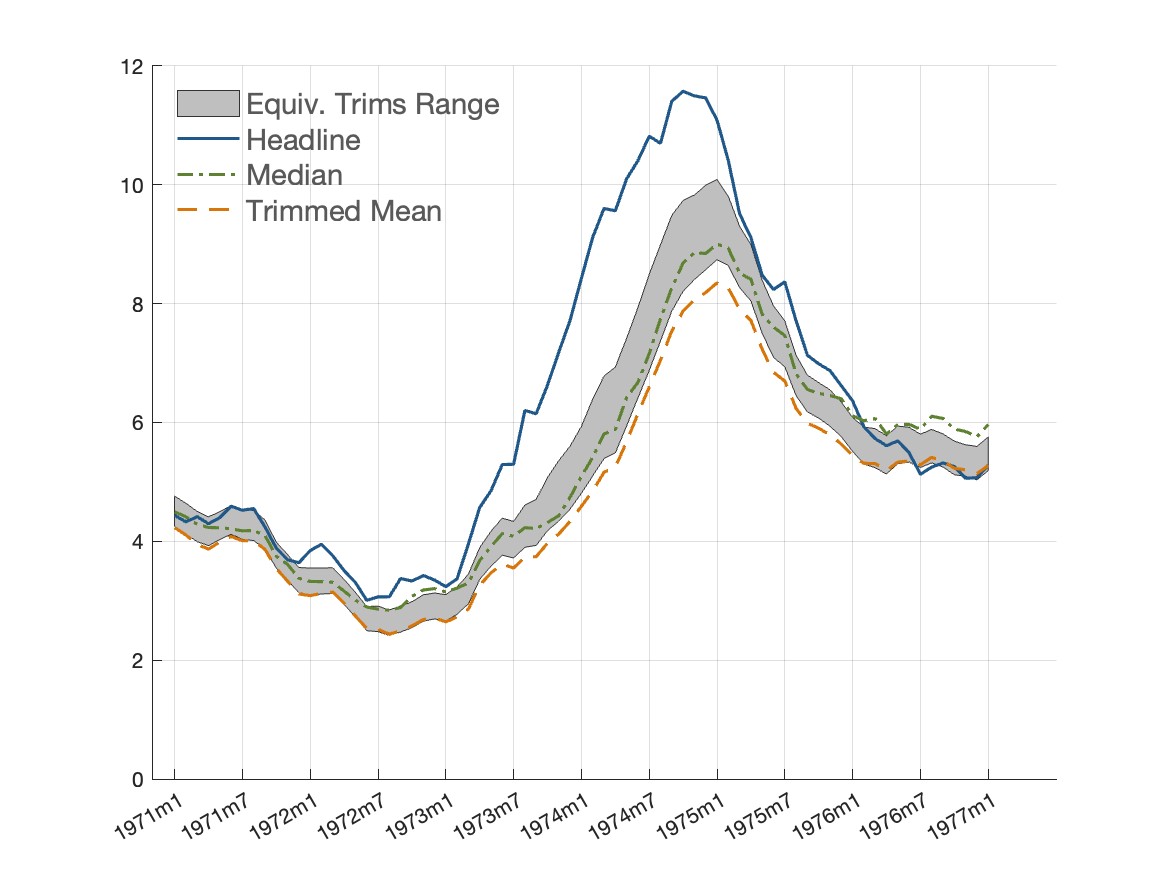}
    \end{subfigure}
    ~
    \begin{subfigure}[t]{0.48\textwidth}
    \caption{1971--1976}\label{fig:robust_range_2008}
    \includegraphics[width=1.0\textwidth]{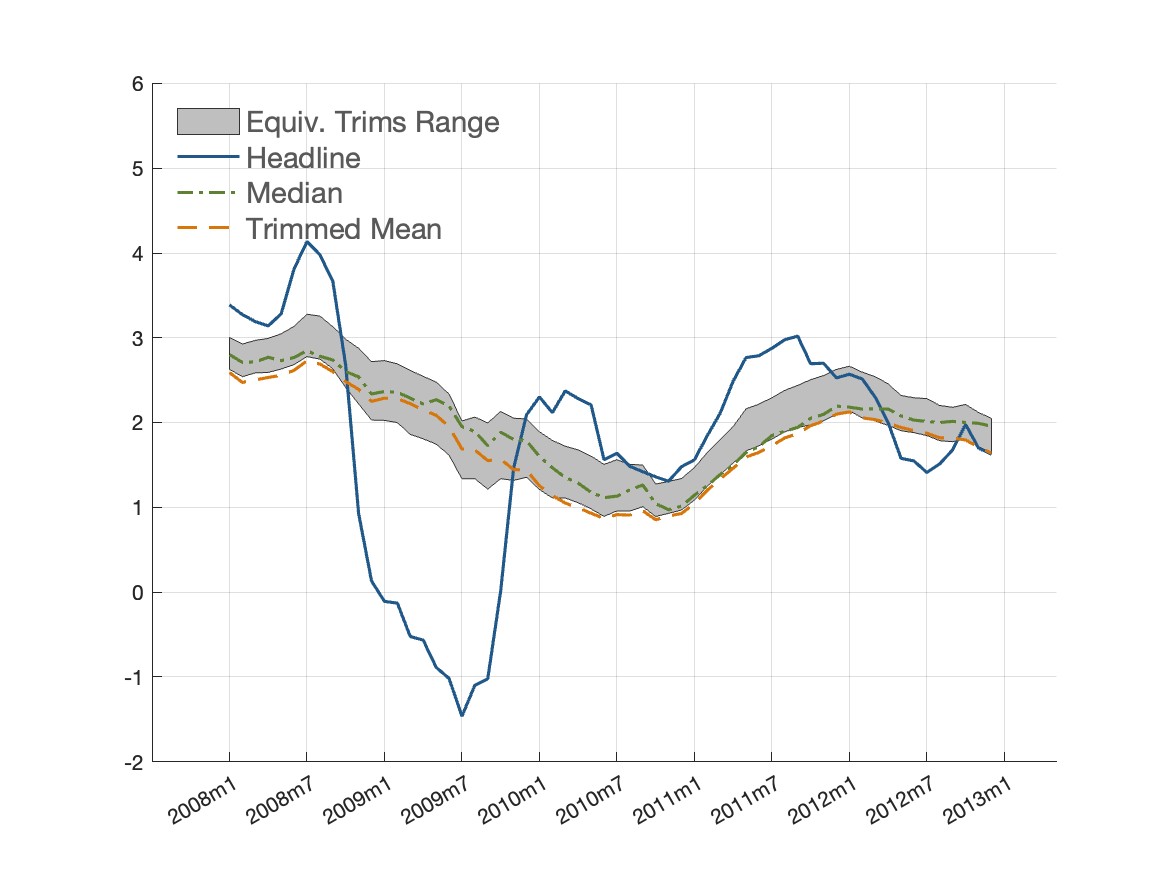}
    \end{subfigure}
    
    \end{centering}
    
    \vspace{-0.2cm}
    \singlespacing
    \footnotesize{}\textit{Notes:}
    The figures shows the range (in gray) of inflation implied by the set of trimmed mean inflation measures whose RMSE with respect to current trend inflation are statistically equivalent to the best trim at the 5 percent level.
    The figure also includes trimmed mean personal consumption expenditure (PCE) and median PCE as calculated by the authors using the methodologies of \citet{CleMedian} and \citet{DallasTrimmed}, 
    as well as headline PCE inflation, taken directly from the PCE data published by the Bureau of Economic Analysis.
    
\end{figure}

\begin{figure}
    \begin{centering}
        
    \caption{Prediction across Best Trims - February 2024}\label{fig: Prediction_HM}
    
    \begin{subfigure}[t]{0.48\textwidth}
    \caption{Centered Trend}
    \includegraphics[width=1.0\textwidth]{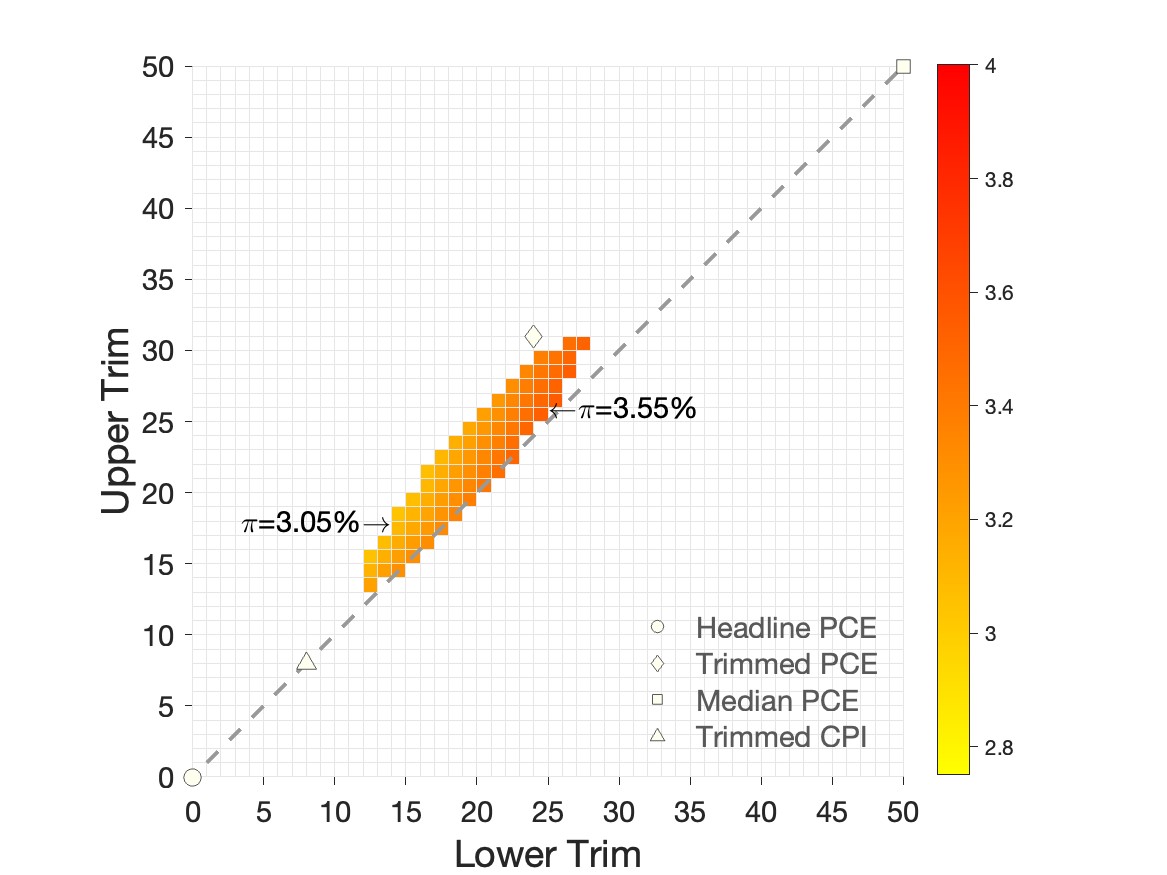}
    \end{subfigure}
    ~
    \begin{subfigure}[t]{0.48\textwidth}
    \caption{Future Trend}
    \includegraphics[width=1.0\textwidth]{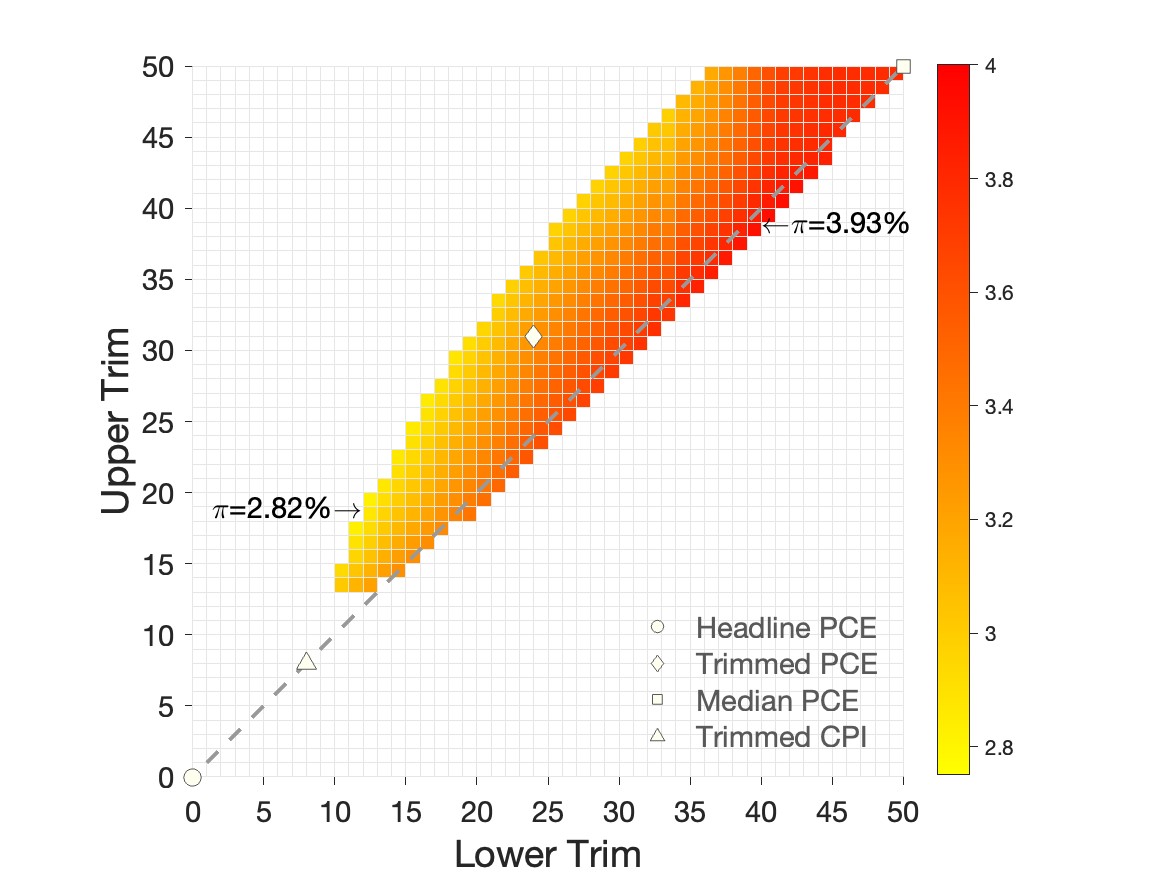}
    \end{subfigure}
    
    \begin{subfigure}[t]{0.48\textwidth}
    \caption{Band-Pass Trend}
    \includegraphics[width=1.0\textwidth]{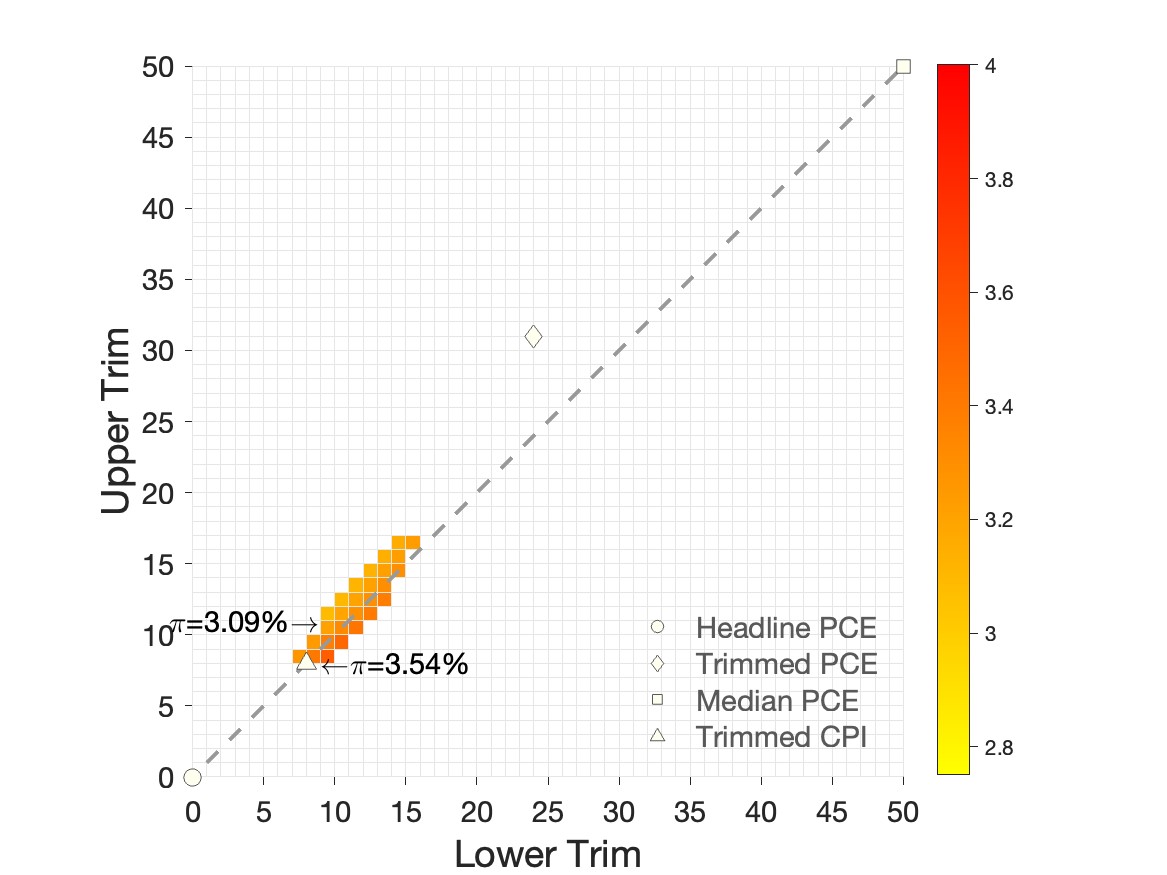}
    \end{subfigure}
    ~
    \begin{subfigure}[t]{0.48\textwidth}
    \caption{Forward Trend}
    \includegraphics[width=1.0\textwidth]{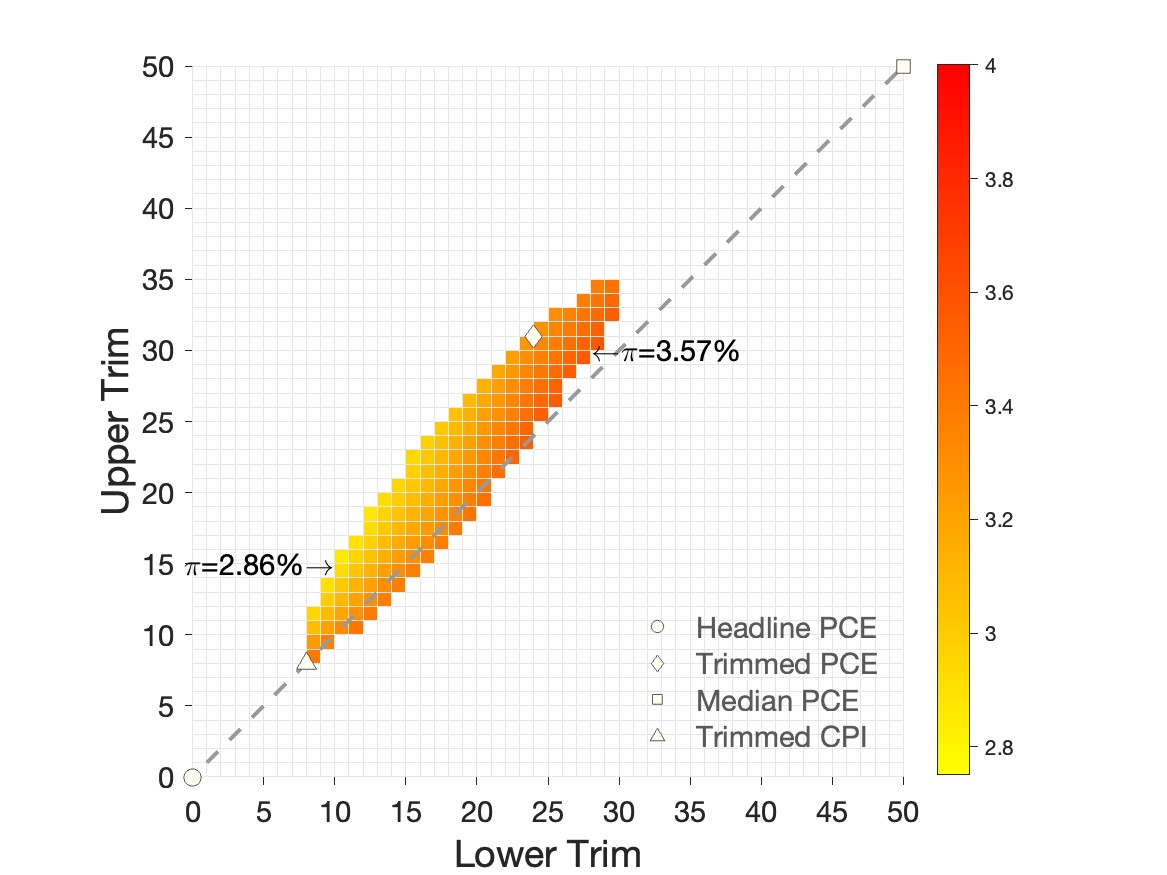}
    \end{subfigure}
    
    \end{centering}
    
    \vspace{-0.2cm}
    \singlespacing
    \textit{\footnotesize{}Note:}\footnotesize{} 
    The figures show heat maps of the prediction level for February 2024 across the best 50 trim combinations, ranked according to their RMSE when targeting band-pass or forward trend inflation over the sample 1970--2024.
    The best trims vary according to the trend inflation series being targeted.
    The set of best trims is defined as those with an RMSE statistically indistinguishable at the 5 percent significance level from the lowest RMSE across all trims.
    
    \label{fig: Prediction_All}
\end{figure}

%% file: Appendix/a02q_Housing.tex
\FloatBarrier
\subsection{Excluding Housing from Trimmed-Mean Inflation}\label{app: Housing}

Housing is the single largest expenditure category and may have different dynamics from the remainder of the consumption basket \citep{adams2022}. 
It is also one of the series most commonly included in the official trimmed mean and median inflation measures (see Table \ref{tab:Excluded_Categories}). 
This can raise questions over housing having too large of a role in the behavior of trimmed mean inflation measures. 
That is not the case. 
We now reproduce our main results while excluding owner occupied housing from the construction of trimmed mean measures and re-weighting the remaining categories accordingly. 
All of our results are preserved, with the obvious exception of the level of the optimal trim cutoffs that changes to reflect the exclusion of housing from the set of expenditure categories. 

\begin{table}[tbh!]
    \caption{Best Trims for Trimmed-Mean Inflation without Housing}
    \begin{center}
    \begin{threeparttable}
    \begin{tabular}{ll cScS cS}
        \hline \hline
        \multirow{2}{*}{Target} & \multirow{2}{*}{Sample} &  & \multicolumn{5}{c}{Best Trim} \\
         \cline{4-8}
         &  &  & {Lower Trim} & & {Upper Trim} & & {RMSE}  \\
        \hline 
                          & {1970--2024} & & 27 & & 28 & & 1.20 \\
        {Centered Trend}  & {1970--1989} & & 26 & & 23 & & 1.65 \\
                          & {2000--2024} & & 27 & & 30 & & 0.88 \\
        \hline
                          & {1970--2024} & & 20 & & 20 & & 1.22 \\
        {Band-Pass Trend} & {1970--1989} & & 22 & & 19 & & 1.47 \\
                          & {2000--2024} & & 18 & & 19 & & 1.09 \\
        \hline
                          & {1970--2024} & & 47 & & 50 & & 2.19 \\
        {Future Trend}    & {1970--1989} & & 24 & & 25 & & 3.02 \\
                          & {2000--2024} & & 47 & & 49 & & 1.65 \\
        \hline
                          & {1970--2024} & & 24 & & 26 & & 1.70 \\
        {Forward Trend}   & {1970--1989} & & 24 & & 23 & & 2.37 \\
                          & {2000--2024} & & 27 & & 28 & & 1.26 \\
        \hline
    \end{tabular}
    \begin{tablenotes}
        \item {\footnotesize \textit{Notes:} 
        The table reports the best trims for different targets of trend inflation over different samples as determined by the predictive performance across trims. 
        The root-mean-square error (RMSE) of the best trim is also reported. 
        }
    \end{tablenotes}
    \end{threeparttable}
    \end{center}
    \label{tab:best_trim_housing}
\end{table}

\begin{figure}
    \begin{centering}
        
    \caption{RMSE across Trims: 1970--2024 (Sample without Housing)}
    
    \begin{subfigure}[t]{0.48\textwidth}
    \caption{Centered Trend}
    \includegraphics[width=1.0\textwidth]{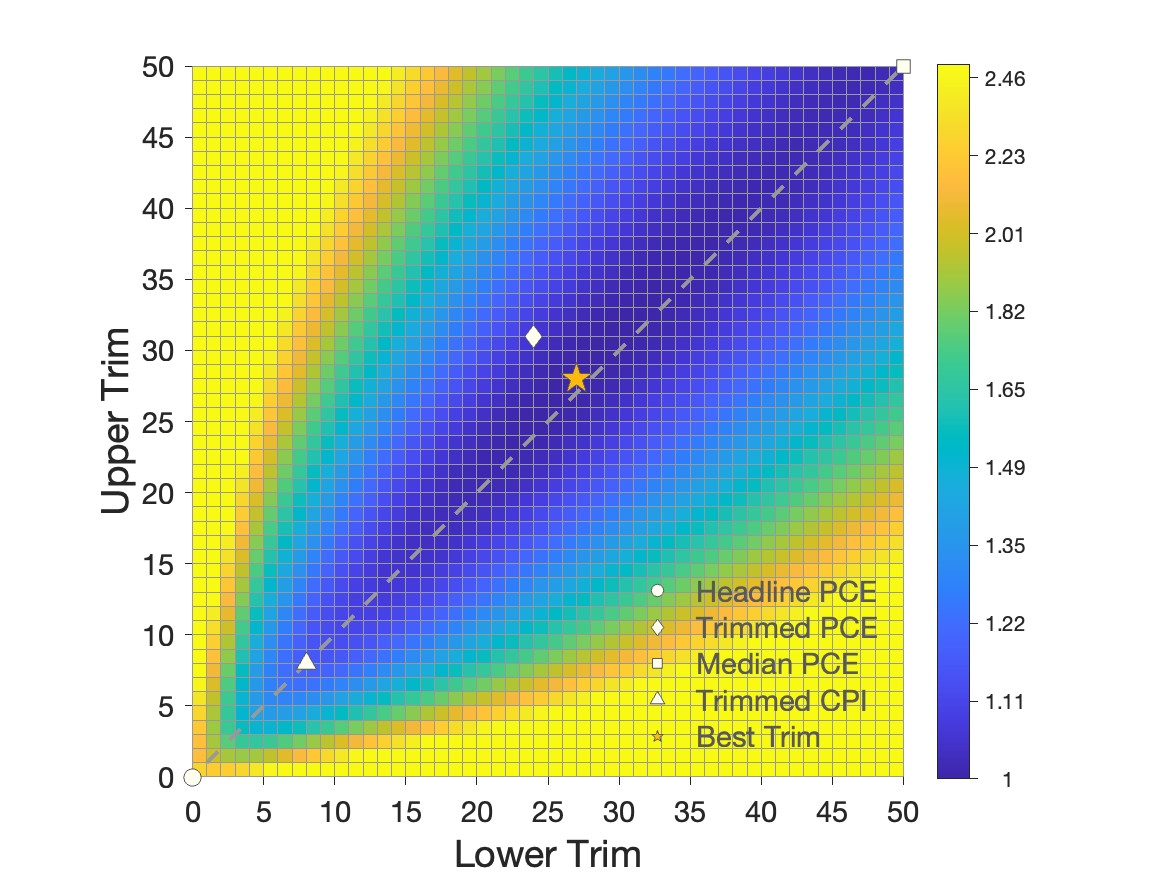}
    
    \end{subfigure}
    ~
    \begin{subfigure}[t]{0.48\textwidth}
    \caption{Forward Trend}
    \includegraphics[width=1.0\textwidth]{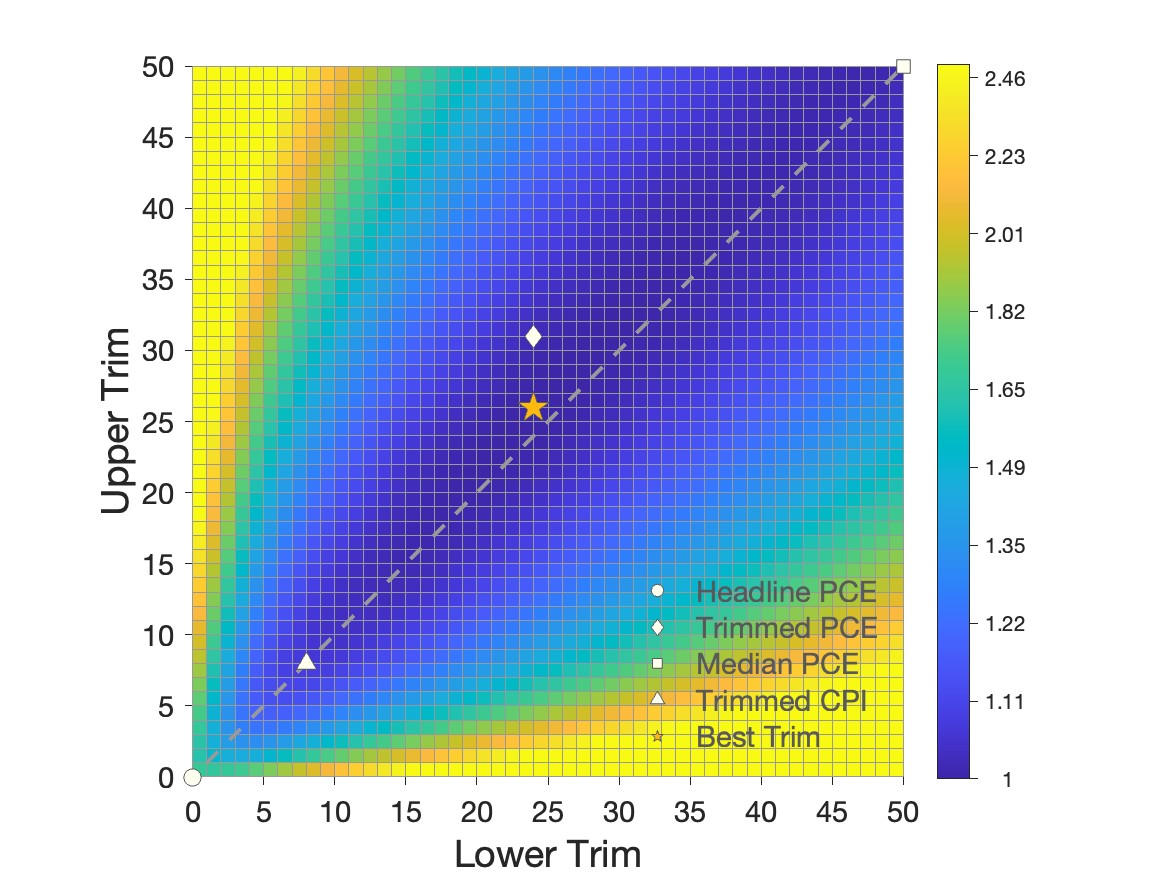}
    \end{subfigure}
    
    \begin{subfigure}[t]{0.48\textwidth}
    \caption{Band-Pass Trend}
    \includegraphics[width=1.0\textwidth]{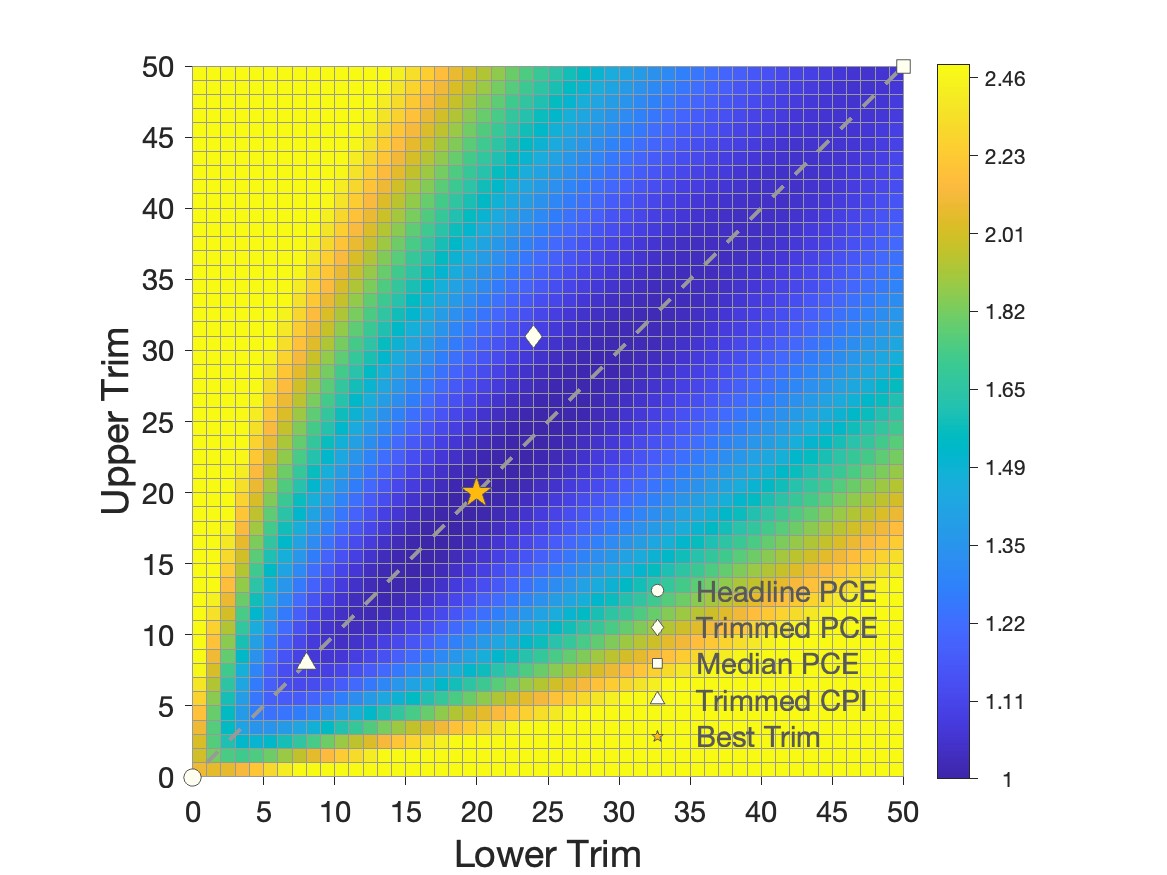}
    \end{subfigure}
    ~
    \begin{subfigure}[t]{0.48\textwidth}
    \caption{Future Trend}
    \includegraphics[width=1.0\textwidth]{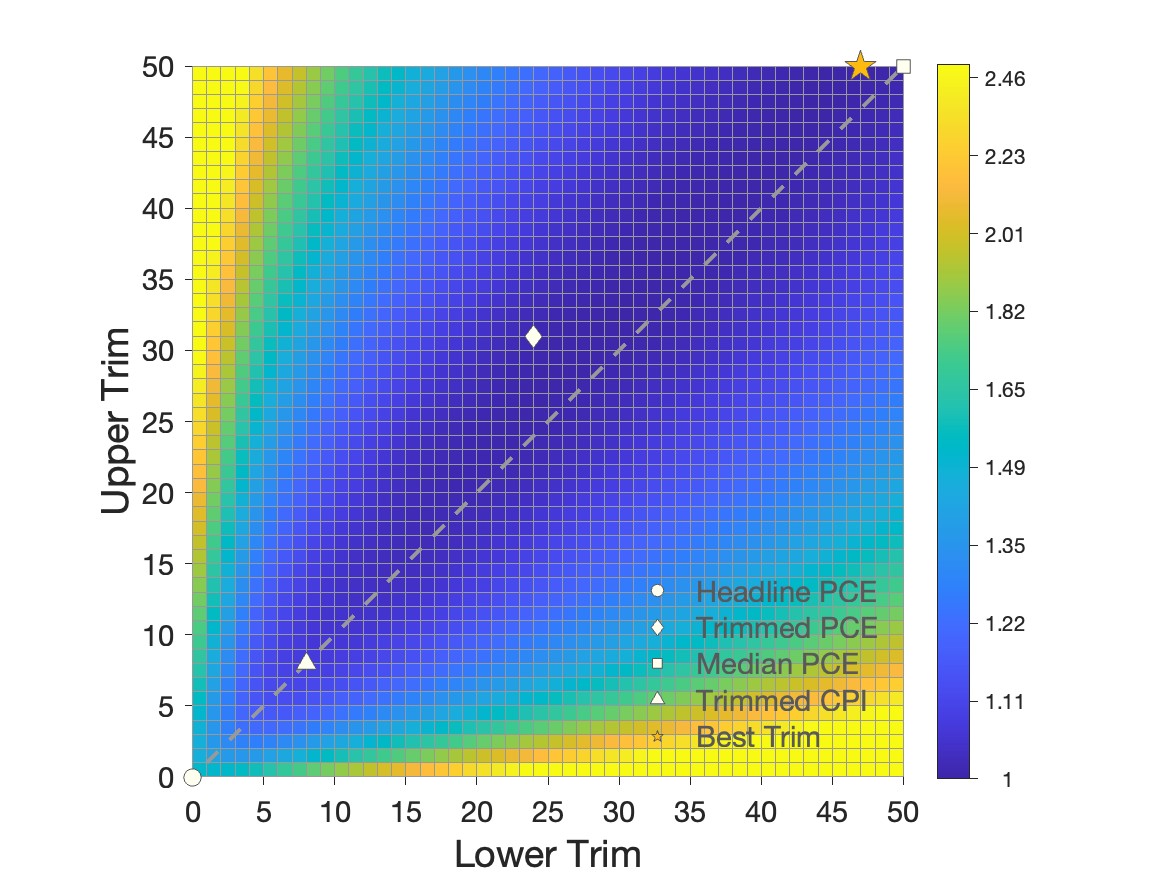}
    \end{subfigure}
    
    \end{centering}
    
    \vspace{-0.2cm}
    \singlespacing
    \textit{\footnotesize{}Note:}\footnotesize{} 
    The figures show heat maps of the RMSE when targeting trend inflation with different combinations of trimmed mean inflation measures.
    Four measures of trend inflation are considered. 
    To ensure comparability across plots, the RMSE numbers are reported relative to the RMSE of the best trim reported in Table \ref{tab:best_trim_housing}.
    
    \label{fig: RMSE_Long_housing}
\end{figure}

\begin{figure}
    \begin{centering}
        
    \caption{RMSE across Trims: 2000--2024 (Sample without Housing)}
    
    \begin{subfigure}[t]{0.48\textwidth}
    \caption{Centered Trend}
    \includegraphics[width=1.0\textwidth]{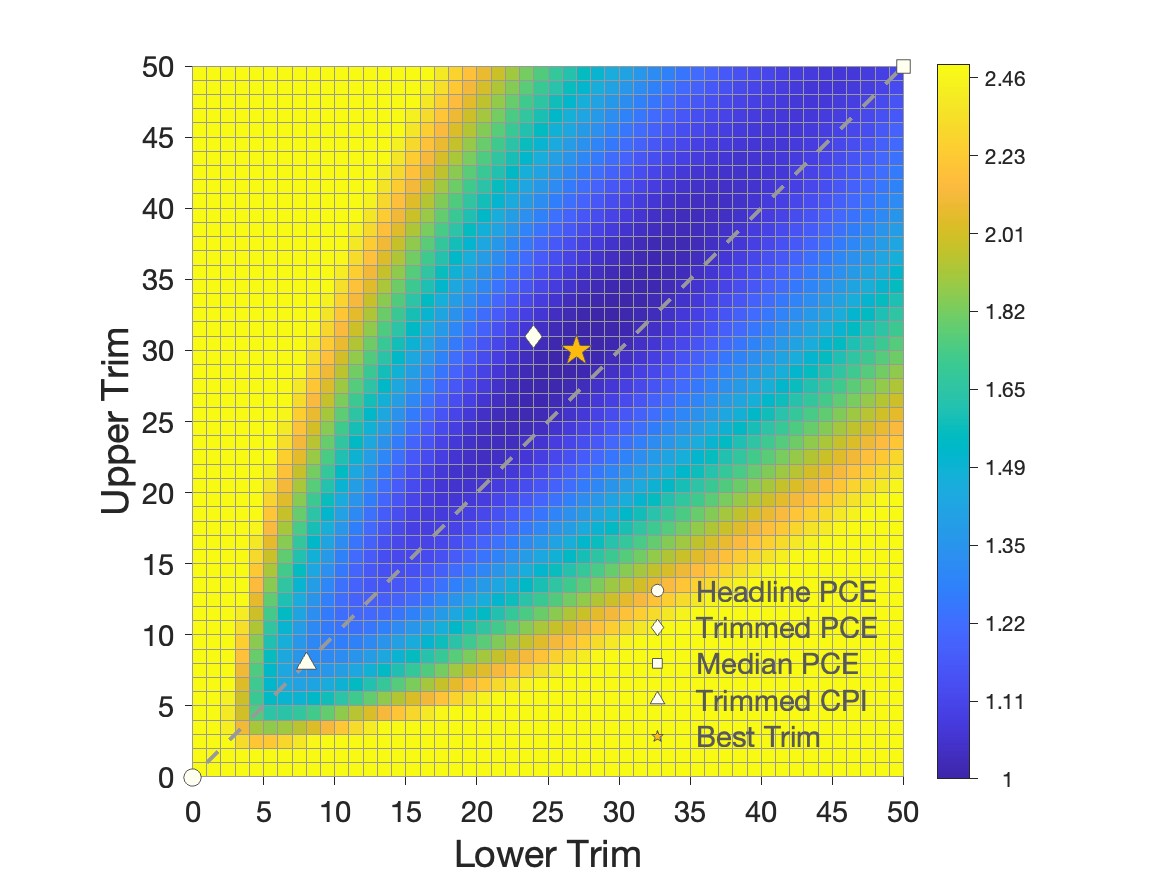}
    \end{subfigure}
    ~
    \begin{subfigure}[t]{0.48\textwidth}
    \caption{Forward Trend}
    \includegraphics[width=1.0\textwidth]{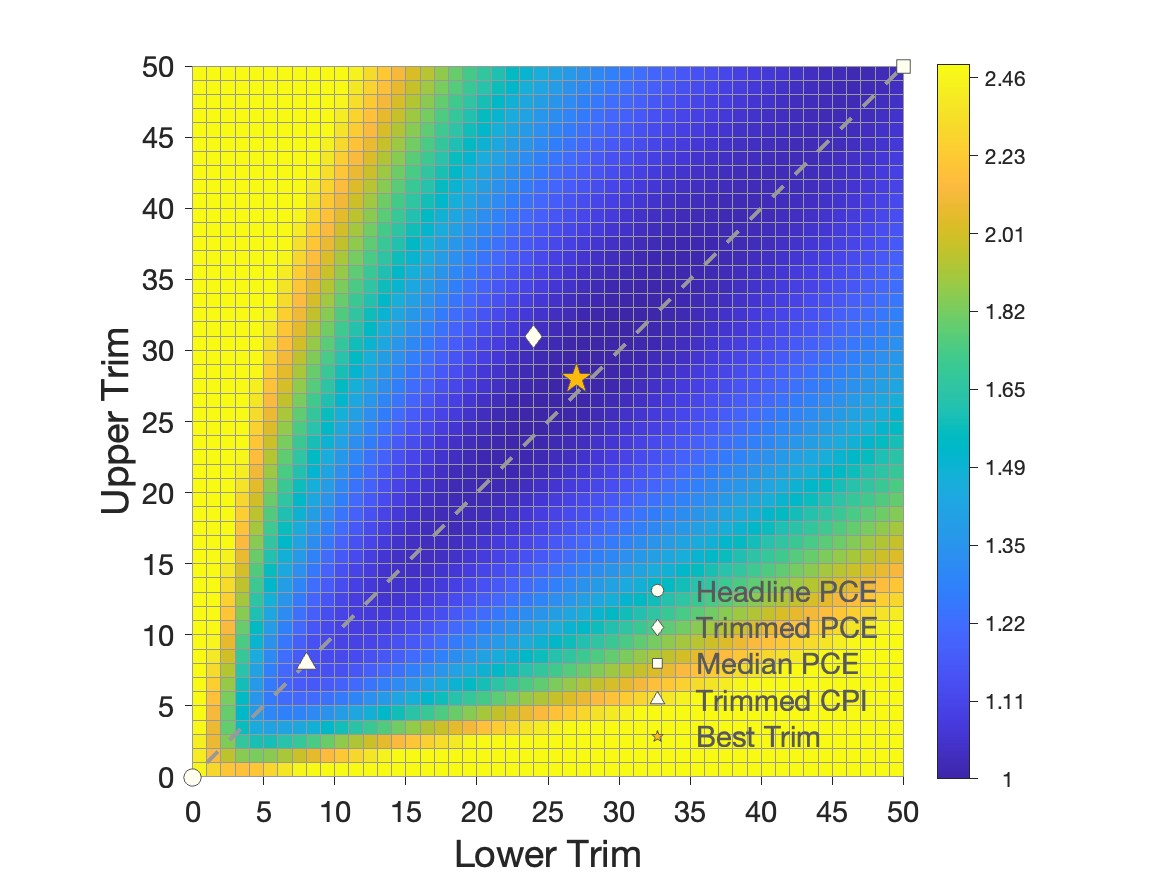}
    \end{subfigure}
    
    \begin{subfigure}[t]{0.48\textwidth}
    \caption{Band-Pass Trend}
    \includegraphics[width=1.0\textwidth]{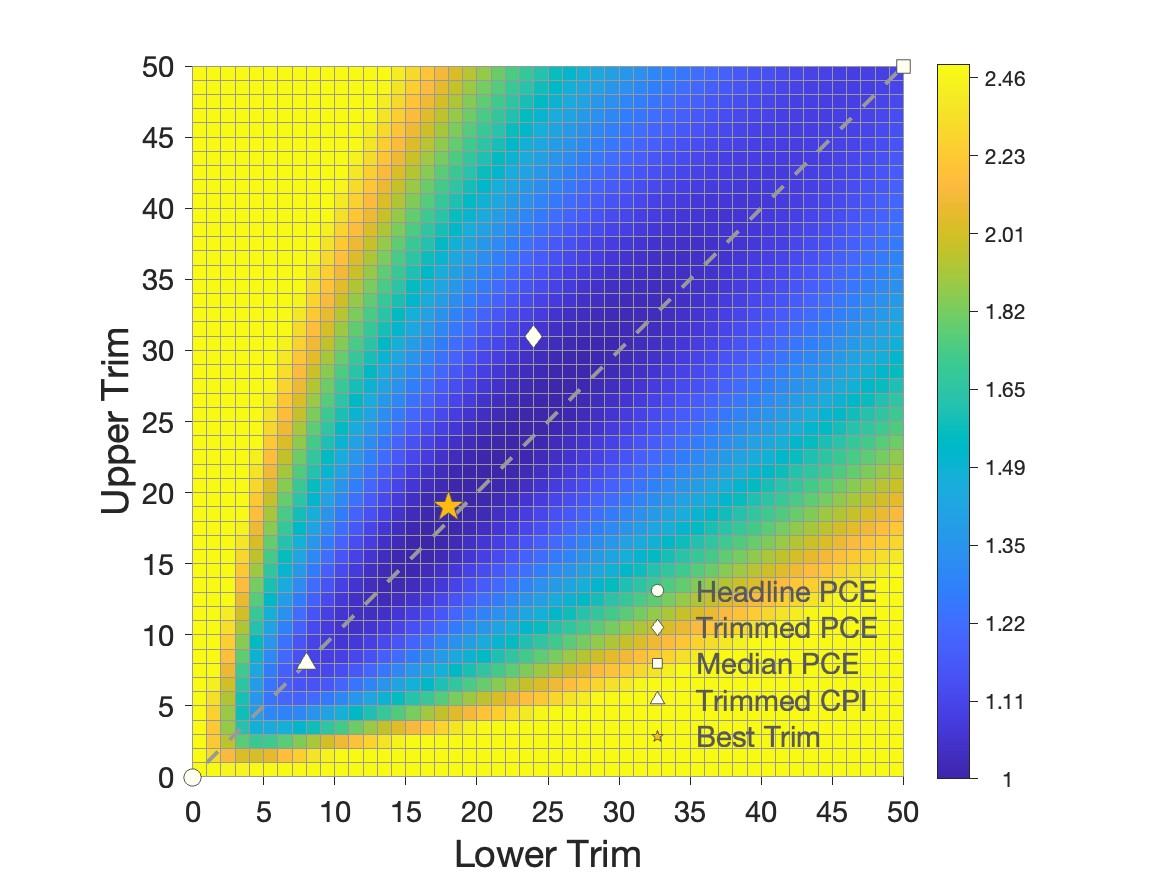}
    \end{subfigure}
    ~
    \begin{subfigure}[t]{0.48\textwidth}
    \caption{Prediction}
    \includegraphics[width=1.0\textwidth]{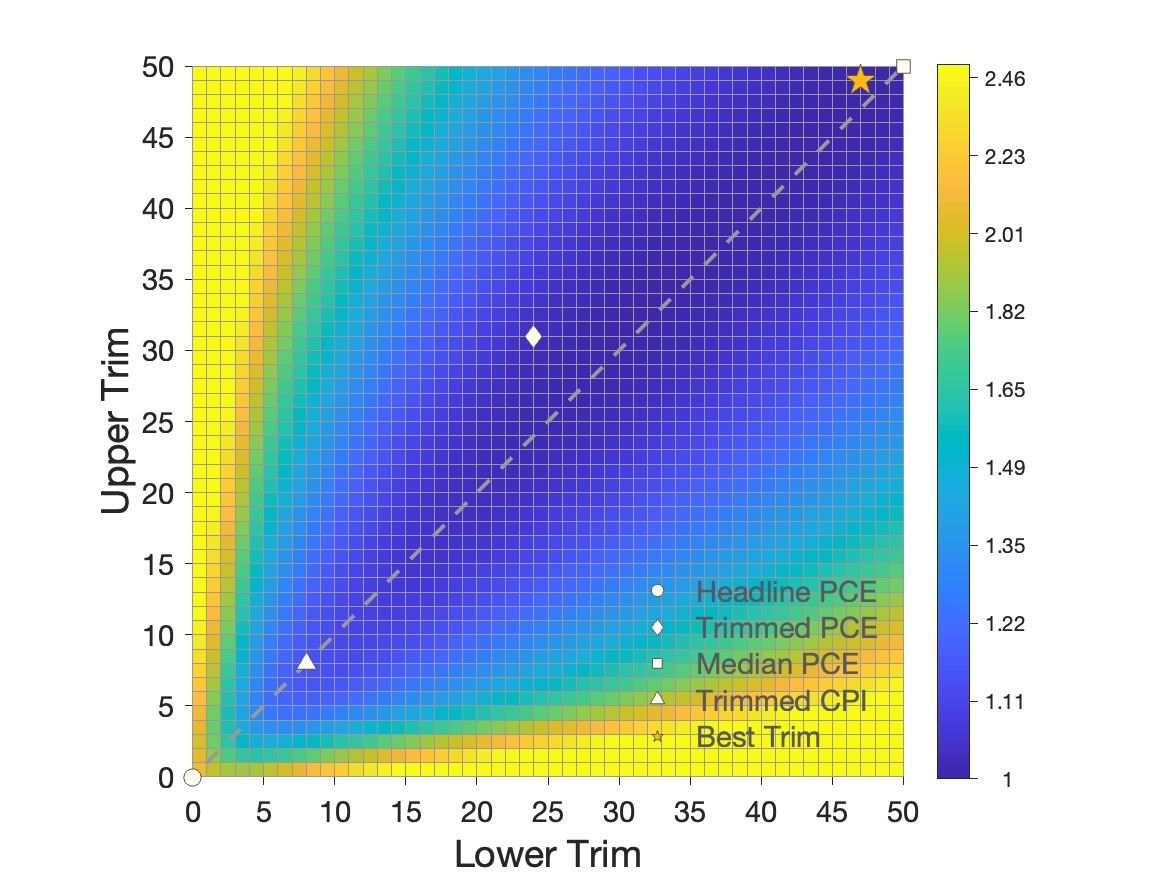}
    \end{subfigure}
    
    \end{centering}
    
    \vspace{-0.2cm}
    \singlespacing
    \textit{\footnotesize{}Note:}\footnotesize{} 
    The figures show heat maps of the RMSE when targeting trend inflation with different combinations of trimmed mean inflation measures.
    Four measures of trend inflation are considered. 
    To ensure comparability across plots, the RMSE numbers are reported relative to the RMSE of the best trim reported in Table \ref{tab:best_trim_housing}.
    
    \label{fig: RMSE_00s_housing}
\end{figure}

\begin{figure}
    \begin{centering}
        
    \caption{RMSE across Trims: 1970--89 (Sample without Housing)}
    
    \begin{subfigure}[t]{0.48\textwidth}
    \caption{Centered Trend}
    \includegraphics[width=1.0\textwidth]{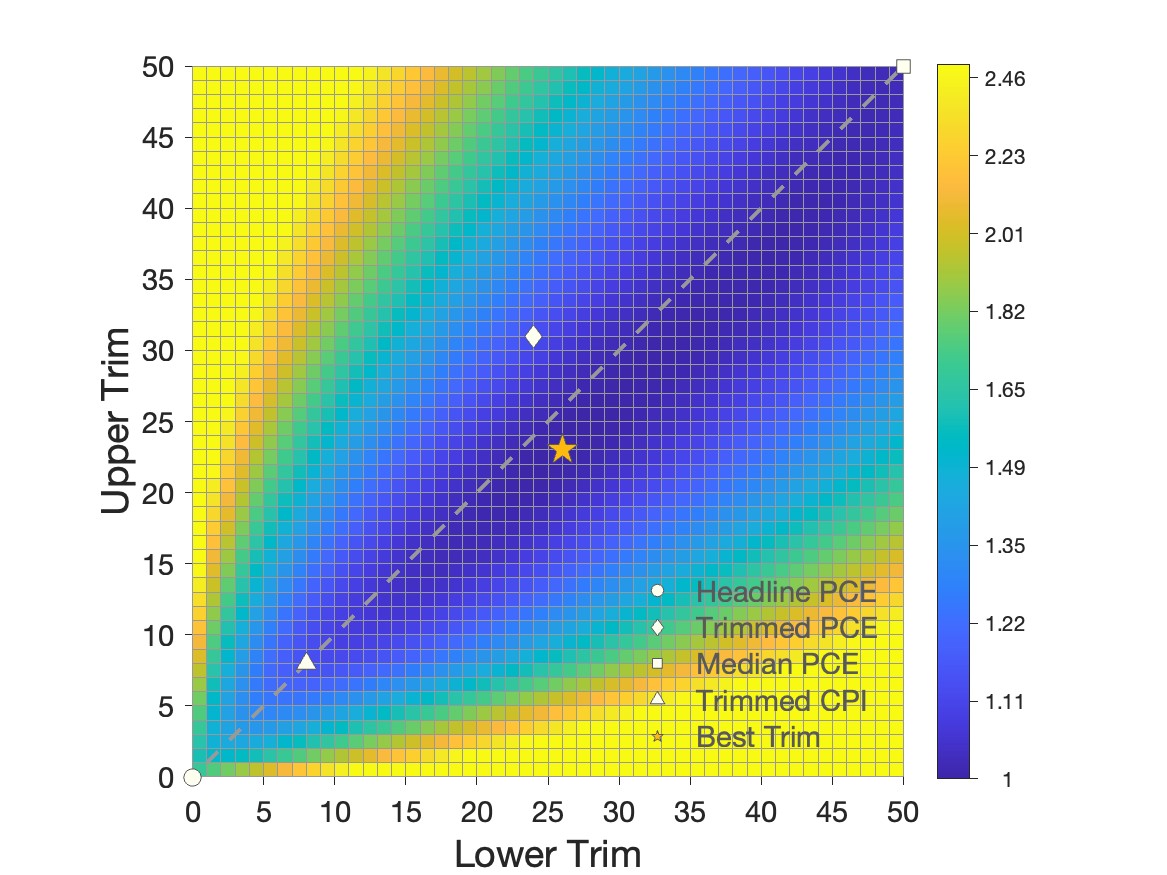}
    \end{subfigure}
    ~
    \begin{subfigure}[t]{0.48\textwidth}
    \caption{Forward Trend}
    \includegraphics[width=1.0\textwidth]{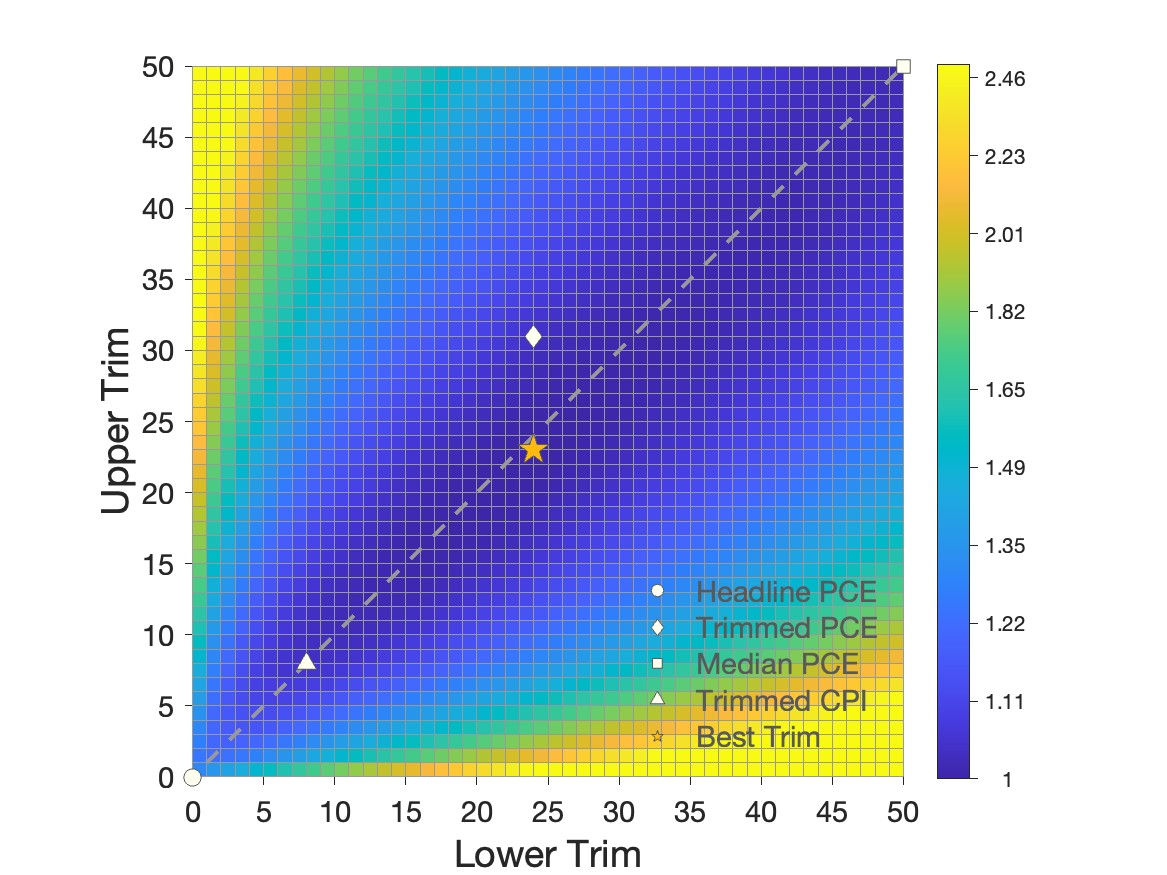}
    \end{subfigure}
    
    \begin{subfigure}[t]{0.48\textwidth}
    \caption{Band-Pass Trend}
    \includegraphics[width=1.0\textwidth]{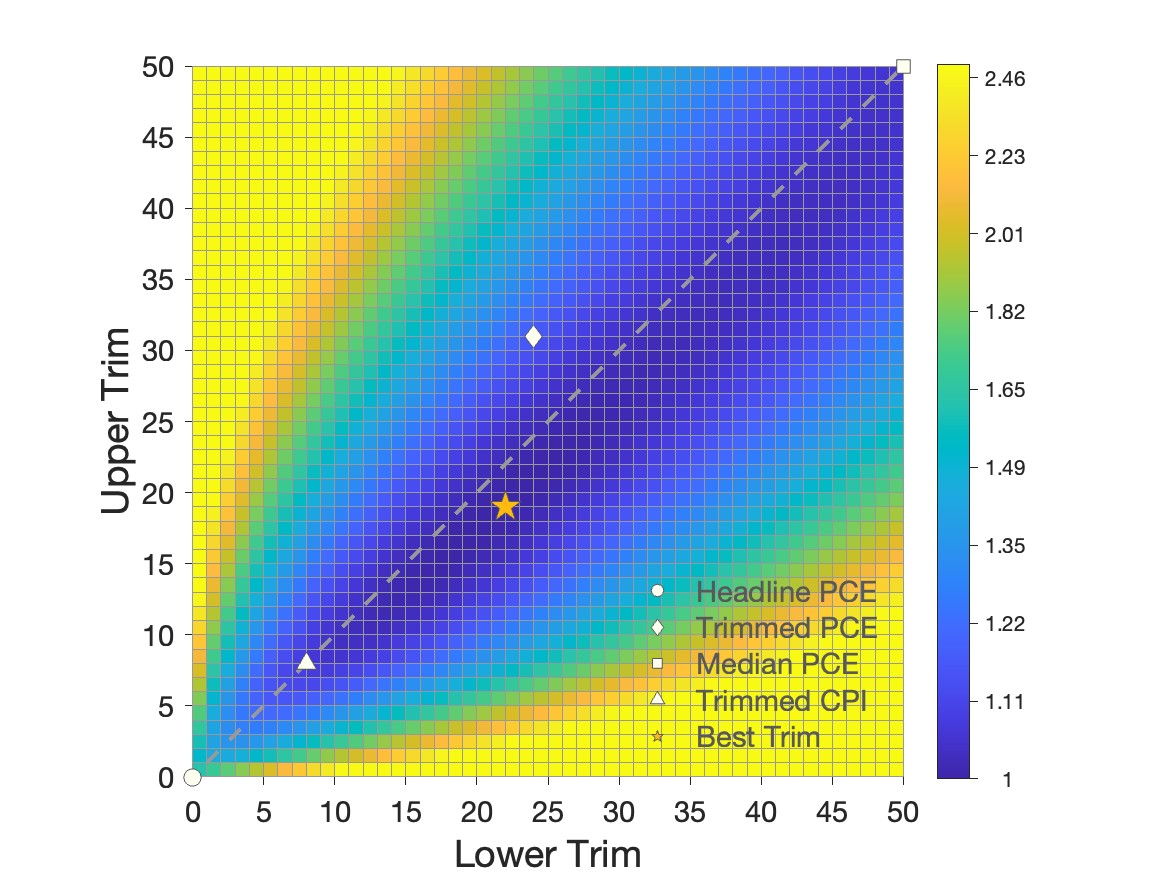}
    \end{subfigure}
    ~
    \begin{subfigure}[t]{0.48\textwidth}
    \caption{Prediction}
    \includegraphics[width=1.0\textwidth]{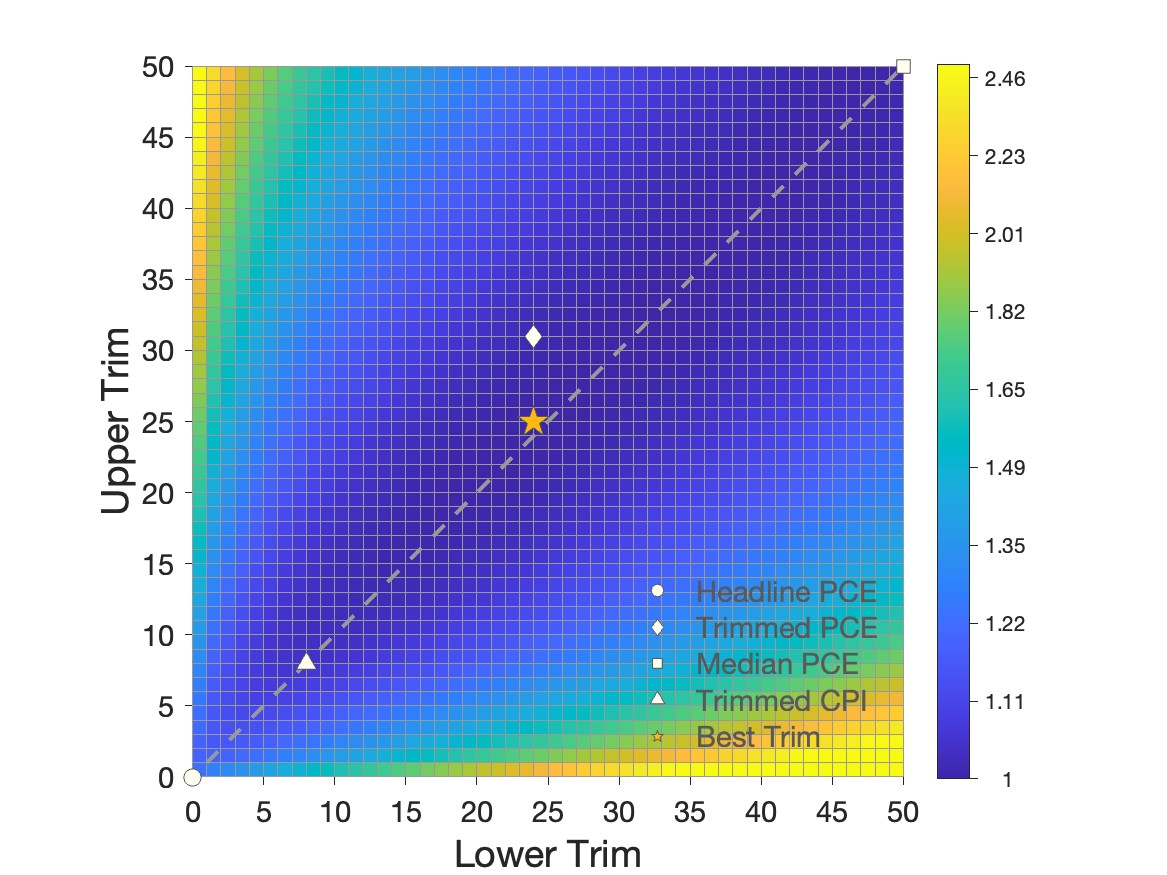}
    \end{subfigure}
    
    \end{centering}
    
    \vspace{-0.2cm}
    \singlespacing
    \textit{\footnotesize{}Note:}\footnotesize{} 
    The figures show heat maps of the RMSE when targeting trend inflation with different combinations of trimmed mean inflation measures.
    Four measures of trend inflation are considered. 
    To ensure comparability across plots, the RMSE numbers are reported relative to the RMSE of the best trim reported in Table \ref{tab:best_trim_housing}.
    
    \label{fig: RMSE_80s_housing}
\end{figure}